\documentclass[aps,prb,twocolumn,superscriptaddress,superscriptaddress]{revtex4-2}

\usepackage{braket}
\usepackage{amsmath}
\usepackage{amsfonts}       
\usepackage{graphicx}
\usepackage{dcolumn}
\usepackage{epsfig}
\usepackage{bm}
\usepackage{array}
\usepackage{hyperref}
\hypersetup{
colorlinks=true,
citecolor=blue,
linkcolor=blue,
urlcolor=blue,
}
\usepackage{xcolor}			

\usepackage{amssymb}
\usepackage{bbm}
\usepackage[english]{babel}
\usepackage{setspace}
\usepackage{wasysym}
\usepackage{adjustbox}
\usepackage{framed}
\usepackage{latexsym} 
\usepackage{multirow}    
\usepackage{booktabs}    
\usepackage{calc}        
\usepackage[normalem]{ulem}

\usepackage{calrsfs}								
\DeclareMathAlphabet{\pazocal}{OMS}{zplm}{m}{n}	

\definecolor{DeepSkyBlue}{RGB}{0,104,139}
\colorlet{MySky}{white!40!blue}
\colorlet{MyViolet}{red!45!blue}
\colorlet{MyBlue}{black!40!blue}
\colorlet{MyRed}{black!40!red}
\colorlet{MyOrange}{red!70!yellow}
\colorlet{MyGreen}{black!60!green}
\colorlet{MyBrown}{black!70!brown}
\colorlet{MyGray}{black!60!white}

\newcommand{\be}{\begin{equation}}
\newcommand{\ee}{\end{equation}}

\newcommand{\beq}{\begin{eqnarray}}
\newcommand{\eeq}{\end{eqnarray}}

\newcommand{\bracket}[2]{\langle#1|#2\rangle}

\UseRawInputEncoding

\begin{document}

\title{Strange correlators for topological quantum systems from bulk-boundary correspondence}

\author{Luca Lepori}
\email[correspondence at: ]{llepori81@gmail.com}
\affiliation{Dipartimento di Scienze Matematiche, Fisiche e Informatiche, Universit\`a  di Parma, Parco Area delle Scienze, 53/A, I-43124 Parma, Italy.}
\affiliation{INFN, Gruppo Collegato di Parma, Parco Area delle Scienze 7/A, 43124, Parma, Italy.}
\affiliation{QSTAR and INO-CNR, Largo Enrico Fermi 2, 50125 Firenze, Italy.}

\author{Michele  Burrello}
\affiliation{Center for Quantum Devices and Niels Bohr International Academy, Niels Bohr Institute, University of Copenhagen, DK–2100 Copenhagen, Denmark.}

\author{Andrea Trombettoni}
\affiliation{Dipartimento di Fisica, Universit\`a di Trieste, Strada Costiera 11, I-34151 Trieste, Italy.}
\affiliation{SISSA and INFN, Sezione di Trieste, Via Bonomea 265, I-34136 Trieste, Italy.}

\author{Simone Paganelli}
\affiliation{Dipartimento di Scienze Fisiche e Chimiche, Universit\`a dell'Aquila, via Vetoio,
I-67010 Coppito-L'Aquila, Italy.}

\begin{abstract}
  ``Strange" correlators provide a tool to detect  topological phases arising in many-body models by
  computing the matrix elements of suitably defined
  two-point correlations between the states under investigation and trivial reference states. Their effectiveness depends on the choice
  of the adopted operators.
  In this paper we give a systematic procedure 
  for this choice, 
  discussing the advantages of choosing 
  operators 
  using the bulk-boundary correspondence of the systems under scrutiny.
  Via the scaling exponents, we directly relate the algebraic decay of the strange correlators with the scaling dimensions of gapless edge modes operators. 
  We begin our analysis with 
  lattice models hosting symmetry-protected topological phases and 
  we analyze the sums of the strange correlators,
  pointing out that integrating
  their moduli substantially reduces cancellations and
  finite-size effects.
  We also analyze 
  instances of systems hosting intrinsic topological
  order,  as well as  strange correlators between states with different nontrivial topologies. 
  Our results for both translational and non-translational invariant cases, and in presence  of on-site disorder and long-range couplings, extend the validity of the strange correlators approach for the diagnosis of topological phases of matter, and indicate a general procedure for their optimal choice.
\end{abstract}

\maketitle

\section{Introduction}

In the last decades, topological phases of matter have been at the
heart of condensed matter and quantum many-body physics \cite{bernevigbook,kotetesbook,wen2019}.
Since the discovery of the quantum Hall effect, a 
plethora of paradigmatic topological models with different characteristics
has been proposed and studied to exemplify the variety of topological phenomena 
occurring in quantum many-body systems. The goal of this research line is to drive the technological design of 
quantum systems of increasing complexity, in the quest to realize
topological quantum features stable against disorder and noise.

Despite the 
progress in the study of topological states, their
identification in realistic scenarios is not in general an easy task.
For instance, 
physical setups display finite
size effects and disorder which may hinder the observation of the most common
topological features, especially when they are extrapolated from
numerical results obtained for small systems. 

The development of diagnostic tools for topological phases of matter is therefore an 
important task 
for the study of some of the fundamental features of quantum many-body
systems. 

Among the 
tools for the detection of the onset of topological
states, the so-called \textit{strange} correlators \cite{xu2014} have proved especially useful for theoretical and numerical investigations \cite{sengupta2014,sengupta2014bis,lu2015,beach2016,scaffidi2016,takayoshi2016,lu2016,luo2017,bultinck2018,chen2020,ellison2021,vanhove2021,xu2022,zhou2022}; they follow an approach which, in spirit, is related to the characterization of off-diagonal long range order (ODLRO) \cite{penrose1956}. In systems with ODLRO the one-body density matrix is determined from the correlation functions, evaluated in the ground-state or at equilibrium at finite temperature or more generally in a given state. Once the one-body density matrix is known, in translationally invariant systems one can find whether the system exhibits ODLRO by examining how the one-body density matrix decays as a function of the distance \cite{penrose1956,colcelli2018}. In absence of translational invariance, one has to determine whether the largest eigenvalue of the one-body density matrix scales with the number of particles, and if it does so, then there is ODLRO.

In this work, we will show that strange correlators generalize the notion of two-point correlation functions and one-body density matrix and provide similar scaling features to detect topological states of matter belonging to phases with protected gapless edge modes. 

Let us label by $\ket{\Psi}$ a state that one desires to understand whether it is topological or not. Strange correlators, denoted by $s$, are defined by estimating suitable two-point matrix elements between  $\ket{\Psi}$ and {\color{black} a topologically trivial state} $\ket{\Omega}$. $\ket{\Omega}$ is used as reference
on the same Hilbert space, and must be carefully chosen based on the system symmetries. The strange correlators are defined by:
\beq
s[ \hat{o} , \hat{o}^{\prime}]_{{\bf r}, {\bf r^{\prime}}} \equiv \frac{\bra{\Omega} \hat{o} ({\bf r}) \hat{o}^{\prime}({\bf r^{\prime}}) \ket{\Psi}}{\bra{\Omega} \Psi \rangle} \,,
\label{dec}
\eeq
where we explicitly evidenced the dependence on the chosen operators
$\hat{o}, \hat{o}^{\prime}$, whereas the dependence on the states $\ket{\Psi}$ and $\ket{\Omega}$ is left implicit. These are two-point functions taken in the bulk of the system, such that, differently from other diagnostic tools for topological phases, they do not rely on the presence of physical boundaries or peculiar boundary conditions.\\
\indent Eq. (\ref{dec}) makes clear the difference with the ODLRO criterion, which is instead based on
expectation values
of the form $g^{(1)} \equiv \bra{\Psi} \hat{o}^\dag ({\bf r}) \hat{o} ({\bf r^{\prime}}) \ket{\Psi}$. To stress the parallel with the strange correlators \eqref{dec}, we observe that $g^{(1)}=s[ \hat{o}^\dag , \hat{o}]_{{\bf r}, {\bf r^{\prime}}}$ when $\ket{\Omega} = \ket{\Psi}$: if $g^{(1)}$ converges to a constant for large distances $\left|{\bf r}-{\bf r^{\prime}}\right|$ in translationally invariant systems, then there is ODLRO; when instead $g^{(1)}$ decays as a power law, then there is quasi-ODLRO, and when it decays faster than a power law, the system displays neither.\\
\indent Similarly, the main property of the strange correlators $s$ is that, for short-range entangled topological phases of matter \cite{chen2013}, they display a power law decay (or saturate to a finite value) as a function of the distance when $\ket{\Psi}$ is a topological state. On the contrary, if both $\ket{\Psi}$ and $\ket{\Omega}$ are trivial states, all the strange correlators $s$ decay exponentially in $\left|{\bf r}-{\bf r^{\prime}}\right|$. In Ref. \cite{xu2014}, it was argued that this behavior holds for several topological phases in one and two spatial dimensions. The discussed phases include both short-range entangled states whose topological features give rise to protected gapless modes at their boundaries, and a large class of symmetry-protected bosonic phases of matter whose wavefunctions can be derived by suitable Wess-Zumino-Witten models \cite{bi2013,senthil2013}. In these systems, the power law decay of the strange correlators can be qualitatively understood by means of their mapping into the two-point correlation function of a gapless model in a system with a lower space dimension \cite{xu2014}. When the system under investigation is two-dimensional, this mapping translates the strange correlators in the bulk into suitable correlation functions of the corresponding (1+1)-dimensional conformal field theory at the boundary.

In this work, we address strange correlators  
in one-, two-, and three-dimensional systems, 
including also long-range couplings, finite size and boundary condition effects, as well as on-site disorder.

We discuss the bulk-boundary correspondence at the basis of their scaling properties and we show that the optimal choice of the operators $\hat{o}$ to use strange correlators as a diagnostic tool for topological phases is related to the creation of particle-hole excitations in the bulk of the related systems. For the sake of clarity, we decided to first present and discuss in detail
  some concrete examples, and then to extend our observations by presenting
  a general procedure: for strange correlators used to detect topological phases,
  we propose to systematically exploit the information about the boundary theory
  to implement a correct choice of the operators entering the strange correlators.
  More precisely, in Section \ref{choice} we discuss the following points:
  {\it a)} For non-interacting systems, the optimal operators entering the strange correlators
  must have overlaps with the bulk operators describing both filled and empty bands. In particular, the strange correlators can be decomposed in terms corresponding to the creation of particle-hole excitations over the ket ground state. {\it b)} For a general –- possibly interacting –- theory,
  one has to know the boundary theory for the edge modes. 
  Once the  latter is known, and the operators creating the edge modes have been identified as well, one has to determine the corresponding lattice operators. {\it c)} The operators must be then  continuously connected to bulk operators.

We discuss how to efficiently extract information from strange correlators, studying their summation, possibly taken in modulo. 
These sums provide a more efficient diagnosis than the decay of the single strange correlator even for the small system sizes
typically achievable in numerical simulations. Furthermore, they are more efficient also in systems with disorder, as in many cases relevant for experiments, where 
distinguishing the power-law from the exponential decays may be difficult.
We show that the finite-size scaling of suitable strange correlators allows us to identify the anomalous scaling dimensions of the operators related to the protected topological massless edge modes of the systems under analysis. Finally, generalized strange correlators between states with different nontrivial topologies are addressed.
Our findings 
allow us to extend the validity of the strange correlators approach,
leading to a general strategy for their optimal choice and numerical evaluation.

In the next section we present the main scaling features of strange correlators, then, in Sec. \ref{sec:transl} we discuss several paradigmatic examples of topological superconductors and insulators based on systems with translational invariance. Sec. \ref{sec:dis} numerically verifies the robustness of the scaling of strange correlators in the presence of disorder for two-dimensional Chern insulators. In Sec. \ref{choice} we discuss the relation between the bulk operators $\hat{o}({\bf r}),\hat{o}^{\prime}({\bf r^{\prime}})$, the boundary modes of the topological phases under analysis and their particle-hole excitations; Sec. \ref{LRE} extends the results obtained for symmetry-protected topological phases to the case of fractional quantum Hall states with intrinsic topological order. The appendices provide detailed calculations for several of the presented examples and discuss the negative results obtained for the Kitaev surface code, which exemplifies a system with intrinsic topological order without protected gapless edge modes.

\section{Scaling of the sums of strange correlators}
\label{scalings}

To enhance the visibility of the scaling of the strange correlators with the finite size of the considered system, we introduce the following functions of the linear lattice size $L$:
\begin{align}
&{S}[\hat{o} , \hat{o}^{\prime}]_L = \frac{1}{L^d} \,  \sum_{{\bf r},{\bf r^{\prime}}} \, s[ \hat{o} , \hat{o}^{\prime}]_{{\bf r}, {\bf r^{\prime}}}\,, \\
&\bar{S}[ \hat{o} , \hat{o}^{\prime}]_L = \frac{1}{L^d} \,  \sum_{{\bf r},{\bf r^{\prime}}} \, \left|s[ \hat{o} , \hat{o}^{\prime}]_{{\bf r}, {\bf r^{\prime}}}\right|\, ,
\label{sumstrange}
\end{align}
 where $d$ is the dimension of the lattice, so that the number of sites scales as $\sim L^d$.
These functions constitute respectively the sums of the strange correlators $s[ \hat{o} , \hat{o}^{\prime}]_{{\bf r}, {\bf r^{\prime}}}$ and of their moduli $|s [\hat{o} , \hat{o}^{\prime}]_{{\bf r}, {\bf r^{\prime}}}|$ over the whole system under analysis, rescaled by the volume.
In the following, we will consider, in particular, lattice models, such that ${\bf r}$ and ${\bf r^{\prime}}$ label the lattice site vectors, and $d$ is the lattice space dimension. 

A scaling result that we will exploit in the following, inspired by the Penrose-Onsager criterion for the ODLRO \cite{penrose1956}, is that, when the modulus of the strange correlator, $|s[ \hat{o} , \hat{o}^{\prime}]_{{\bf r}, {\bf r^{\prime}}}|$ decays asymptotically as $|{\bf r} - {\bf r^{\prime}}|^{-2\alpha}$, for large system sizes the sum in Eq. \eqref{sumstrange} behaves as:
\beq
\bar{S}[ \hat{o} , \hat{o}^{\prime}]_L \sim \left\{
\begin{array}{cc}
 L^{d-2  \alpha}    &  \mathrm{for} \, \alpha<\frac{d}{2} \\
 \ln L    & \mathrm{for}\, \alpha=\frac{d}{2}  \\
 \mathrm{const} + O(L^{d-2\alpha})\, & \mathrm{for}\, \alpha>\frac{d}{2} \, .
\end{array}
\right.\,
\label{qfiscal}
\eeq 
The saturation to an asymptotic constant characterizes also the case in which, instead, $s[ \hat{o} , \hat{o}^{\prime}]_{{\bf r}, {\bf r^{\prime}}}$ decays exponentially. However, in this situation, the constant is reached exponentially rather than algebraically, see App. \ref{appscal} for more details.
The previous scaling behavior does not hold, in general, for $S[ \hat{o} , \hat{o}^{\prime}]_L$. Without the modulus, indeed, $s[ \hat{o} , \hat{o}^{\prime}]_{{\bf r}, {\bf r^{\prime}}}$ may display non-monotonic (typically oscillating) character: its envelope may still decay as $|{\bf r} - {\bf r^{\prime}}|^{-2\alpha}$, but the non-monotonic behavior yields a faster decay of $S[ \hat{o} , \hat{o}^{\prime}]_L$ and complicates the derivation of formulas similar to \eqref{qfiscal}. 
Such a complication is not present for the standard case of repulsively interacting bosons with ODLRO, due to the non oscillating nature of their one-body density matrix.
In general, the relations in Eq. \eqref{qfiscal} are valid for $S[ \hat{o} , \hat{o}^{\prime}]_L$ if $s[ \hat{o} , \hat{o}^{\prime}]_{{\bf r}, {\bf r^{\prime}}}$ is monotonic.

In the traditional ODLRO criteria, one performs the integral of the one-body density matrix, which is (in translational invariant systems) nothing but the momentum distribution at ${\bf k}=0$, which
gives the indication of the presence of ODLRO or, which is the same, Bose-Einstein condensation \cite{PS}. The advantage of integrating the correlation functions is that it makes more convenient to detect ODLRO, since their non-vanishing value for large distances is multiplied by the volume when they are integrated. That is the reason why the first signature of ODLRO in Bose-Einstein condensation has been obtaining by measuring the peak at zero momentum of the momentum distribution \cite{PS}. 
In the case of standard one-body density matrices in superfluid bosonic systems, the decay of the two-point function $g^{(1)}$ can be expressed in terms of the anomalous dimension critical exponent associated with the considered bulk operators $\hat{o}$ \cite{colcelli2018}.
As a consequence, this exponent determines the scaling of the largest eigenvalue of the one-body density matrix on the state $\ket{\Psi}$ with the particle number or the size of the system \cite{colcelli2018}.

In the case of strange correlators,  when $\ket{\Omega}$ is a topologically trivial state and $\ket{\Psi}$ lies in a topological phase characterized by gapless edge modes, $s[ \hat{o} , \hat{o}]_{{\bf r}, {\bf r^{\prime}}}$ can be associated to the two-point correlation function of suitable operators of the theory describing the boundary between the two phases, through a suitable space-time rotation. The exponent $\alpha$ is then associated to the anomalous scaling dimension of a boundary operator related to $\hat{o}$ through bulk-boundary correspondence, as we will discuss in Sec. \ref{choice}. For instance, in the case of non-interacting topological insulators with gapless Dirac modes on their boundaries, when $\hat{o}$ is a free fermionic field, $\alpha = (d-1)/2$, as expected by the corresponding scaling dimensions.
If, instead, $\ket{\Psi}$ and $\ket{\Omega}$ belong to the same phase, no gapless boundary separates them in the rotated space-time picture, and the behavior of $s[\hat{o} , \hat{o}^{\prime}]_{{\bf r}, {\bf r^{\prime}}}$ mimics the corresponding two-point correlation functions in the related gapped phase. In this case, $s[\hat{o} , \hat{o}^{\prime}]_{{\bf r}, {\bf r^{\prime}}} \propto \exp\left[-|{\bf r} - {\bf r^{\prime}}|/\xi\right]$ decays exponentially for large distances, such that the sum $\bar{S}[ \hat{o} , \hat{o}^{\prime}]_L$ exponentially converges to a constant.

\section{Translationally invariant models}
\label{sec:transl}
In this Section, we analyze several lattices hosting nontrivial symmetry-protected topological phases, in the case of translational invariant systems, such that we can derive exact results in the momentum space basis. 
Notably, in this condition, the sum (of the moduli) of the strange correlators, $S \, (\bar{S})[ \hat{o} , \hat{o}^{\prime}]_{{\bf r}, {\bf r^{\prime}}}$ is equal to the
largest eigenvalue $\lambda_0$ of the matrix $s \, (|s|) [ \hat{o} , \hat{o}^{\prime}]_{{\bf r}, {\bf r^{\prime}}} $ in Eq. \eqref{dec}, for varying ${\bf r}$ and ${\bf r^{\prime}}$. This fact is a rephrased version of the famous Penrose-Onsager criterion for the Bose-Einstein condensation \cite{penrose1956,huangbook}, here applied to topological quantum systems. In Section \ref{choice}, we will elaborate more on this point.

\subsection{The Kitaev chain}
\label{kitaevsec}
We analyze first the paradigmatic model for one-dimensional topological superconductivity, namely, the Kitaev chain \cite{kitaevorig}. Its Hamiltonian is given by: 
\begin{equation}
\begin{array}{c}
H_{\mathrm{Kit}} = -\frac{t}{2}\sum_x\left(c^\dag_x c_{x+1} + c^\dag_{x+1}c_x\right) - \\ 
{}\\
- \mu \sum_x c_x^{\dagger} c_x - \frac{\Delta}{2}\sum_x\left(c^\dag_x c^\dag_{x+1} + c_{x+1}c_x\right) \, ,
\end{array}
\label{kitaev}
\end{equation}
where $x$ labels the sites of the chain. The model in Eq. \eqref{kitaev} represents a one-dimensional p-wave superconductor with pairing $\Delta$. The parameter $\mu$ is the chemical potential of the system and $t>0$ represents a hopping amplitude along the chain. The model is characterized by particle-hole symmetry, and it belongs to the D class of the 
"tenfold-way" classification of the topological insulators and superconductors \cite{altland1997,ludwig2009,bernevigbook}. In particular, the Kitaev chain displays a topological phase for $|\mu|<t$ and a trivial phase for $|\mu| > t$. For open boundary conditions, the topological phase is characterized by the well-known zero energy Majorana modes localized at its edges. Correspondingly, the system displays two degenerate ground-states. These ground-states, that we label by $\ket{\Psi_+}$ and $\ket{\Psi_-}$, are distinguished by their total fermionic parity. The transport properties of the chain are affected qualitatively by these modes, therefore transport is a useful diagnostic for topology, see for instance the recent reviews \cite{Zhang2019,Prada2020} and the analysis in Ref. \cite{giuliano2018,Chung2020}, concerning the introduction of long-range pairings and interactions.

For superconducting systems in the BCS mean-field description, such as $H_{\mathrm{Kit}}$, the conservation of the particle number is violated by the pairing term. This causes arbitrariness in the choice of $\hat{o},\hat{o}'$ in the strange correlators: several choices are possible, with the constraint that the total fermionic parity must be preserved by $\hat{o} \hat{o}'$.

In order to identify the most efficient choice of the strange correlators, for a reliable detection of the topological phase transition in this model, it is instructive to consider first the limit $\mu \to 0$, with open boundary conditions, which constitute the paradigmatic example for the topological phase. In this limit, the two ground-states are expressed in terms of equal-amplitude linear superpositions of all the possible occupied Cooper pair states \cite{iemini2015}:
\begin{align}
&\ket{\Psi_+} = \mathcal{N}^{-1/2}\sum_n {(-1)^n} \sum_{{\bf j}_{2n}}\ket{{\bf j}_{2n}}\,, \label{psiplus}\\
&\ket{\Psi_-} = \mathcal{N}^{-1/2}\sum_n {(-1)^n} \sum_{{\bf j}_{2n+1}}\ket{{\bf j}_{2n+1}} \, ,
\end{align}
where ${\bf j}_m$ is an ordered $m$-plets of sites, such that $\ket{{\bf j}_{m}}=c^\dag_{j_1}c^\dag_{j_2}\ldots c^\dag_{j_m}\ket{0}$, and the norm is $\mathcal{N} = \sum_n \begin{pmatrix} L \\ 2n \end{pmatrix}= \sum_n \begin{pmatrix} L \\ 2n +1\end{pmatrix}=2^{L-1}$. In this situation, it is straightforward to study the strange correlators associated with the structure of the Cooper pairs. In particular, for the even ground-state, we consider:
\begin{equation}
s[ c , c]_{x,y} = \frac{\bra{\Omega} c_{x}c_{y} \ket{\Psi_+}}{\bra{\Omega}\Psi_+\rangle} \, ,
\label{oks}
\end{equation}
where the Fock vacuum state, corresponding to the ground-state in the (trivial) $\mu\to-\infty$ limit of Eq. \eqref{kitaev}, can be adopted as the trivial reference state $\ket{\Omega}$. For this specific choice of $\ket{\Omega}$, $s$ in Eq. \eqref{oks} is a constant, independent on $x$ and $y$ (when we restrict to $y>x$): the operator $c_xc_y$ selects indeed the components of $\ket{\Psi_+}$ with two particles only, and their amplitude is independent on their distance, given the construction in Eq. \eqref{psiplus}. Similar results can be obtained for $\ket{\Psi_-}$,  replacing $\ket{\Omega}$ with a generic single-particle state. This constant behavior of the strange correlator in Eq.  \eqref{oks} constitutes a strong indication of the topological character of the state, and it suggests that the optimal choice for the operators $o$ and $o'$ must include terms probing the Cooper pair spatial profile.\\
\indent The systematic and general calculation of the strange correlators of the Kitaev chain can be performed in the translational invariant case. In particular, we consider a periodic chain with $L$ sites and antiperiodic boundary conditions, $c_x = -c_{x+L}$. 
The Kitaev Hamiltonian in Eq. \eqref{kitaev} assumes in momentum space the Bogoliubov - de Gennes form: 
\begin{equation} \label{Hkitmom}
H_{\mathrm{Kit}}= \sum_{0\le k < \pi} (c^\dag_k,c_{-k}) \left[\left(\mu -t\cos k\right) \tau_z -\left(\Delta \sin k\right)\tau_y\right] \begin{pmatrix} c_k \\ c^\dag_{-k}\end{pmatrix},
\end{equation}
where $\tau_j$ are the Pauli matrices in the Nambu-Gorkov formalism. 
The Bogoliubov energies are:
\begin{equation}
E_{\pm}(k) = \pm \sqrt{\left(\mu -t\cos k\right)^2 + \Delta^2\sin^2k} \, .
\end{equation}
The positive energy eigenmodes can be written in the form
\begin{equation}
 \label{eigenmodes}
\eta_k = u_k c_k - v_{k} c^\dag_{-k}\,,
\end{equation}
with normalized coefficients
\beq
\begin{array}{c}
u_k = \cos\frac{\theta(k)}{2} \, , \quad \quad v_{k}= -i \sin\frac{\theta(k)}{2}\, ,\\
{}\\
 \quad \quad \theta(k) = \arctan \Big[\frac{-\Delta \sin k}{\mu -t\cos k} \Big] \,,
\end{array}
\label{Kitaevtheta}
\eeq
where $\theta(k)$ is meant in the interval $\left(-\pi,\pi\right]$ and its winding number across the same interval distinguishes topological and trivial phase.
The general normalized BCS ground-state is derived by the condition $\eta_k \ket{GS}=0$ and it reads:
\begin{equation} \label{Gauss}
\ket{GS} = \prod_{0\le k < \pi} \left[u_{k} + v_{k}c^\dag_k c^\dag_{-k} \right]\ket{0} = \mathcal{N} \prod_{0\le k < \pi} e^{g_k c^\dag_k c^\dag_{-k}}\ket{0},
\end{equation}
where $\ket{0}$ is the Fock vacuum state and: 
\begin{equation}
g_k = u_k^{-1}v_k\,, \qquad \mathcal{N} = \prod_{0\le k < \pi} u_{k}\,. 
\end{equation}

Starting from the equations above, the strange correlators can be evaluated (see App. \ref{app:Kitaev} for a detailed derivation).
The overlap between two generic normalized ground-states $\ket{\Omega}$ and $\ket{\Psi}$ of the Kitaev chain reads:
\begin{equation} \label{kitaevoverlap}
\bracket{\Omega}{\Psi}= \prod_{0\le k < \pi} \left[\cos\frac{\theta_\Psi(k) - \theta_\Omega(k)}{2} \right],
\end{equation}
where $\theta_{\Psi/\Omega}$ labels the $\theta$ parameters in Eq. \eqref{Kitaevtheta} for the BCS ground-states $\ket{\Psi}$ and $\ket{\Omega}$ respectively.
The simplest quadratic strange correlators result:
\begin{align} 
&\frac{\bra{\Omega}c^\dag_x c_y \ket{\Psi}}{\bracket{\Omega}{\Psi}} = \sum_{q>0} \frac{2\cos q(x-y) \sin \frac{\theta_\Psi(q)}{2}\sin\frac{\theta_\Omega(q)}{2}}{L\cos \frac{\theta_\Psi(q)-\theta_\Omega(q)}{2}}\,,
\label{Kcdc}\\
&\frac{\bra{\Omega}c^\dag_x c^\dag_y \ket{\Psi}}{\bracket{\Omega}{\Psi}} =  -\sum_{q>0} \frac{2\sin q(x-y) \cos \frac{\theta_\Psi(q)}{2}\sin\frac{\theta_\Omega(q)}{2}}{L \cos \frac{\theta_\Psi(q)-\theta_\Omega(q)}{2}}\,.
\label{Kcdcd}
\end{align}
Similar results can be obtained also for $c_xc_y$. Moreover, analogous expressions of Eqs. \eqref{Kcdc} and \eqref{Kcdcd} in real space are given in App. \ref{app:BCS}. 

\begin{figure*}[t]
\includegraphics[width=5.5cm]{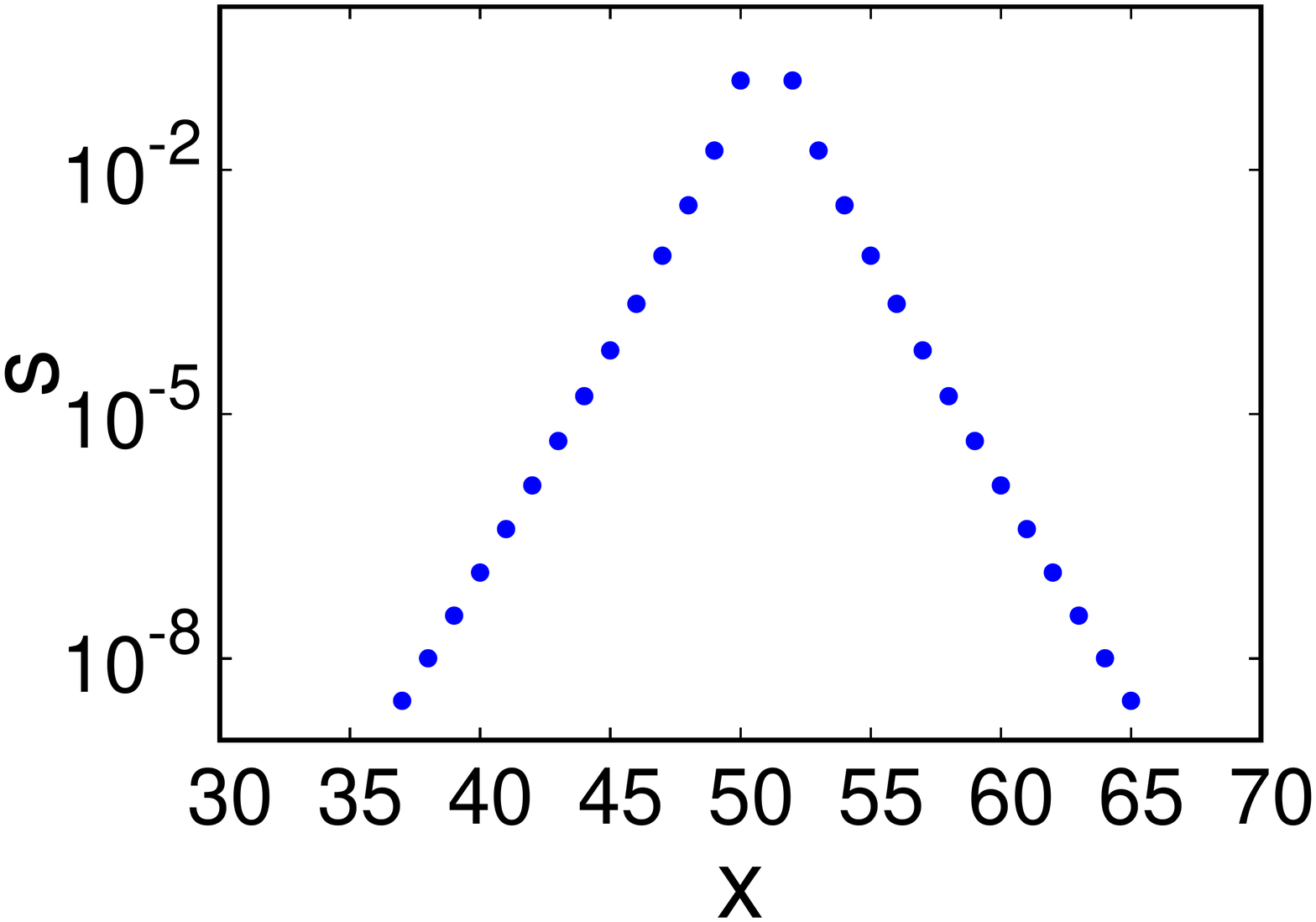}
\llap{\parbox[b]{1.9cm}{\color{black}(a)\\\rule{0ex}{3.2cm}}}
\includegraphics[width=5.5cm]{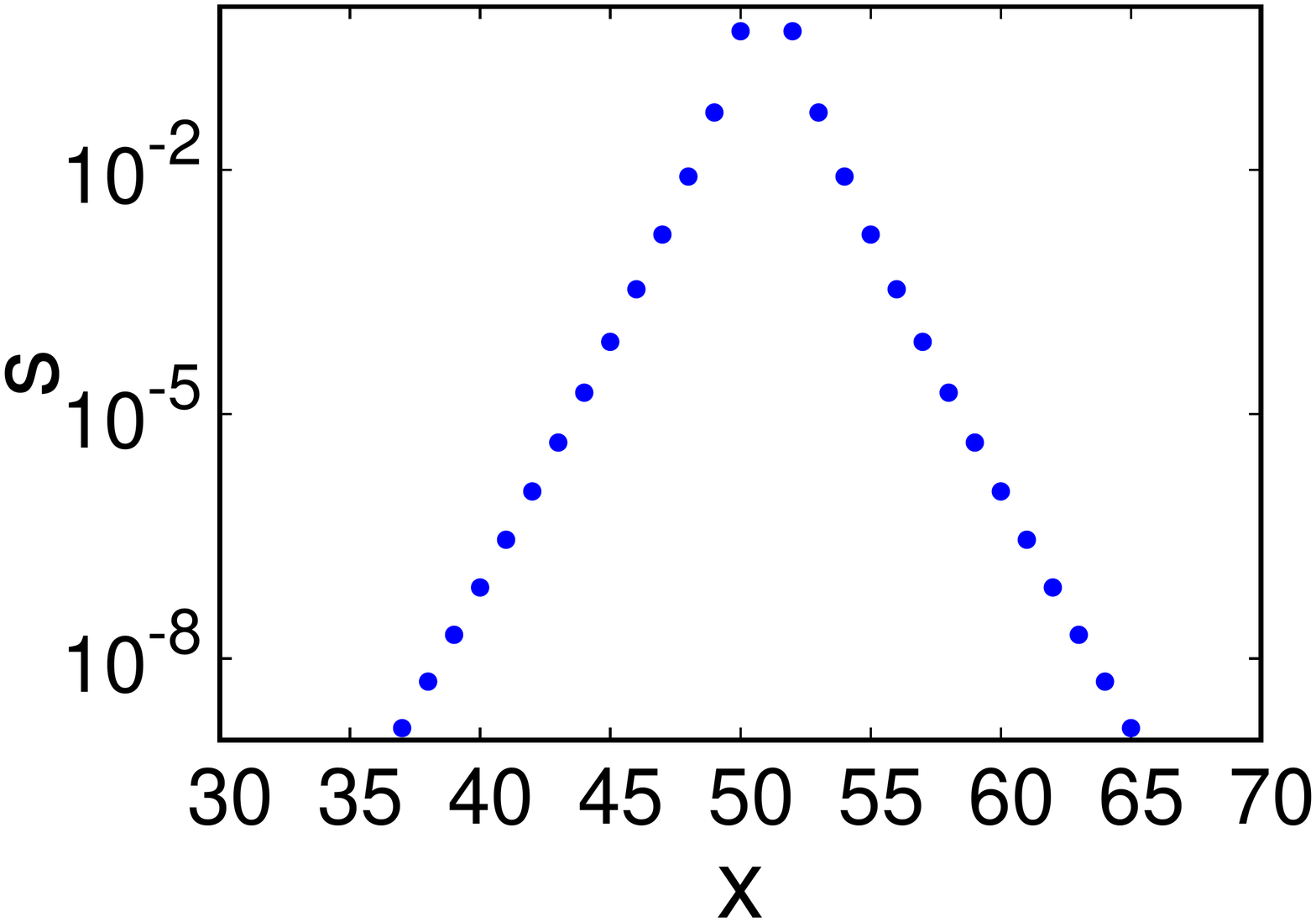}
\llap{\parbox[b]{1.9cm}{\color{black}(b)\\\rule{0ex}{3.2cm}}}
\includegraphics[width=5.5cm]{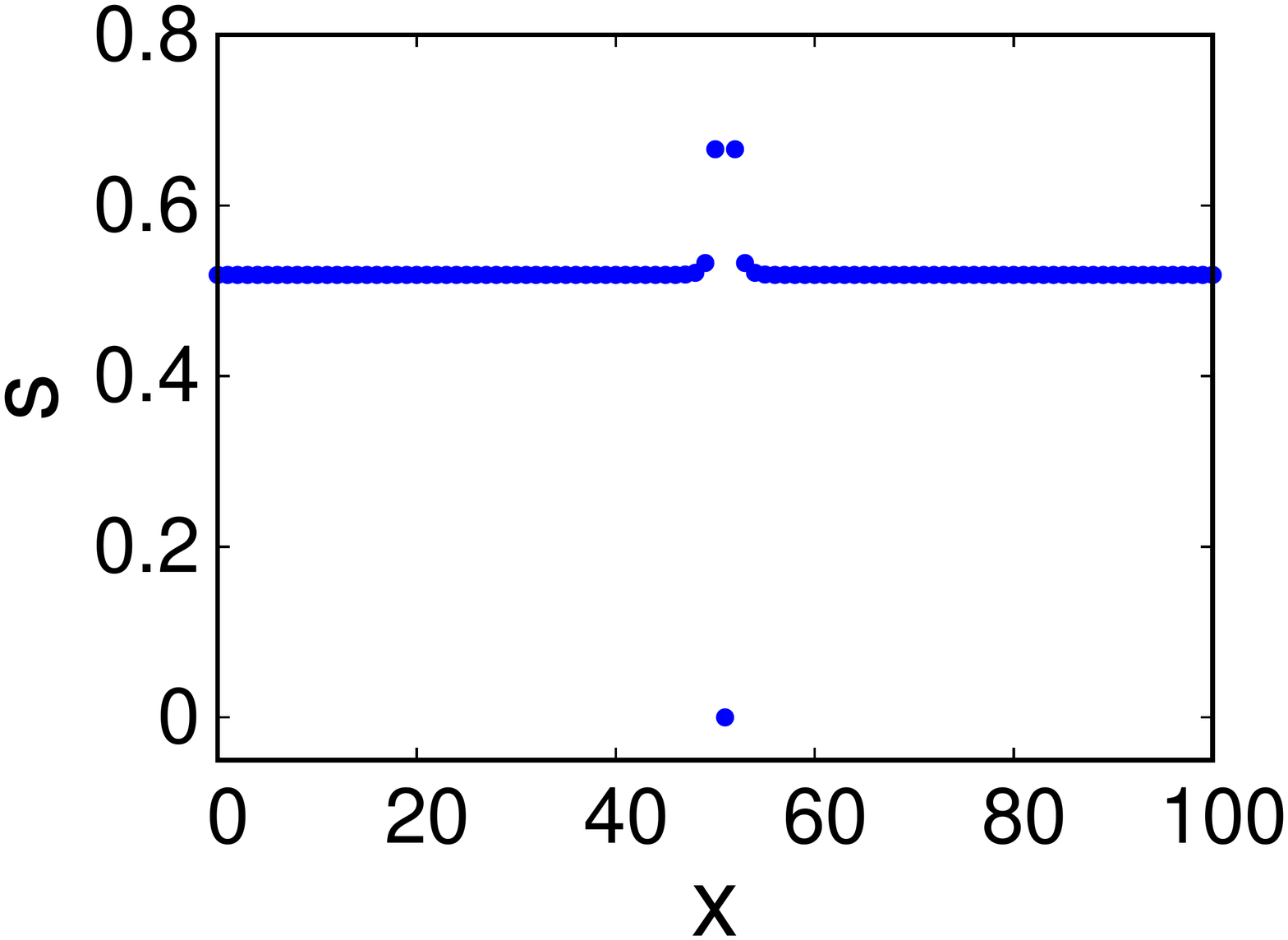}
\llap{\parbox[b]{1.9cm}{\color{black}(c)\\\rule{0ex}{3.2cm}}}
\caption{Majorana strange correlators $|s\left[\gamma,\gamma \right]_{x,y}|$ in Kitaev chains with antiperiodic boundary conditions of length $L=101$ and $\Delta=t$ plotted as a function of $x$ for $y=51$.  (a) Both ground-states $\ket{\Omega}$ and $\ket{\Psi}$ are taken in the trivial phase with $\mu_\Psi/t=3$ and $\mu_\Omega/t=6$. The pair strange correlator decays exponentially as expected. (b) Both the ground-states are taken in the topological phase with $\mu_\Psi/t=0.1$ and $\mu_\Omega/t=0.3$; the pair strange correlator decays exponentially also in this case. (c) Pair strange correlator for ground-states in different phases with $\mu_\Psi/t=0.3$ and $\mu_\Omega/t=3$: in this case $|s\left[\gamma,\gamma \right]_{x,y}|$ does not decay to zero and displays constant behavior for $x\neq y$.}
\label{fig:strangekitaevp}
\end{figure*}

The previous strange correlators, however, do not provide a satisfactory diagnosis of the topological phases of the model in the generic case: Eq. \eqref{Kcdc} does not display a qualitatively different behavior to distinguish when $\ket{\Omega}$ and $\ket{\Psi}$ are in the same or different topological phases; it typically decays exponentially and it may present staggering; Eq. \eqref{Kcdcd}, instead,  generally breaks the symmetry between $\ket{\Omega}$ and $\ket{\Psi}$ and, for instance, it vanishes when $\ket{\Omega}$ is the Fock vacuum state, thus it is not reliable for a full diagnosis of the system phases. 

To overcome these limitations, we consider a more symmetric strange correlator, which is inspired by the onset of Majorana zero modes at the edges of the Kitaev chain with open boundaries. 
 Indeed,  the optimal choice for the strange operator approach is identified by bulk-boundary correspondence: since the interfaces between trivial and topological domains of the Kitaev chains are characterized by protected Majorana modes, the boundary theory suggests to take 
$\hat{o}=c+c^\dag$ and  $\hat{o}=\hat{o}'$.
 Therefore,  we define a Majorana strange correlator, based on Majorana operators $\gamma_x=c_x+c^\dag_x$ in the bulk of the model:
\begin{equation}
\label{corrMajo}
\begin{array}{c}
|s\left[\gamma,\gamma\right]_{x,y}|  \equiv  \left|\frac{\bra{\Omega} \left(c_{x} + c_{x}^\dag\right) \left(c_{y} + c^\dag_y\right)  \ket{\Psi}}{\bra{\Omega}\Psi\rangle}\right| = \\
{}\\
= \left|\sum_{0\le k < \pi} \frac{2\sin k(x-y) \sin \left(\frac{\theta_\Psi(k)-\theta_\Omega(k)}{2}\right)}{L \cos \frac{\theta_\Psi(k)-\theta_\Omega(k)}{2}}\right| \,.
\end{array}
\end{equation}
The same results would be recovered by assuming  the other Majorana operator $i(c_x-c_x^\dag)$.
The second identity in Eq. \eqref{corrMajo} is valid for $x\neq y$ and is obtained by observing that Eq. \eqref{Kcdc} is symmetric under the exchange $x\leftrightarrow y$. Therefore the terms stemming from $c^\dag_x c_y + c_x c^\dag_y$ vanish for $x \neq y$. The strange correlator $|s\left[\gamma,\gamma \right]_{x,y}|$ is thus related to the combination $c^\dag_x c^\dag_y + c_xc_y$ which returns the result in Eq. \eqref{corrMajo}. Furthermore, given the absolute value, the result is symmetric under the exchange $\ket{\Omega} \leftrightarrow \ket{\Psi}$.

In the following, we fix $\Delta=t$, and we label with $\mu_\Omega$ and $\mu_\Psi$ the chemical potentials respectively adopted to obtain the Kitaev ground-states $\ket{\Omega}$ and $\ket{\Psi}$. In Fig. \ref{fig:strangekitaevp}, we plot typical examples of the strange correlator $|s\left[\gamma,\gamma \right]_{x,y}|$. Panel (a) displays $|s|$ for two ground-states in the trivial phase (obtained for $\mu_\Psi/t= 3$ and $\mu_\Omega/t=6$ respectively); in this case $|s\left[\gamma,\gamma \right]_{x,y}|$ decays exponentially with the distance $|x-y|$. Panel (b) displays $|s|$ for two ground-states both in the topological phase ($\mu_\Psi/t= 0.1$ and $\mu_\Omega/t=0.3$): also in this case, the strange correlator decays exponentially. Panel (c) shows $|s| $ for ground-states taken in different phases ($\mu_\Psi/t= 0.3$ and $\mu_\Omega/t=3$); in this case, the  strange correlator returns, for $x\neq y$, an approximately constant value different from zero, and it is reminiscent of the limiting case with $\mu_\Psi=0$ and $\mu_\Omega=-\infty$, discussed above for open boundaries.

\begin{figure}[t]
\includegraphics[width=0.8\columnwidth]{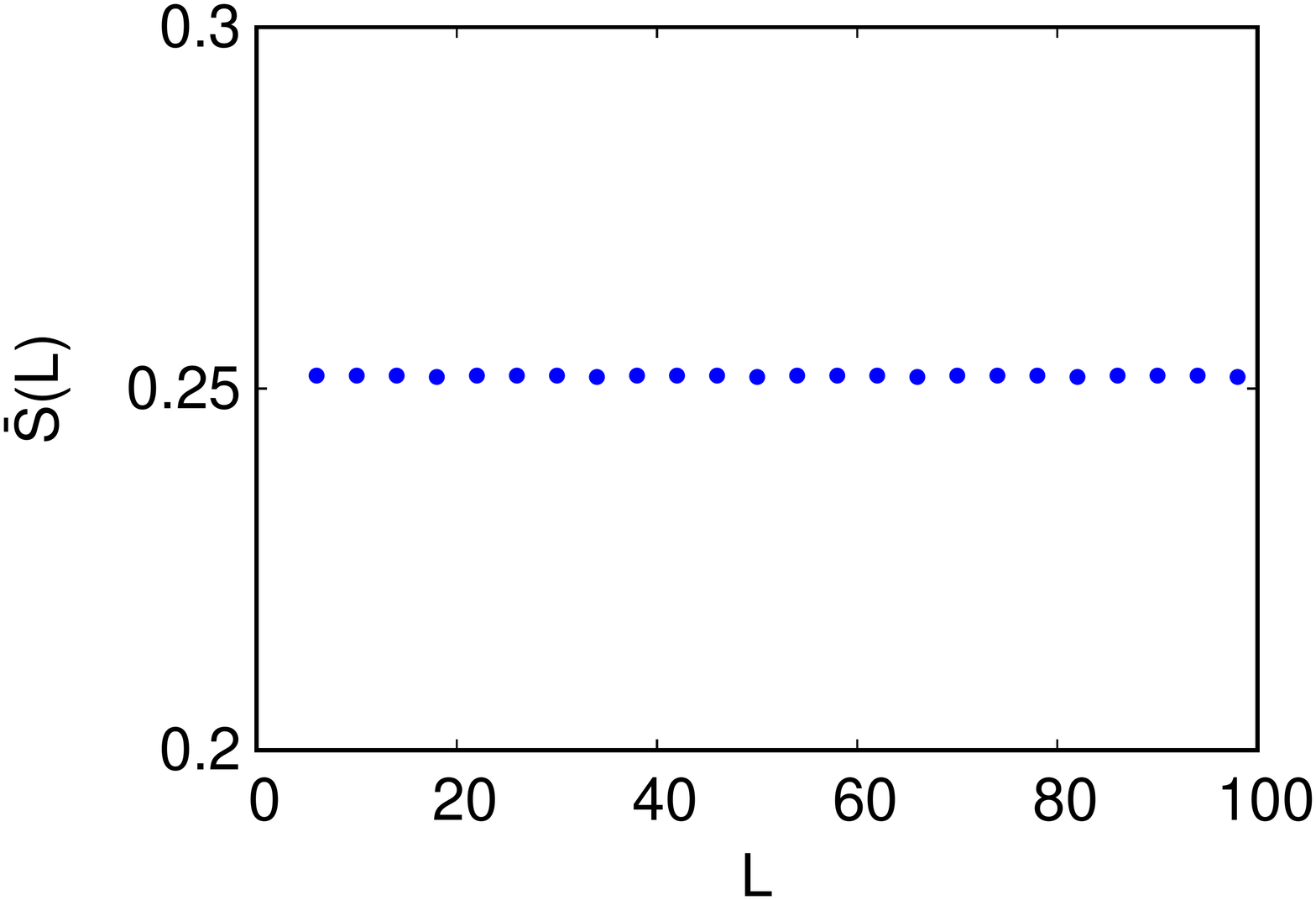}
\llap{\parbox[b]{9.5cm}{\color{black}(a)\\\rule{0ex}{4cm}}}
\vspace{-0.7cm}
\includegraphics[width=0.8\columnwidth]{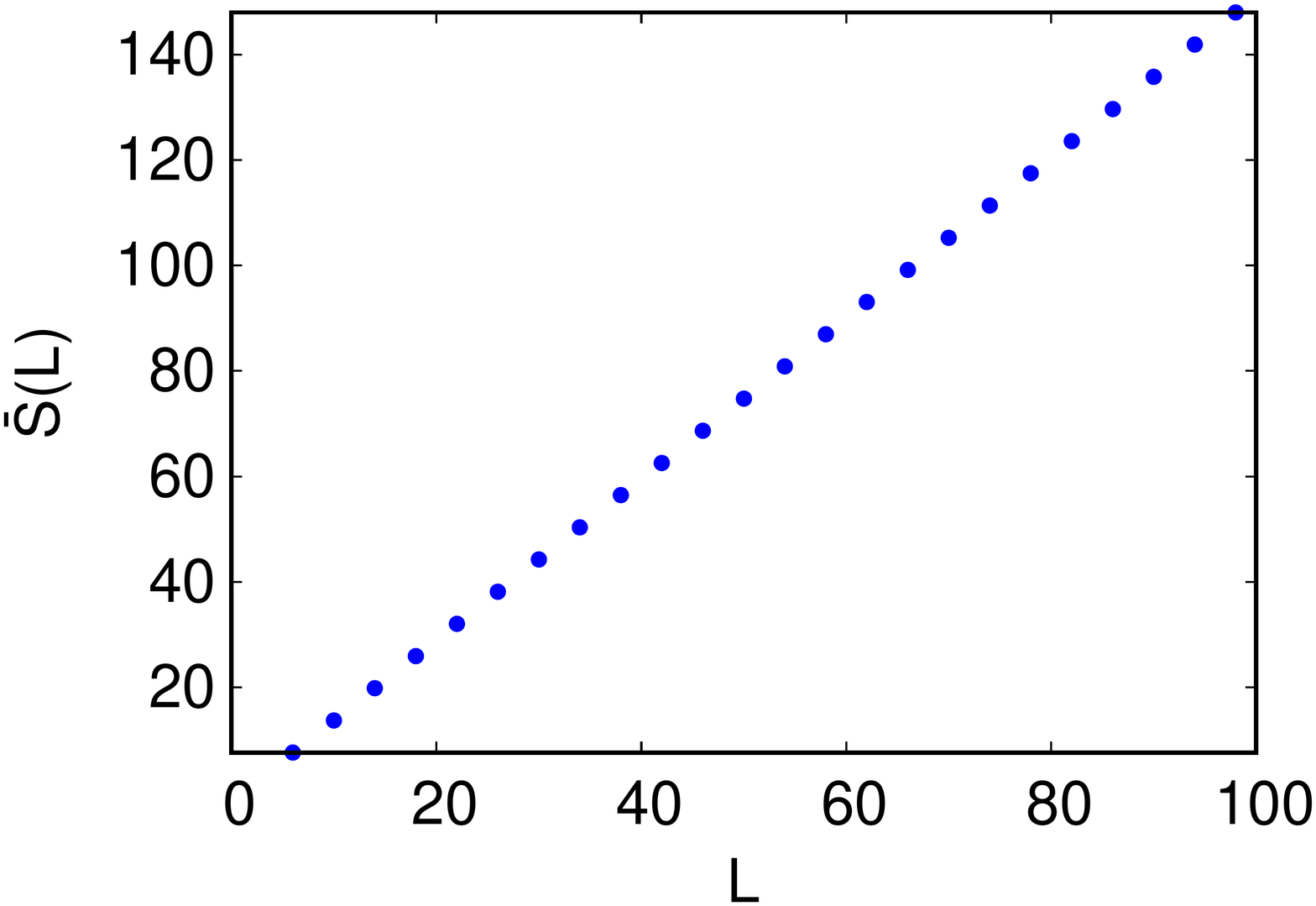}
\llap{\parbox[b]{9.5cm}{\color{black}(b)\\\rule{0ex}{4cm}}}
\caption{Scaling in $L$ of the sum of Majorana strange correlators $\bar{S} \left[\gamma,\gamma \right]_L $, referred to the Kitaev chain $H_{\mathrm{Kit}}$ in Eq. \eqref{kitaev}. We assumed  antiperiodic boundary conditions and we set $\Delta=t$ with chemical potentials specified as follows. (a) $\mu_{\Omega} /t=0.3$ and $\mu_{\Psi} /t=0.1$: both states are in the topological phase and $\bar{S}$ displays a constant behavior. (b) $\mu_{\Psi} /t =0.1$ and $\mu_{\Omega} / t=6$: $\bar{S} \left[\gamma,\gamma \right]_L$ displays a linear behavior consistent with $\ket{\Psi}$ being in a topological phase with Majorana edge modes with scaling dimension $\alpha=0$, and $\ket{\Omega}$ being a trivial state.}
\label{kitaev_scal}
\end{figure}

In Fig. \ref{kitaev_scal}, we display instead the scaling behaviour with the system size of the sum $\bar{S} \left[\gamma,\gamma \right]_L =\frac{1}{L} \sum_{x,y} |s\left[\gamma,\gamma \right]_{x,y}|$, obtained for 
$\mu_\Psi/t=0.1$ with $\mu_\Omega/t = 0.3$ [panel (a)] and $\mu_\Omega/t = 6$ [panel (b)]. A different qualitative behaviour is clearly observable: when both states are in the same phase, $\bar{S} \left[\gamma,\gamma \right]_L $ rapidly converges to a constant [in Fig. \ref{kitaev_scal} $\bar{S}$ is practically a constant given the very rapid exponential decay of the corresponding strange correlator in Fig. \ref{fig:strangekitaevp}(b)]. Instead, when the states are in different phases, panel (b), the linear growth is clearly visible and it corresponds to the predicted behavior in Eq. \eqref{qfiscal}: indeed, $(0+1)$-dimensional Majorana modes are characterized by the scaling dimension $\alpha=0$. 
In both cases, the modulus in Eq. \eqref{corrMajo} is essential, since otherwise a vanishing value for the sum, $S \left[\gamma,\gamma \right]_L  =\frac{1}{L} \sum_{x,y} s\left[\gamma,\gamma\right]_{x,y}$, is obtained. 

\subsection{Long-range Kitaev chain}
\label{kitaevLRs}
\begin{figure} [t]
\includegraphics[width=0.8\columnwidth]{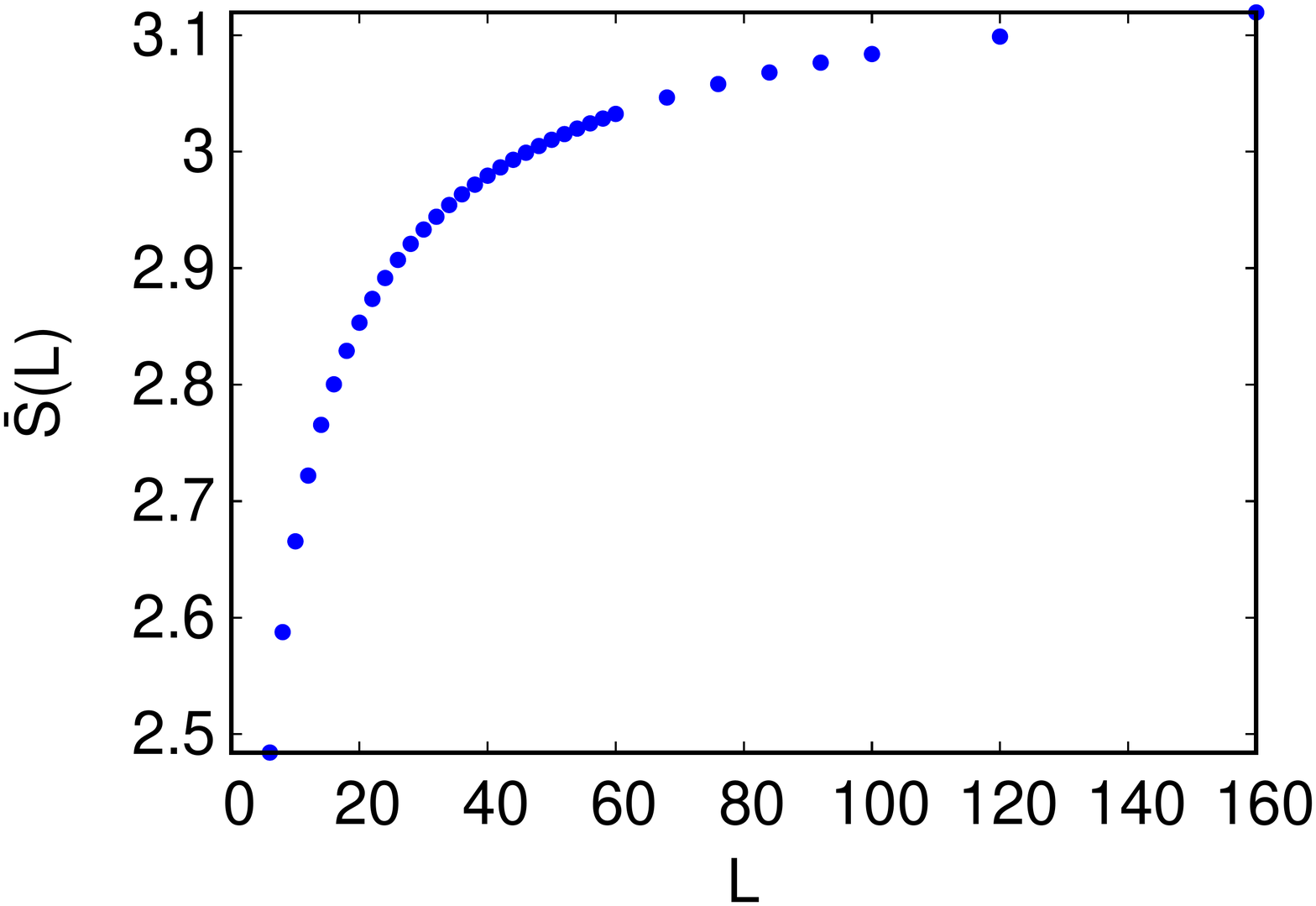}
\llap{\parbox[b]{9.5cm}{\color{black}(a)\\\rule{0ex}{4cm}}}
\vspace{-0.7cm}
\includegraphics[width=0.8\columnwidth]{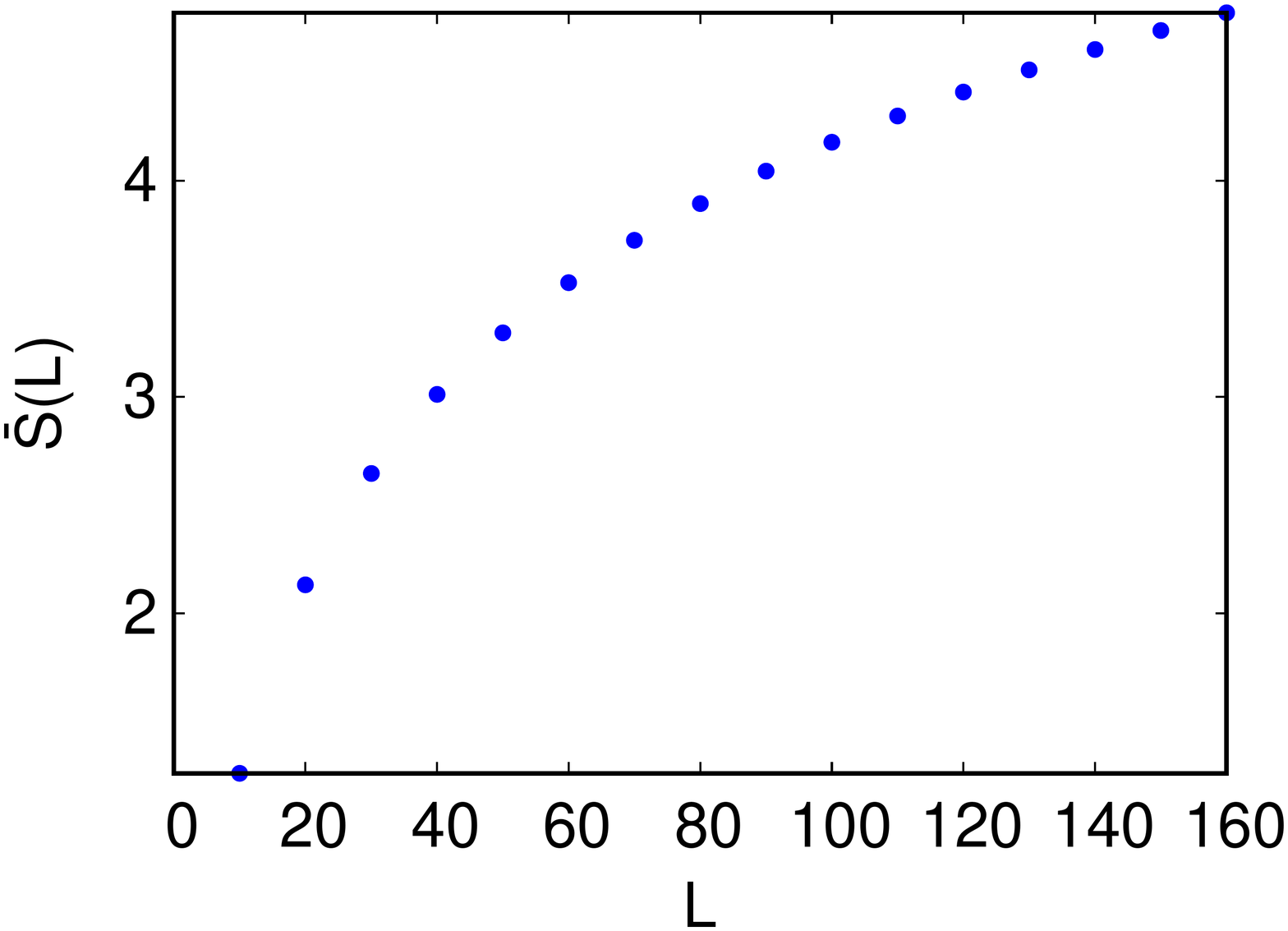}
\llap{\parbox[b]{9.5cm}{\color{black}(b)\\\rule{0ex}{4cm}}}

\caption{Scaling in $L$ of the sum of Majorana strange correlators $\bar{S} \left[\gamma,\gamma \right]_L $, referred to the Kitaev chain $H_{\mathrm{KLR}}$ with a long range {\color{black} superconducting pairing, ruled by a power-law exponent $\delta$, in turn defined in Eq. \eqref{kitaevLR}}. We assumed antiperiodic boundary conditions and we set $\Delta=t$ with chemical potentials specified as follows. (a)  $\mu_{\Omega} / t = 6$, $\mu_{\Psi} /t= 0.1$ and $\delta=0.5$: the sublinear behavior is consistent with the absence of a topological phase transition for $\delta<1$, {\color{black} such that the system displays two distinct trivial phases, without edge modes at short-range pairing and with gapped edge modes at long-range pairing.} (b)  $\mu_{\Omega} / t = \mu_{\Psi} /t= 0.1$, {\color{black} $\delta_{\Omega}=10$,} and $\delta_{\Psi}=0.5$: {\color{black} the states $\ket{\Omega}$ (topological phase at short-range pairing) and $\ket{\Psi}$ (gapped edge-modes phase at long-range pairing) are separated by a topological transition. However,} the resulting sum $\bar{S} \left[\gamma,\gamma \right]_L $ still scales sub-linearly.}
\label{kitaev_LR_scal}
\end{figure}

In Refs. \cite{vodola2014,lepori2016eff,leporiLR}, an extension of the Kitaev chain has been considered, involving long-range pairing:
\begin{equation}
\begin{array}{c}
H_{\mathrm{KLR}} = -\frac{t}{2}\sum_x\left(c^\dag_x c_{x+1} + c^\dag_{x+1}c_x\right) -\\
{}\\\color{black}
- \mu \sum_x c_x^{\dagger} c_x -\frac{\Delta}{2}\sum_{x,l} \frac{1}{l^{\delta}} \left(c^\dag_x c^\dag_{x+l} + c_{x+l}c_x\right) .
\end{array}
\label{kitaevLR}
\end{equation}
Here, the exponent $\delta$ dictates how fast the superconducting pairing decays algebraically with the distance. The limit $\delta\to \infty$ corresponds to the nearest-neighbor pairing appearing in Eq. \eqref{kitaev}. It is known \cite{vodola2014,lepori2016eff,leporiLR} that the Hamiltonian $H_{\rm KLR}$ for $\delta>1$ displays a physics qualitatively similar to the Kitaev model in Eq. \eqref{kitaev}. In particular, we verified that the Majorana strange correlator $|s\left[\gamma,\gamma \right]_{x,y}|$, defined in Eq. \eqref{corrMajo}, is still able to capture the topological transition for $\delta>1$.  
However, for $\delta<1$ the situation changes completely, since the pairing term in momentum space diverges, as well as the Bogoliubov dispersion.
Correspondingly, purely long-range (insulating) phases appear, which are not included in the standard tenfold classification \cite{altland1997} of the short-range topological insulators and superconductors. In particular for $\mu/t < 1$ a phase with massive edge modes \cite{vodola2014},  and related to nontrivial long-range topologies \cite{leporiLR}, is realized. {\color{black}These modes are indeed still localized at the edges, although to a minor extent than in the short-range limit (they display algebraic tails in the wavefunction decay) and they possess a nonvanishing energy also in the thermodynamic limit. They can be thought as the result of the hybridization, induced by the long-range pairing, of the Majorana edge modes characterizing the short-range $(\delta >1)$ topological phase at $|{\mu}/{t} | < 1$ ,  see \cite{leporiLR} and references therein.} 
Therefore, at least for ${\mu}/{t}  < 1$, a topological phase transition is expected when varying the exponent $\delta$ across the critical value $\delta=1$.  
{\color{black} 
We mention that in very recent literature, several other properties of the edge modes in presence of long-range couplings have been discussed,
 see the review \cite{defenu2021R} and references therein.
A notable example is the case of Goldstone excitations from
the spontaneous symmetry breaking of continuous symmetries \cite{diessel2022,meng2023,defenu2023}.
}\\
\indent The phase transition at $\mu=t$ is known to evolve, for $\delta < 1$, into a transition between the topological phase with massive edge modes, at $\mu < t$, and a phase with trivial long-range topology and without edge states \cite{vodola2014, leporiLR}. In Fig. \ref{kitaev_LR_scal} (a), we display $\bar{S} \left[\gamma,\gamma \right]_L $ for $\delta=0.5$, $\mu_{\Omega} / t = 6$ and $\mu_{\Psi} /t= 0.1$; its growth is clearly sublinear. In particular, a fit reveals a growth compatible with a logarithm as a leading term.  This behaviour {\color{black} is compatible} with a finite mass-gap  \cite{vodola2014,lepori2016eff,leporiLR}, both in the bulk and for the edge modes, when the open chain is considered and still with vanishing scaling dimension. 
 In Fig. \ref{kitaev_LR_scal} (b), we report $\bar{S} \left[\gamma,\gamma \right]_L $ for  $\mu_{\Psi} /t = \mu_{\Omega} / t = 0.1$ , $\delta_{\Psi}=10$, and $\delta_{\Psi}=0.5$: different nontrivial topologies, short and long-range, are considered. The resulting sum $\bar{S} \left[\gamma,\gamma \right]_L $ still appears to scale sub-linearly. Overall, both the panels suggest
 the breakdown of the picture in \cite{xu2014}, in the presence of long-range couplings and interactions. More comment will be given in Section \ref{choice}.\\
\indent We finally notice that, analogously to the standard Kitaev chain, the modulus is essential in the analysis of the strange correlator sums: in its absence strong cancellation effects may emerge, resulting in the vanishing of $S \left[\gamma,\gamma \right]_L $, independently on the state choices.

\subsection{ A two-dimensional Chern insulator}
\label{iqhe}

The previous scaling analysis of the strange correlator sums can be extended to higher-dimensional topological systems. In the following, we address both a two-dimensional and a three-dimensional topological insulator example. 

Let us first address the Bernevig-Hughes-Zhang (BHZ) two-dimensional model, which constitutes a minimal model for two-dimensional Chern insulators displaying anomalous Hall response \cite{bernevigbook,kotetesbook,notaqsh1}. 
The corresponding Hamiltonian reads:
\beq
H^{\mathrm{(2D)}} (\mathbf{k}) = 
\begin{pmatrix}
 \tilde{M} (\mathbf{k}) & \sin k_x + i \sin k_y  \\
\sin k_x - i \sin k_y & -\tilde{M}(\mathbf{k}) 
\end{pmatrix}\, ,
\label{BHZ_2D}
\eeq
where $\tilde{M}(\mathbf{k}) \equiv  M - 4 + 2 \, (\cos k_x+ \cos k_y) $.
The two-component spinorial structure of Eq. \eqref{BHZ_2D} can either be originating from a sublattice or orbital degree of freedom, or be an effective representation of an inner spin degree of freedom in the case of spin-orbit coupled systems. In the following we will label with $A$ and $B$ fermionic operators referring to the two spinor components.

The Hamiltonian $H^{\mathrm{(2D)}}$ does not possess time-reversal invariance and, when considering a generic chemical potential, it belongs to the A class of the  ("tenfold-way") classification of topological insulators and superconductors \cite{altland1997,ludwig2009,kotetesbook}. 
Instead, at half filling the same Hamiltonian is charge-conjugation invariant, 
$H^{\mathrm{(2D)}} (\mathbf{k}) = - \sigma_1 \, H^{\mathrm{(2D)} \, *} (- \mathbf{k}) \, \sigma_1$, so that
its energy bands display opposite energies at opposite momenta, and $H^{\mathrm{(2D)}} (\mathbf{k})$ is in the D symmetry class \cite{altland1997,ludwig2009,kotetesbook}.

\begin{figure*} [t]
\includegraphics[scale=0.6]{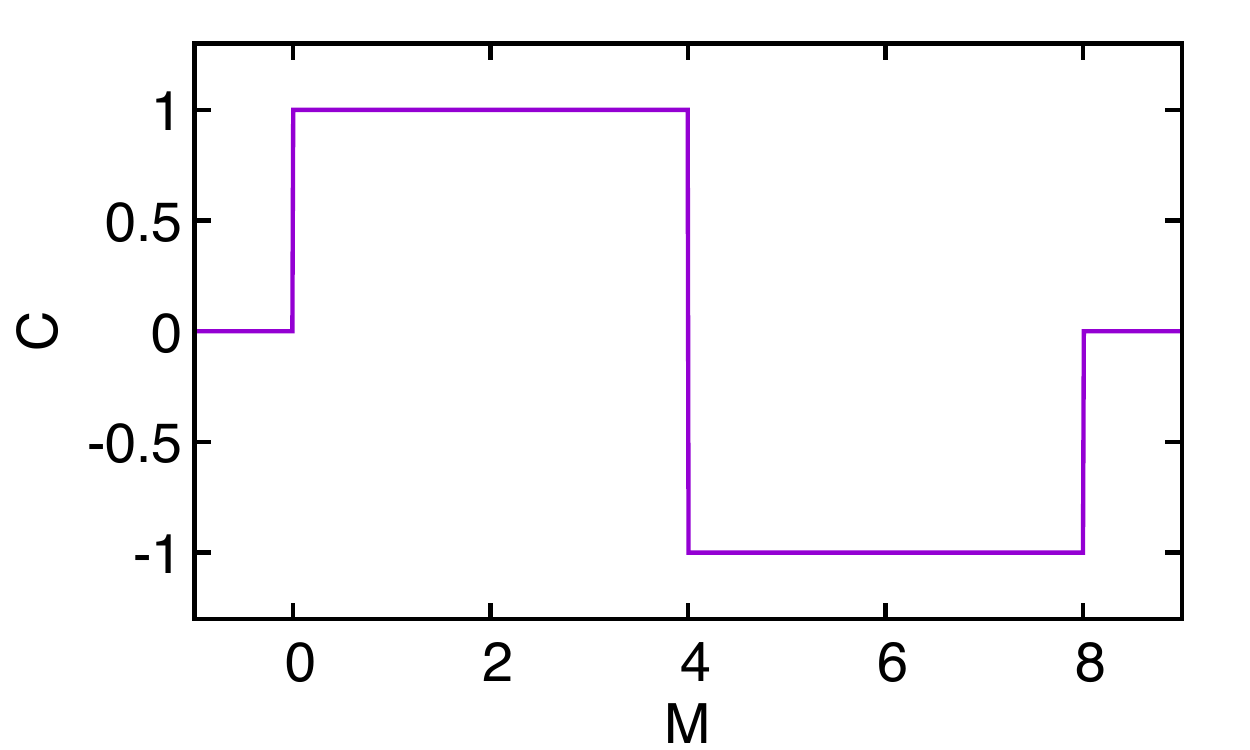}
\llap{\parbox[b]{2.25cm}{\color{black}(a)\\\rule{0ex}{3.5cm}}}
\includegraphics[scale=0.6]{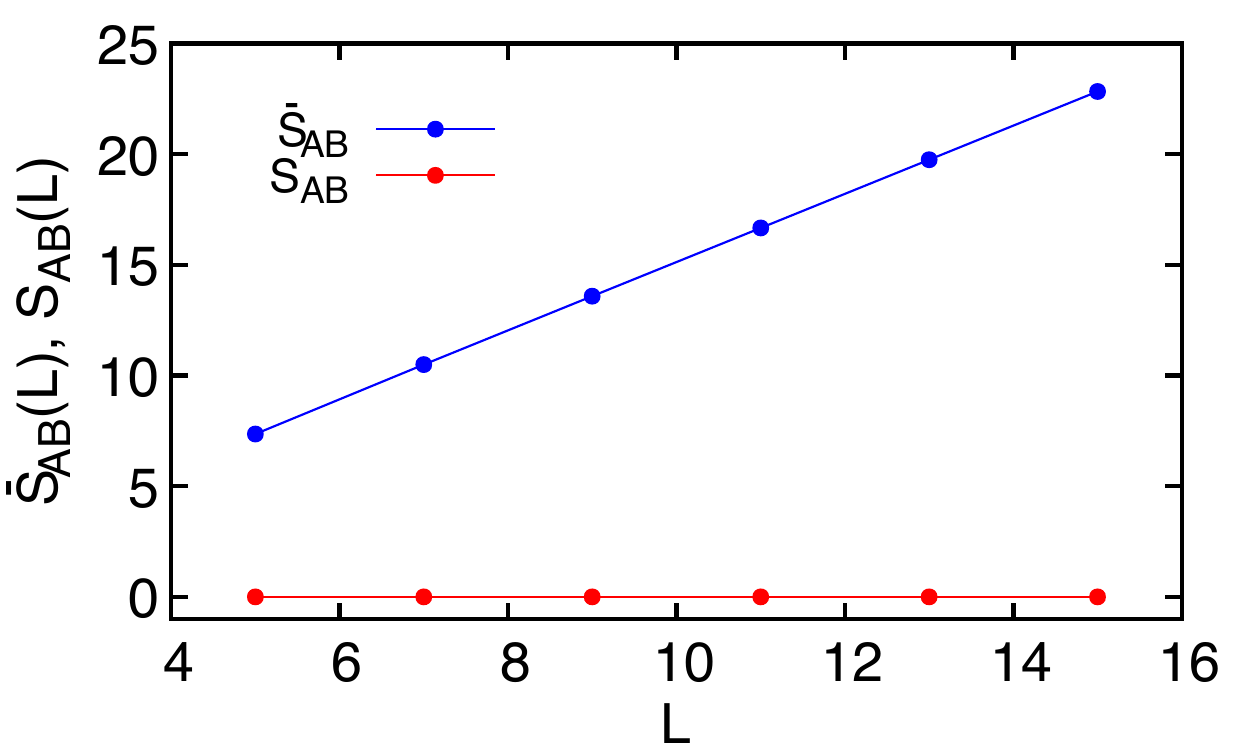}
\llap{\parbox[b]{2.25cm}{\color{black}(b)\\\rule{0ex}{3.5cm}}}
\caption{Panel (a): Different phases of the 2D BHZ model in Eq. \eqref{BHZ_2D} and related Chern numbers, as a function of $M$. Panel (b): $\bar{S}[ c^{\dagger}_A , c_B]_L$ and ${S}[ c^{\dagger}_A , c_B]_L$ for $M_{\Omega} = 2$ and $M_{\psi} = 6$: only the sum of the moduli display a linear behavior, signaling the different topological phase of $\ket{\Omega}$ and $\ket{\Psi}$.} 
\label{mista2D}
\end{figure*}

 This model displays different phases depending on the value of $M$, as shown in the left panel of Fig. \ref{mista2D}. For $M<0$ and $M>8$, the system is fully gapped and it corresponds to a standard band insulator with vanishing Chern number \cite{bernevigbook}. For $0<M<4$ and $4<M<8$, instead, two topological gapped phases appear with Chern number $\pm 1$, thus displaying opposite Hall conductivities.  These phases are separated by critical points at $M=0,4,8$, characterized by Dirac band-touching points.
In the following,  we will denote by $\ket{M = a}$ ($a$ being a real number) the ground-state of the insulating phase of the Hamiltonian in Eq. \eqref{BHZ_2D} with $M = a$.

Differently from the superconducting Kitaev chain in Eq. \eqref{kitaev}, the model defined by Eq. \eqref{BHZ_2D} conserves the particle number. Therefore, the simplest choice for the definition of the strange correlators is given by the operators $o=c^\dag_{f}$ and $o'=c_{f'}$, with $f=A,B$. This choice is also consistent with the bulk-boundary correspondence, related to the onset of chiral edge modes at the boundaries of the topological insulator phases (see Sec. \ref{choice} for more detail).

In particular, we evaluate the strange correlators between many-body ground-states of the Hamiltonian \eqref{BHZ_2D} obtained for different values of the parameter $M$, $M_\Omega$ and $M_\Psi$. The ground-states are generically defined as
\begin{equation}
\ket{\Psi_j} = \prod_{{\bf k}} \gamma^{(-)\dag}_{{\bf k}} \ket{0}\, ,
\end{equation}
where $\ket{0}$ is the fermionic vacuum, ${\bf k}$ spans the full BZ, and  we label with $\gamma^{(+)\dag}_{{\bf k}}$ and $\gamma^{(-)\dag}_{{\bf k}}$ the creation operators of the single-particle eigenstates with positive and negative energies, respectively. In particular, we set:
\begin{equation}
\gamma^{(\pm)\dag}_{{\bf k}} = a^{(\pm)}_{A,{\bf k}} \, c^\dag_{A, {\bf k}} + a^{(\pm)}_{B,{\bf k}} \, c^\dag_{B, {\bf k}}\,,
\label{trasf}
\end{equation} 
where  $a_{f,{\bf k}}^{(\pm)}$ are suitable coefficients and  $c_{f,{\bf k}}^\dag$ are the usual fermionic creation operators.
\begin{figure*} [t]
\includegraphics[scale=0.6]{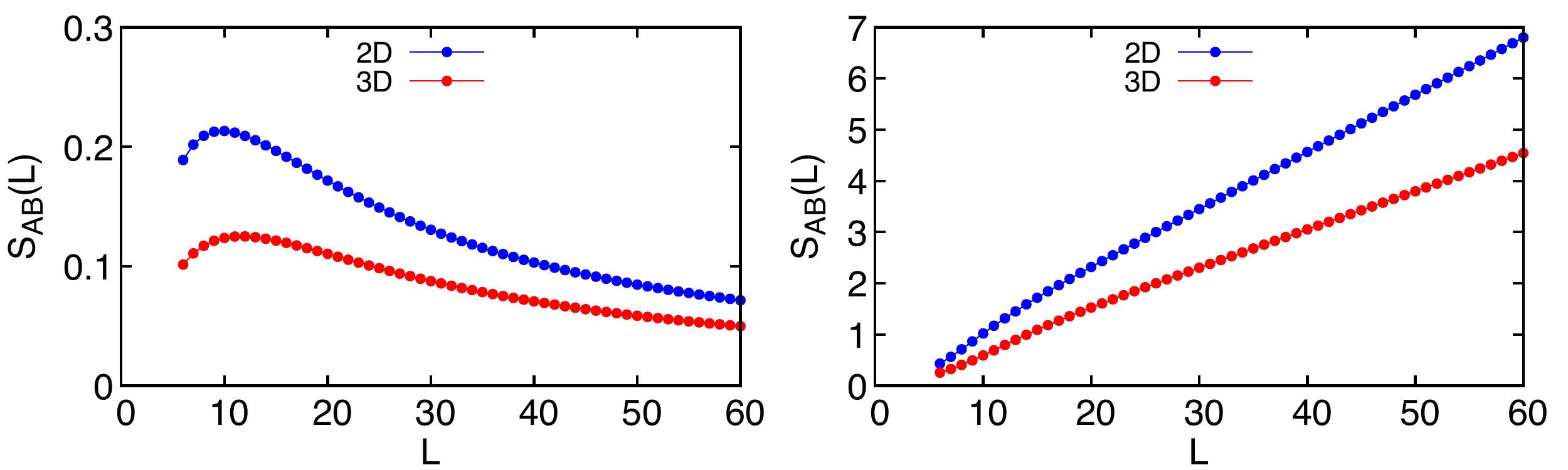}
\llap{\parbox[b]{17.0cm}{\color{black}(a)\\\rule{0ex}{3.5cm}}}
\llap{\parbox[b]{2.0cm}{\color{black}(b)\\\rule{0ex}{3.5cm}}}
\caption{Scaling in $L$ of $S[ c^{\dagger}_A , c_B]_L$, for the two-dimensional models in Eqs. \eqref{BHZ_2D} and \eqref{BHZ_2D_lat} (blue points) and
for the three-dimensional models in Eqs. \eqref{BHZ_3D} and \eqref{BHZ_3D_lat} (red points), both for {\color{black} $M_{\Omega} = M_{\Psi} = -1$} [panel (a)] and  {\color{black}$M_{\Omega} = - M_{\Psi} = -1$} [panel (b)]. Periodic boundary conditions are assumed here; open boundaries yield the same qualitative trends (see the main text).
 }
\label{2-3D}
\end{figure*}
\begin{figure} [t]
\includegraphics[width=0.8\columnwidth]{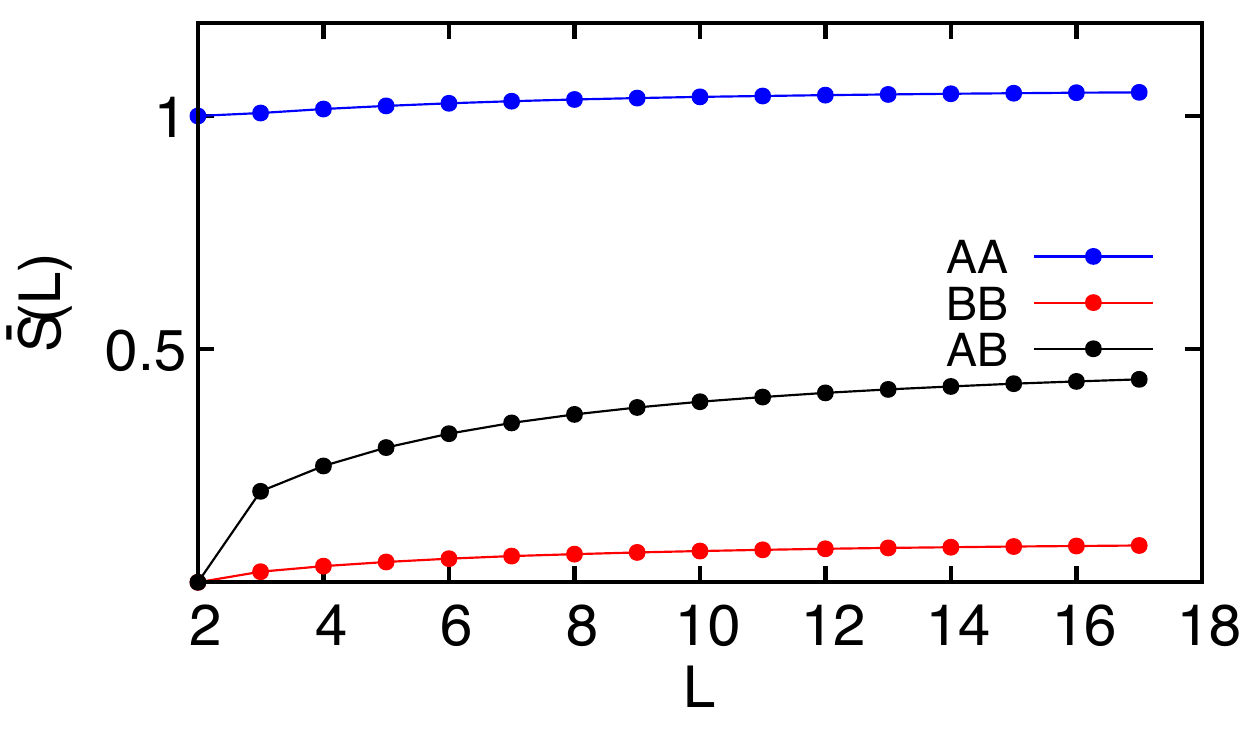}
\llap{\parbox[b]{2.cm}{\color{black}(a)\\\rule{0ex}{3cm}}}

\includegraphics[width=0.8\columnwidth]{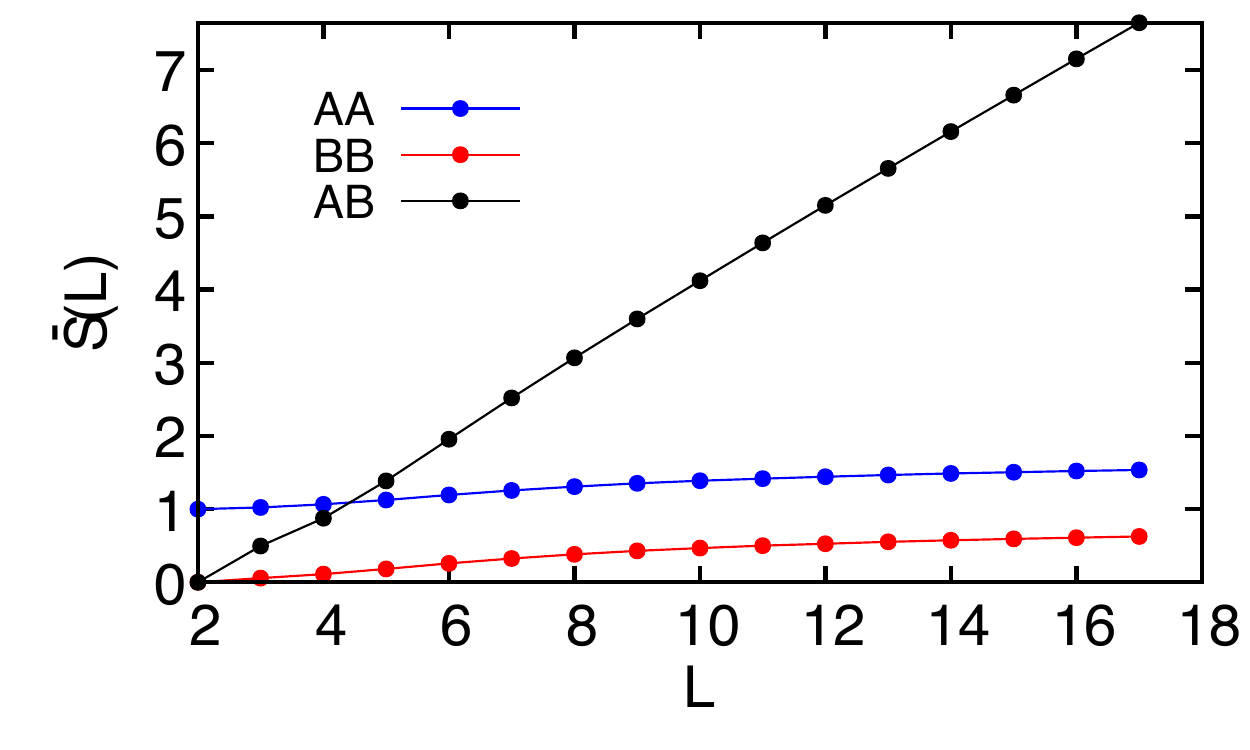}
\llap{\parbox[b]{2.cm}{\color{black}(b)\\\rule{0ex}{3cm}}}

\caption{Scaling in $L$ of  $\bar{S}[ c^{\dagger}_m , c_n]_L$ with $n,m=A,B$, for the two-dimensional model in Eqs. \eqref{BHZ_2D} and \eqref{BHZ_2D_lat}, for {\color{black} $M_{\Omega} = M_{\Psi} = -1$ (a) and  $M_{\Omega} = - M_{\Psi} = - 1$ (b).}
Open boundary conditions are assumed. }
\label{2D_open}
\end{figure}

The Hamiltonian \eqref{BHZ_2D} can  be written in real space as 
\begin{equation}
\begin{array}{c}
H_{\mathrm{lat}}^{(\mathrm{2D})} = \sum_{\mathbf{r}}
\Big[c^{\dagger} (\mathbf{r} + \hat{x}) \left( \sigma_3 -\frac{i}{2}\sigma_1 \right) c (\mathbf{r})  \, + \\
{}\\
+ \, c^{\dagger} (\mathbf{r} + \hat{y}) \left( \sigma_3 +\frac{i}{2}\sigma_2 \right) c (\mathbf{r}) + \frac{M-4}{2} \, c^{\dagger} (\mathbf{r}) \, \sigma_3 \, c (\mathbf{r}) + \mathrm{H. \, c.} \Big],
\end{array}
\label{BHZ_2D_lat}
\end{equation}
where $\sigma_i$ labels the Pauli matrices and the operators $c^\dag(\mathbf{r})$ and $c(\mathbf{r})$ are meant as two-component spinors (with $A$ and $B$ components).
The strange correlators are defined by
\beq
s[c^{\dagger}_f , c_{f^{\prime}}]_{{\bf r}, {\bf r^{\prime}}} =  \frac{\bra{\Omega} c^\dag_{f}({\bf r}) c_{f^{\prime}}({\bf r^{\prime}})\ket{\Psi}}{\bracket{\Omega}{\Psi}}\, .
\label{matper}
\eeq
It is convenient to consider at first the sums $S[ c^{\dagger}_f , c_{f^{\prime}}]_L $. Details about the calculations of these quantities 
are given in App. \ref{strasl}.
For periodic boundary conditions,  we obtain:
\beq
S[ c^{\dagger}_f , c_{f^{\prime}}]_L  =   \tilde{s}[ c^{\dagger}_f , c_{f^{\prime}}]_{{\bf 0}} \, ,
\label{sumper}
\eeq
where $\tilde{s}[ c^{\dagger}_f , c_{f^{\prime}}]_{{\bf k}}$ denotes the Fourier transform of $s[c^{\dagger}_f , c_{f^{\prime}}]_{{\bf r}, {\bf r^{\prime}}}$.
We notice that, since the Hamiltonian in Eq. \eqref{BHZ_2D_lat} with periodic boundary conditions is translationally invariant, then $\tilde{s}[ c^{\dagger}_f , c_{f^{\prime}}]_{{\bf 0}}$ in Eq. \eqref{sumper} is the largest eigenvalue of the matrix, in ${\bf r}$ and ${\bf r^{\prime}}$, defined by Eq. \eqref{matper} (see Ref. \cite{shi2003} and App. \ref{strasl}). Therefore, the scaling in $L$ of $\tilde{s}[ c^{\dagger}_f , c_{f^{\prime}}]_{{\bf 0}}$ yields a criterion for topology at zero temperature, analogous to the  well-celebrated Onsager-Penrose criterion for Bose-Einstein condensation and ODLRO \cite{PS}.  The same analogy holds also in the absence of translational invariance, provided that 
  the largest eigenvalue of the matrix in Eq. \eqref{matper} is directly calculated by  diagonalization: its scaling in $L$ is still related, as $\tilde{s}[ c^{\dagger}_f , c_{f^{\prime}}]_{{\bf 0}}$, to the decay behaviour of the strange correlators in Eq. \eqref{matper}.

The results for $S[ c^{\dagger}_A , c_B]_L$,  with {\color{black} $\ket{\Omega} = \ket{M = -1}$ and $\ket{\Psi} = \ket{M = \pm 1}$,} are shown in Fig. \eqref{2-3D} (left and right panels, respectively).
Periodic boundaries are assumed again. 
To regularize possible divergences in ${\bf k} = 0$, which may arise due to the orthogonality of single-particle states \cite{xu2014}, $S[ c^{\dagger}_A , c_B]_L$ is estimated by evaluating Eq. \eqref{sumper} at ${\bf k} = \frac{2 \pi}{L} (1,1,1)$, instead of imposing ${\bf k} = 0$.
Similarly to the Kitaev chain, different qualitative behaviours are clearly observable in Fig. \eqref{2-3D}: we observe an asymptotic decay for the sum of the standard correlators with {\color{black} $\ket{\Omega} = \ket{\Psi} \left(M_\Omega=M_\Psi=-1\right)$,} whereas a linear growth characterizes the sum of the strange correlators with $M_\Omega=-1$ and $M_\Psi=1$.

Next, we consider the sum of the strange correlator moduli, $\bar{S}[ c^{\dagger}_A , c_B]_L$. Its linear behavior is confirmed when $\ket{\Omega}$ and $\ket{\Psi}$ lie in different topological phases as exemplified by Fig. \ref{2D_open}(b) for $M_\Psi=1$ and $M_\Omega=-1$ in the 2D system. On the contrary, a finite constant is approached when the states belong to the same phase and Fig. \ref{2D_open}(a) illustrates the standard correlator obtained for $\ket{\Omega} = \ket{\Psi} = \ket{M = -1}$. 
We emphasize that the results in Fig. \ref{2D_open}(a) are qualitatively the same for both open and periodic boundary conditions; in particular, no divergences occur when both states are in the same phase. The case with states in different topological phases and periodic boundary conditions, instead, can be affected by divergences due to their orthogonality and the data in Fig. \ref{2D_open}(b) correspond to a system with open boundary conditions; see more details in App. \ref{strasl}.

{\color{black} When fixing $\ket{\Omega}$ in the trivial phase, by decreasing $M_\Psi$ from the topological phase across $0$ (the location of the topological transition), one observes the onset of a bending of $\bar{S}[ c^{\dagger}_A , c_B]_L$, as a function of the system size, as seen also in Fig. \ref{2D_open} (a). The same happens if both $\ket{\Omega}$ and $\ket{\Psi}$ are in the same topological phase. In these situations $\bar{S}[ c^{\dagger}_A , c_B]_L$ evolves from an approximately linear behavior for small $L$ to a constant behavior at larger system sizes. This trend reflects the exponential decay of the strange correlator $s[ c^{\dagger}_A , c_B]_{\bf{r},\bf{r'}}$. Indeed, the value of the system size $L$ at which the bending occurs depends on the decay length of  $s[ c^{\dagger}_A , c_B]_{\bf{r},\bf{r'}}$. This length is in turn approximately related to the gap of edge modes that would appear at the interface between regions characterized by the different values of $M_\Psi$ and $M_\Omega$. Therefore, the transition between linearly growing and saturating $\bar{S}[ c^{\dagger}_A , c_B]_L$ reflects the general behavior of the strange correlators.}

{\color{black} In the light of} Eq. \eqref{qfiscal}, and in agreement with the known relation between algebraic functions and their Fourier transforms \cite{nist},
 $f({\bf r})  = \frac{1}{|{\bf r}|^{2 \alpha}} \Longleftrightarrow  \tilde{f}({\bf k}) \propto \frac{1}{|{\bf k}|^{d-2 \alpha}}$,
 the behaviour of $\bar{S}[ c^{\dagger}_A , c_B]_L$ can be linked to the decay of  $S[ c^{\dagger}_f , c_{f^{\prime}}]_{{\bf r}, {\bf r^{\prime}}}$, which respectively decays sub-algebraically and as the power law $|{\bf r} - {\bf r^{\prime}}|^{-1}$ in the two cases. 
In turn, the latter scaling reflects the (mass) scaling dimension $\alpha=1/2$ of the edge modes (since they are free fermions, the relation $\alpha = \frac{d-1}{2} = \frac{1}{2}$ holds).
Therefore, we infer that the algebraic decay of the strange correlator between $\ket{M_\Omega=-1}$ and $\ket{M_\Psi=1}$ is again deeply connected with the scaling dimensions of the edge modes at the interface between the corresponding phases. In Section \ref{choice}, this correspondence will be investigated in more detail.

 So far, we analyzed $S[ c^{\dagger}_A , c_B]_L$ and $\bar{S}[ c^{\dagger}_A , c_B]_L$
when $\ket{\Omega}$ has a trivial topology, the case discussed in Ref. \cite{xu2014}. It is worth to consider also the case where $\ket{\Omega} $ and $\ket{\Psi}$ have different nontrivial
topologies. In particular, we assume $\ket{\Omega} = \ket{M = 2}$ and $ \ket{\Psi} = \ket{M = 6}$, two states with opposite Chern number, $\pm 1$. Also in this case, with {\color{black}periodic} boundary conditions, 
we find that  $\bar{S}[ c^{\dagger}_A , c_B]_L$ scales linearly with $L$, witnessing the presence of different topologies, corresponding to an algebraic decay of the related strange correlators $s[ c^{\dagger}_A , c_B]_{{\bf r}, {\bf r^{\prime}}}$. Instead, $S[ c^{\dagger}_A , c_B]_L$ vanishes, not providing any scaling. These situations are shown in the right panel of Fig. \ref{mista2D}.

Finally, our results show immediately the relevance of our method, also for sake of evaluation simplicity: the effect of algebraic decay of the strange correlators is explicit already at very small linear sizes $L$, where it is known that to distinguish directly the algebraic and subalgebraic (as exponential) decays of $s$ can be demanding in general.

\subsection{A three-dimensional time-reversal generalization}
\label{BHZ}
We consider now the three-dimensional Bernevig-Hughes-Zhang model \cite{bernevigbook,zhang2009} which extends the two-dimensional topological insulator considered in Eq. \eqref{BHZ_2D}:
\begin{widetext}
\begin{equation}
H^{\mathrm{(3D)}} (\mathbf{k}) = 
\begin{pmatrix}
\bar{M} (\mathbf{k}) & (\sin k_x + i \sin k_y) & 0 &  \sin k_z \\
 (\sin k_x - i \sin k_y) & -\bar{M}(\mathbf{k}) &  \sin k_z & 0 \\
0 &  \sin k_z & \bar{M} (\mathbf{k}) & - (\sin k_x - i \sin k_y) \\
 \sin k_z & 0 & -(\sin k_x + i \sin k_y) & - \bar{M}(\mathbf{k})
\end{pmatrix}\, ,
\label{BHZ_3D}
\end{equation}
\end{widetext}
where $\bar{M}(\mathbf{k}) =  M - 6 + 2\, (\cos k_x+ \cos k_y + \cos k_z) $. The Hamiltonian \eqref{BHZ_3D} can be expressed in terms of the Dirac $\Gamma$ matrices and it can be associated to a cubic lattice model. In particular, it has been proposed as a low-energy effective
description for three-dimensional time-reversal-invariant topological insulator \cite{notaqsh2,bernevigbook,zhang2009,Imura2012} (quantum spin-Hall systems).
The model is time-reversal invariant, since:
\beq
H^{\mathrm{(3D)}} _{\mathrm{lat}}(\mathbf{k}) = U_T \, H^{\mathrm{(3D) \, *}}_ {\mathrm{lat}}(-\mathbf{k}) \, U_T^{\dagger} \,  , \quad U_T = \sigma_2 \otimes \bf{I} \, .
\eeq
The matrix $U_T$ correctly interchanges the upper and lower two-by-two blocks (corresponding to opposite spin-$\frac{1}{2}$ projection), also adding a relative phase. 
The Hamiltonian in Eq. \eqref{BHZ_3D}  belongs to the AII class of the 
"tenfold-way" classification of the topological insulators and superconductors \cite{altland1997,ludwig2009,bernevigbook}.
The model displays four bands, two-by-two degenerate, and symmetric around zero energy. 

As a function of the parameter $M$, the model displays three kinds of phases at half filling (see the detailed analysis of the equivalent model in \cite{Imura2012}): For $M<0$ and $M>12$, the Hamiltonian \eqref{BHZ_3D} defines a trivial insulator; for $0<M<4$ and $8<M<12$ the ground-states correspond to strong topological insulating phases, and their surfaces are characterized by single protected Dirac cone states; for $4<M<8$, the system is characterized by a weak topological insulating phase and its surfaces can present pairs of Dirac cones. Proceeding as in the two-dimensional case, we take $o=c_{f,\sigma}$ with $f=A,B$ and $\sigma=\uparrow, \downarrow$.

The tight-binding model corresponding to Eq. \eqref{BHZ_3D} can  be written in real space as
\beq
\begin{array}{c}
H_{\mathrm{lat}}^{(\mathrm{3D})} = \sum_{\mathbf{r}}
\Big[ c^{\dagger} (\mathbf{r} + \hat{x}) \left( \Gamma_0 -\frac{i}{2}\Gamma_1 \right) c (\mathbf{r}) \, + \\ 
{}\\
+  \, c^{\dagger} (\mathbf{r} + \hat{y}) \left( \Gamma_0 +\frac{i}{2}\Gamma_2 \right) c (\mathbf{r}) \, + \\
{}\\
+  \, c^{\dagger} (\mathbf{r} + \hat{z}) \left( \Gamma_0 -\frac{i}{2}\Gamma_3 \right) c (\mathbf{r})+
\frac{M-6}{2} c^{\dagger} (\mathbf{r})  \Gamma_0 c (\mathbf{r}) + \mathrm{H. \, c.} \Big]\, ,
\end{array}
\label{BHZ_3D_lat}
\eeq
where we introduced the following representation for the Dirac matrices:
\begin{equation}
    \Gamma_0 = \mathbf{I} \otimes \sigma_3 \,,\quad \Gamma_1 = \tau_3 \otimes \sigma_1 \,,\quad
    \Gamma_2 = \mathbf{I} \otimes \sigma_2 \,, \quad
    \Gamma_3 = \tau_1 \otimes \sigma_1\,.
\end{equation}
Here $\mathbf{I}$ is the $2\times 2$ identity matrix, and $\sigma_i$ and $\tau_i$ label Pauli matrices usually associated to electronic spin and orbital degrees of freedom.

The results for the sums $S[ c^{\dagger}_{A\uparrow} , c_{B\downarrow}]_L$,   with {\color{black} $\ket{\Psi} = \ket{M = \pm 1}$ and $\ket{\Omega} = \ket{M = -1}$,} obtained by the expressions reported in App. \ref{strasl}, are shown in  in Fig. \eqref{2-3D} (red dots).
Again, periodic boundary conditions are assumed and, to regularize possible divergences in ${\bf k} = 0$ and $S[ c^{\dagger}_{A\uparrow} , c_{B\downarrow}]_L$ is evaluated  at ${\bf k} = \frac{2 \pi}{L} (1,1,1)$, instead that strictly at ${\bf k} = 0$, as in Eq. \eqref{sumper}. 
 As for the two-dimensional case, a different qualitative behaviour is  observable in Fig. \eqref{2-3D}: an asymptotical decay in the diagonal {\color{black} \big($\ket{\Omega} = \ket{M = -1}$\big)} case, a linear growth in the strange case. {\color{black} In the light of} Eq. \eqref{qfiscal}, these behaviour can be linked to the decays of  $s[ c^{\dagger}_{A \uparrow} , c_{B \downarrow}]_{{\bf r}, {\bf r^{\prime}}}$, sub-algebraic and 
as $\frac{1}{|{\bf r} - {\bf r^{\prime}}|^2}$ in the two cases. In turn, the latter behaviour reflects again the dimension in mass of the edge modes on the $d-1 = 2$ dimensional edge space,  $\alpha = \frac{d-1}{2} = 1$.

We also analyzed the generalized strange correlators between trivial ($\ket{\Omega} = \ket{M = -1}$) and weak ($\ket{\Psi} = \ket{M = 6}$) topological insulators, finding linear growth in $L$ for $S[ c^{\dagger}_{A\uparrow} , c_{B\downarrow}]_L$.

Summing up, all the examined growth behaviours of the strange correlators, linear in $L$, are due to the fact that we are considering free fermionic models and related edge states.  We notice however that the stability of the edge states, then of the power-law decay for the strange correlators, is not even generally guaranteed for three-dimensional topological insulators, if in the presence of strong internal interactions \cite{xu2014}.

\section{Breakdown of translational invariance} \label{sec:dis}
In this Section, 
we focus on the two-dimensional BHZ model in Eq. \eqref{BHZ_2D_lat} with
open boundary conditions, and we add an onsite quenched disorder. Our calculations are performed by the equations in Appendix \ref{app:gen}, and we point out that the same equations also apply to the three-dimensional case in Eq. \eqref{BHZ_3D_lat}.

Disorder is introduced as a random position-dependent offset 
\beq
H_{\mathrm{dis}} =    \sum_{{\bf r}} w_{{\bf r}}  \ c^{\dagger}_{{\bf r}} c_{{\bf r}} \, ,
\eeq
with $w_{{\bf r}}$ uniformly distributed in the range $w_{{\bf r}} \in \big[-\sigma, \sigma]$, where $\sigma$ identifies the disorder strength.
 From the topological phase at $0< M_{\Psi} < 4$, a transition to  the trivial phase is expected for a sufficiently strong disorder, when $\sigma$ becomes comparable to the bulk gap.
 
It is convenient to analyze separately the behaviours of $\bar{S}[ c^{\dagger}_m , c_n]_L$ ($ m$ and  $n$ labelling the sublattices $A$ and $B$), mediated on the various disorder configurations, with $\ket{\Psi}$ both in the trivial and topological regimes.
To properly locate the transition, a number of disorder configurations increasing with $L$ is required, since the configurations for which  $\bar{S}[ c^{\dagger}_n , c_m]_L$ grows linearly tend to dominate at large $L$.

\begin{figure} [t]
\includegraphics[width=0.8\columnwidth]{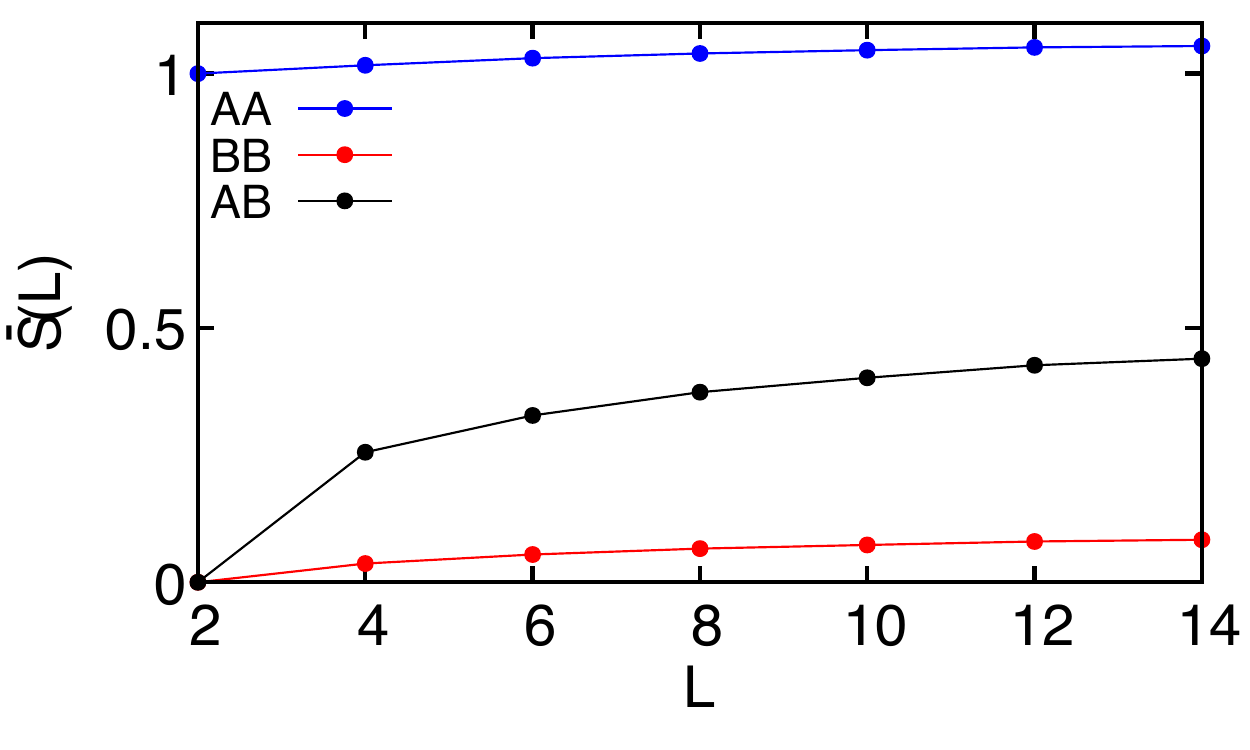}
\llap{\parbox[b]{2.cm}{\color{black}(a)\\\rule{0ex}{3.3cm}}}

\includegraphics[width=0.8\columnwidth]{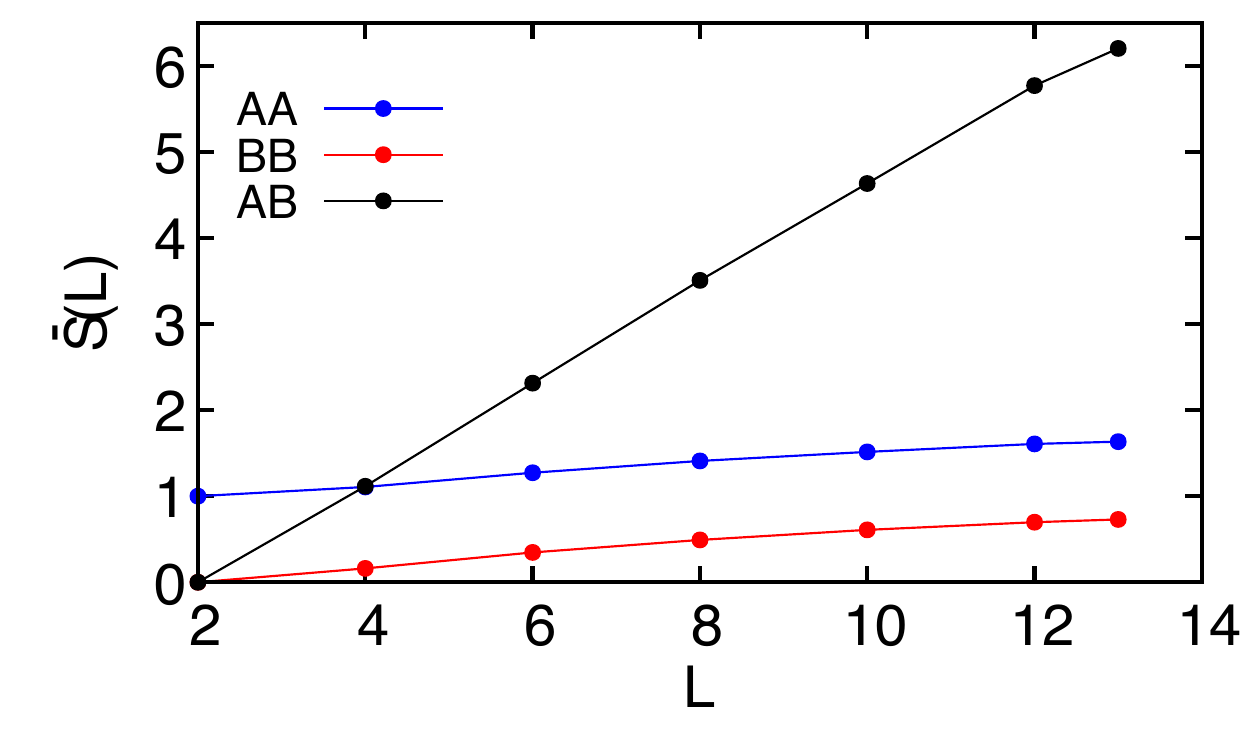}
\llap{\parbox[b]{2.cm}{\color{black}(b)\\\rule{0ex}{3.3cm}}}

\caption{Scaling in $L$ of  $\bar{S}[ c^{\dagger}_m , c_n]_L$ with $n,m=A,B$, for the two-dimensional model in Eqs. \eqref{BHZ_2D} and \eqref{BHZ_2D_lat}, for {\color{black} $M_{\Omega} = M_{\Psi} = -1$ (panel (a)) and  $M_{\Omega} = - M_{\Psi} = -1$ (panel (b)).} Open boundary conditions are assumed and disorder is introduced as described in the main text, with $\sigma=1.2$ and $100$ realizations. Qualitatively similar trends as in the clean case are observed.} 
\label{2D_open_dis}
\end{figure}

\begin{figure} [t]
\includegraphics[width=0.8\columnwidth]{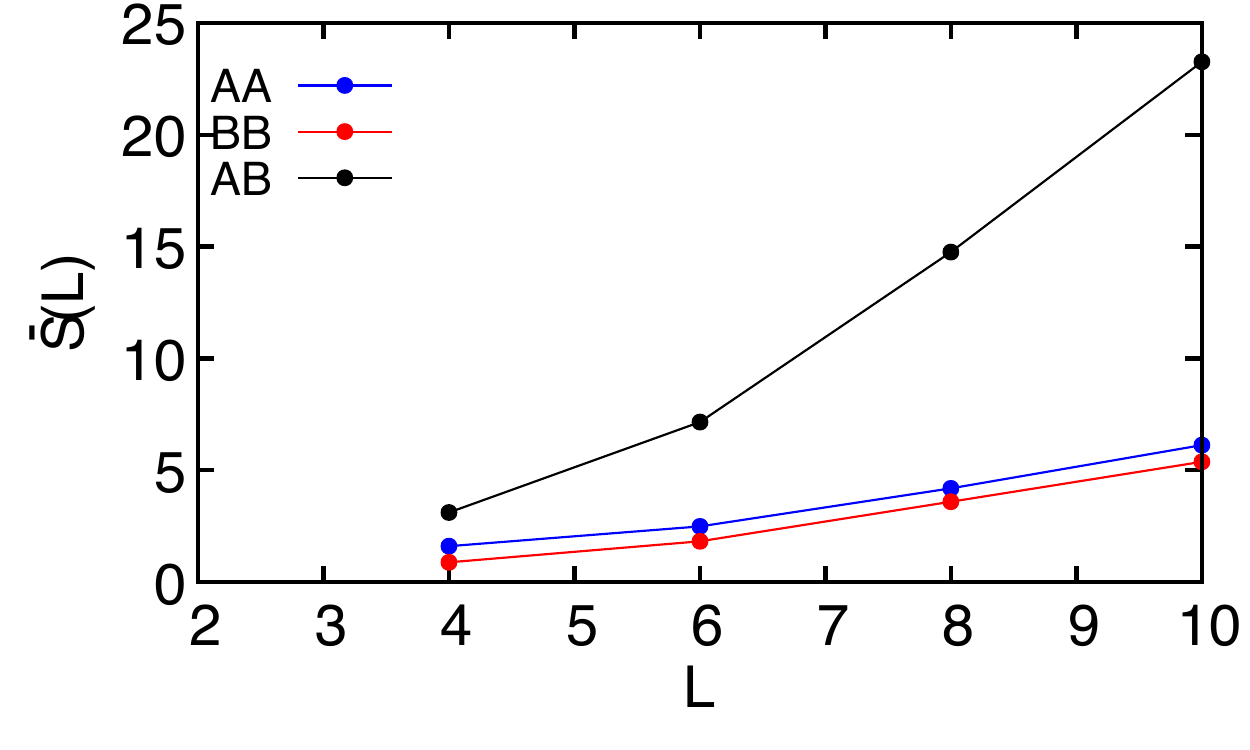}

\caption{Scaling in $L$ of  $\bar{S}[ c^{\dagger}_m , c_n]_L$ with $n,m=A,B$, for the two-dimensional model in Eqs. \eqref{BHZ_2D} and \eqref{BHZ_2D_lat}, for {\color{black}  $M_{\Omega} = - M_{\Psi} = -1$.} Open boundary conditions are assumed and disorder is introduced as described in the main text, with $\sigma=5$ and $100$ realizations.} 
\label{2D_open_largedis}
\end{figure}

Fig. \ref{2D_open_dis} shows strange correlator sums $\bar{S}$ for $\sigma = 1.2$. Results for different numbers of realizations $n_r = 10,100,200$ are produced, showing a satisfactory convergence with $n_r$. This fact is not obvious a priori, since it is expected only at sufficiently small $L$.

The data displayed are obtained by considering the same disorder configuration for both the $\ket{\Psi}$ and $\ket{\Omega}$ states. In both panels, $M_\Omega = -1$, corresponding to a trivial disordere state. The values of $\bar{S}[ c^{\dagger}_m , c_n]_L$ are then mediated on the various configurations.

Analyzing $\bar{S}[ c^{\dagger}_A , c_B]_L$, analogously to the clean cases in Fig. \ref{2D_open},  we clearly distinguish two different behaviors: $\bar{S}[ c^{\dagger}_A , c_B]_L$ tends to a constant if  $M_{\Psi} = -1$, while it grows linearly if $M_{\Psi} = 1$.

These results indicate the stability of the topological phase against local disorder for the moderate disorder corresponding to $\sigma=1.2$, to be compared with the single-particle gap, $\Delta E \approx 1.6$ at $M_{\Psi} = 1$. 

 {\color{black} Furthermore, when considering the space-time rotation of the strange correlators into correlation functions of gapless edge modes \cite{xu2014}, these results provide also an indirect signature of the stability of the edge states arising at the interface between disordered topological and non-topological phases, which are not localized by this moderate disorder \cite{altland1997,ludwig2009,bernevigbook}. 

Fig. \ref{2D_open_largedis} displays the behavior of $\bar{S}[ c^{\dagger}_m , c_n]_L$ for small system sizes when the disorder parameter, $\sigma = 5$, is closer to the transition from the topological insulating phase ($M_\Psi=1$) to a disordered-induced insulating phase. When crossing the phase transition, the linear behavior of the average $\bar{S}[ c^{\dagger}_A , c_B]_L$ as a function of $L$ varies from linear for small disorder, to superlinear when the disorder strength becomes comparable with the single particle gap of the clean system. This is an effect of the normalization $\bracket{\Omega}{\Psi}$ at the denominator of the strange correlators \eqref{dec}. For these intermediate values of the disorder, progressively more configurations display almost orthogonal $\ket{\Psi}$ and $\ket{\Omega}$ states such that the strange correlator denominators are orders of magnitude smaller than the numerators. The strange correlators $s$, in this case, do not display any longer the predicted decay with the distance and this effect changes the predicted qualitative scaling of $\bar{S}[ c^{\dagger}_A , c_B]_L$, whose average value appears to grow faster than linear in $L$.}

Finally, we notice that the present discussion on spatial disorder can be easily adapted to other types of disorder, as on gauge potentials, when optical lattices realizations of topological phases are considered, see e.g. \cite{burrello2013,burrello2016,libroanna}.

\section{Strange correlators and bulk-boundary correspondence}
\label{choice}

In the previous Sections, we probed a link between the linear growth of the sum of suitable bulk strange correlators and the anomalous dimensions of the 
edge mode operators. Here, we elaborate more on this important point, proposing, for a wide family of topological systems, a criterion for an optimal choice of the operators $\hat{o}$ in the strange correlators, that means displaying an algebraic decay if $\ket{\Omega}$ and $\ket{\Psi}$ host different topologies. We focus primarily on the two-dimensional examples.

In the Supplementary Material of Ref. \cite{xu2014}, it has been shown that  strange correlators of general local operators  $\phi_{\alpha}(x_1,y)$ and $\phi_{\alpha}(x_2,y)$, defined in the bulk of a two-dimensional lattice system displaying short-range entanglement and symmetry-protected topological edge modes, are mapped in the thermodynamic limit onto standard correlators of the edge-projected operators $\phi_{\alpha}(x_1)$ and $\phi_{\alpha}(x_2)$, defined on the ground-states of a suitable (1+1)-dimensional conformal theory that describes the related edge modes:
\begin{multline}
\frac{\langle \Omega | \phi_{\beta}(x_1,y) \, \phi_{\alpha}(x_2,y)|\Psi\rangle}{\langle \Omega |\Psi \rangle} = \\
\sum_a \frac{\langle \theta_a | \phi_{\beta}(x_1) \, \phi_{\alpha}(x_2)|\theta_a \rangle}{\langle \theta_a |\theta_a \rangle} ,
\label{map}
\end{multline}
$x_1$ and $x_2$ in the right hand term being located on the edge(s). 
In Eq. \eqref{map}, $a$ labels the different quantum vacua $\theta_a$ of the edge theory, distinguished by symmetries or one-dimensional topologies. In particular, it indicates the different conformal sectors, associated with primary operators, which describe the possible chiral or helical edge excitations of the state $\ket{\Psi}$, as, for instance, in quantum Hall or quantum spin Hall systems. The appearance of the edge modes clearly require systems with open boundaries, but we stress that, in spite of this fact, the map in Eq. \eqref{map} holds strictly if the lattice system has periodic boundary conditions.  \\
\indent This construction is based on the seminal paper \cite{qi2012}, where a direct link is traced between the low-energy Hamiltonian (say $H_{\mathrm{L}}$) of the chiral edge modes,  
described by a conformal field theory, of generic two-dimensional topological fermionic states, and their entanglement Hamiltonian $H_{\mathrm{E}}$, originated by the interaction with their counter-propagating twin edge states. Ref. \cite{qi2012} indicates indeed that the bulk states of topological systems carry the information related to the corresponding edge theory. This information can be extracted from reduced density matrices referring to bulk regions of the systems, as numerically verified also in the framework of fractional quantum Hall states \cite{haldane2008}. These observations can be further extended to fermionic topological systems hosting helical edge states, invariant under time-reversal symmetry.  Furthermore, analogous techniques have been further developed to decompose topological phases of matter in terms of one dimensional Luttinger liquids (see \cite{meng2020} and references therein), thus hinting to the fact that the construction in Ref. \cite{xu2014} is expected to hold also for 
 long-range entangled (intrinsically topologically ordered) system with gapless edge modes, as fractional quantum Hall models, provided to chose correctly the conformal theory related to the edge states. 
The mapping in Eq. \eqref{map} indicates a criterion for the optimal choice of the operators in the strange correlators. Indeed, 
as suggested by the link between the decay exponents of the bulk strange correlators and the anomalous dimensions of the 
edge mode operators, it is useful to consider the right-hand part of Eq. \eqref{map}, and to restrict to the case:
\beq
\sum_a \frac{\langle \theta_a | \eta^{\dagger} (x_1) \, \eta (x_2)|\theta_a \rangle}{\langle \theta_a |\theta_a \rangle} \, .
\label{choicepart}
\eeq
where $\eta^{\dagger}(x)$ is a primary field in the edge conformal theory and, in general, it corresponds to the creation of a quasiparticle belonging to the gapless edge mode.

\begin{figure*} [t]
\includegraphics[scale=0.21]{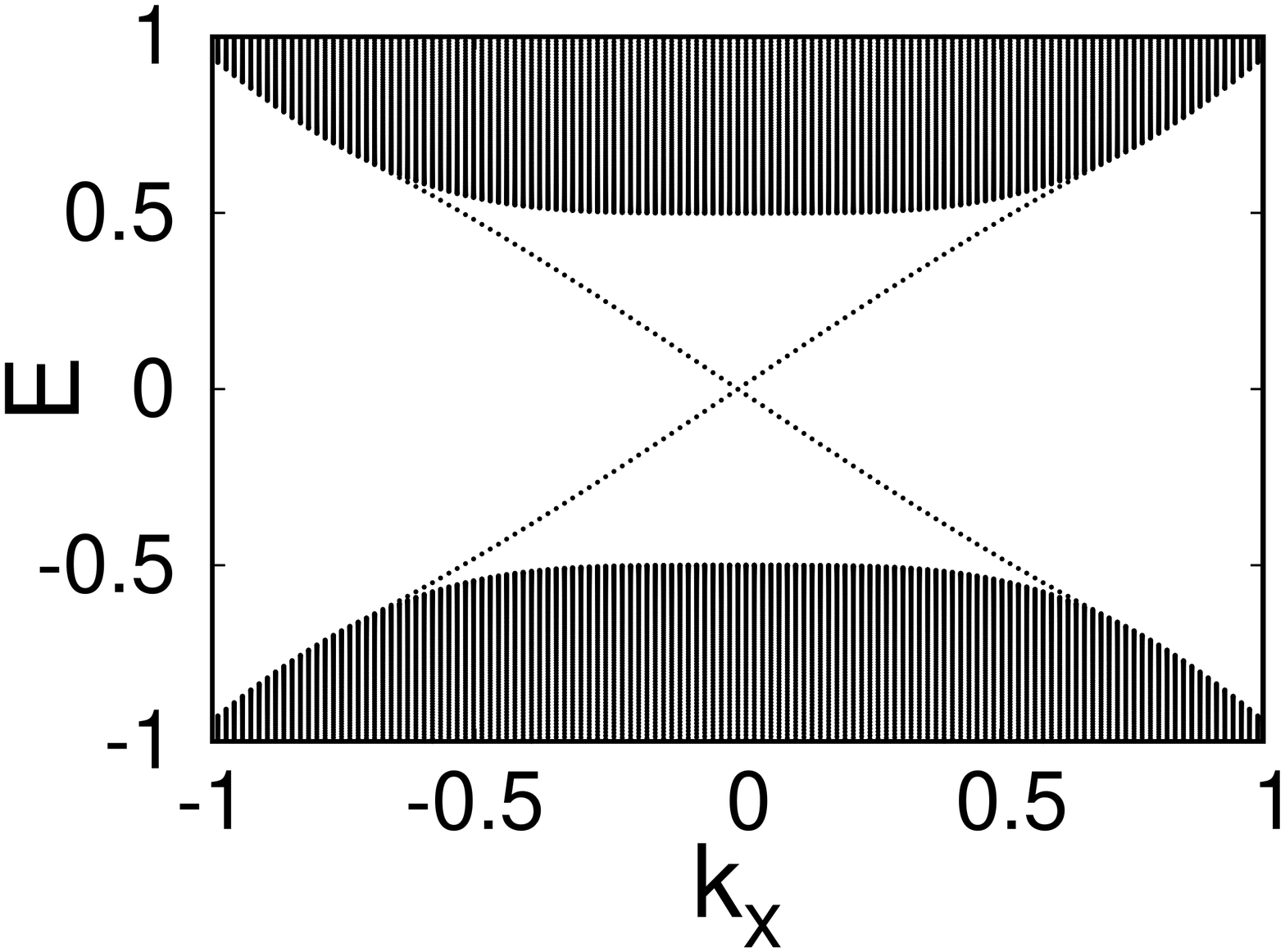}
\includegraphics[scale=0.21]{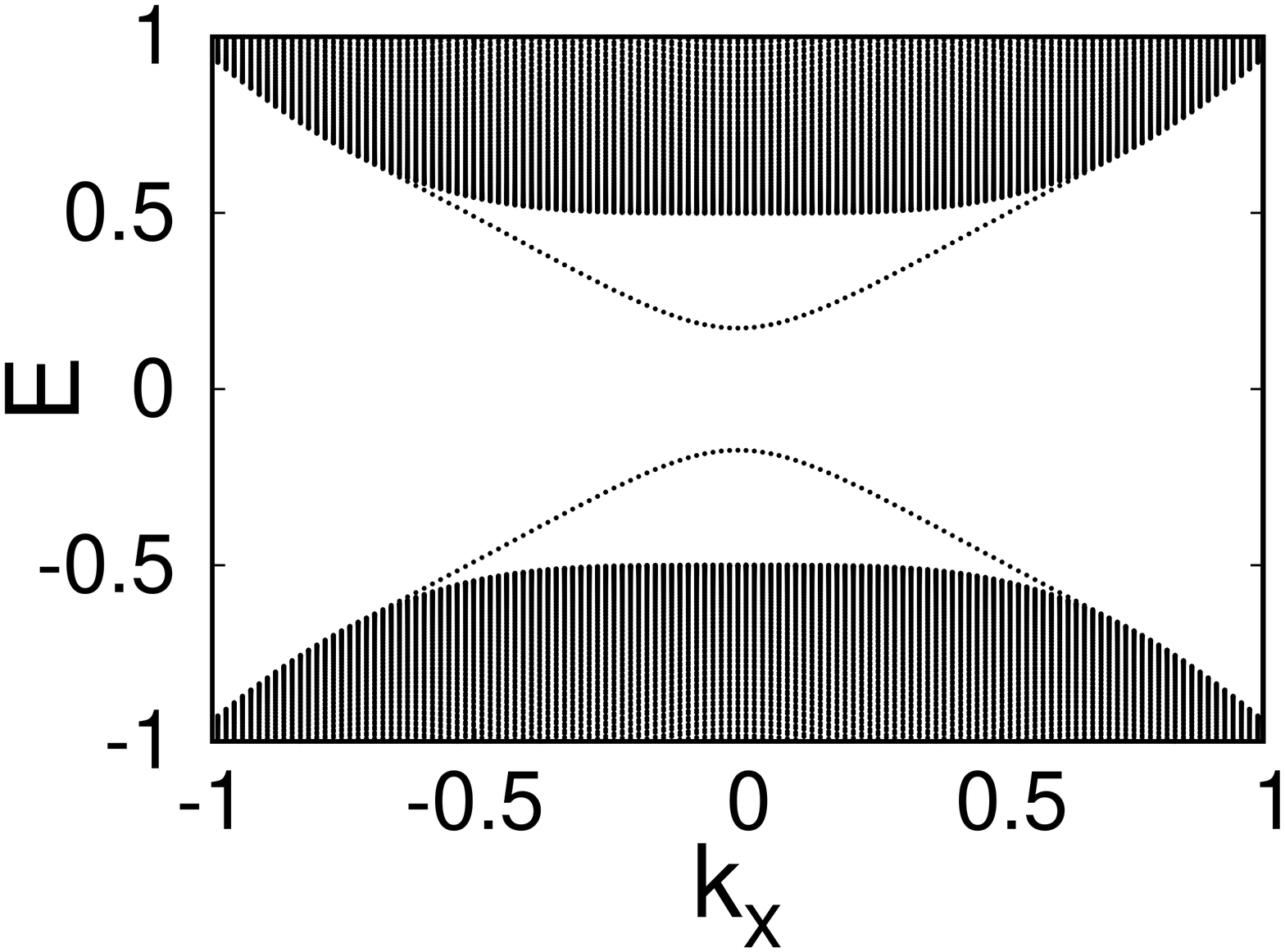}
\includegraphics[scale=0.21]{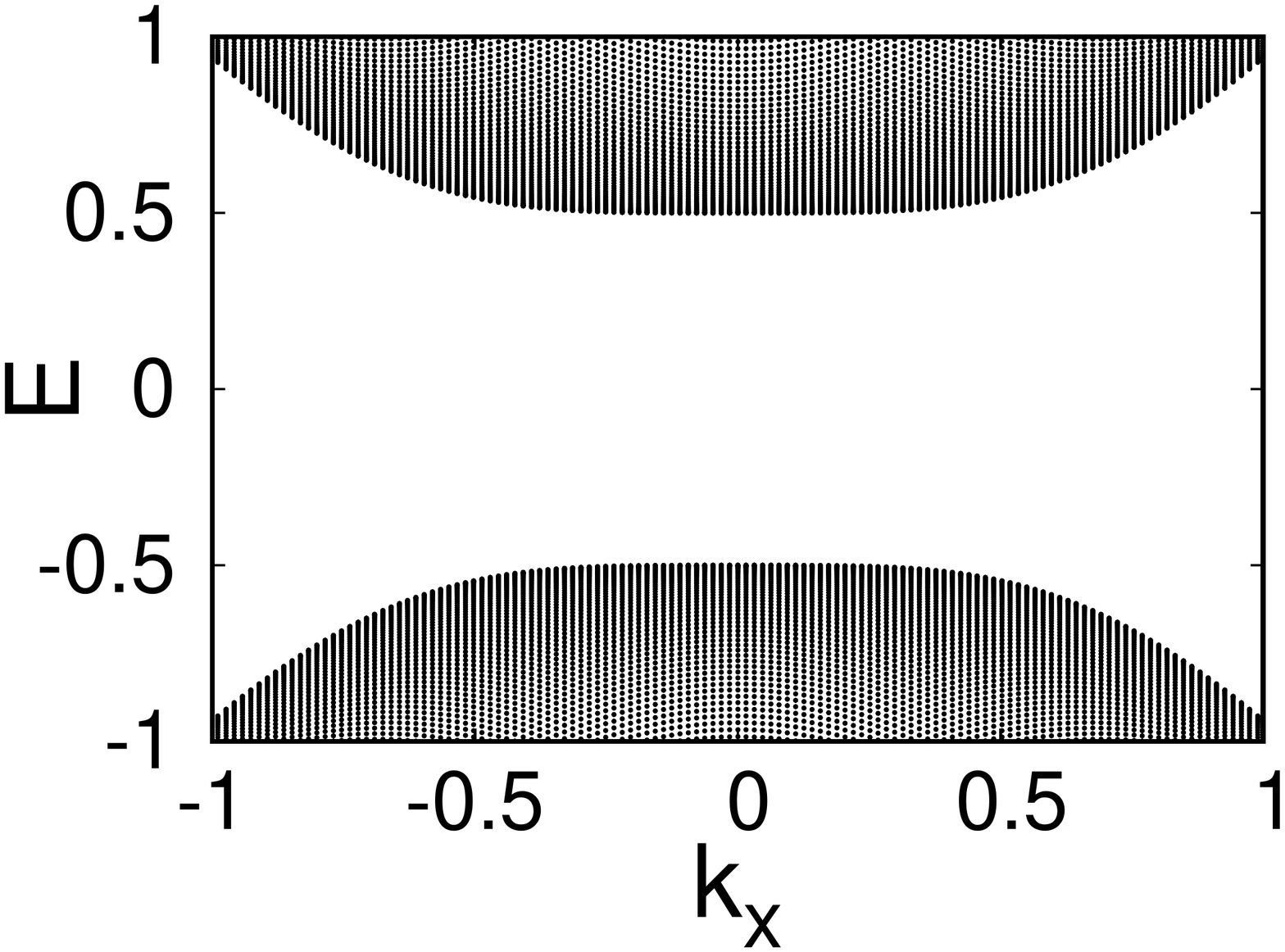}
\caption{Edge state evolution as one hopping along $\hat{y}$, say between the $L$-th and the $1$-st rows of sites, is varied between $0$ and $1$, such to interpolate continuously between open and periodic boundary conditions along $\hat{y}$, keeping periodic the boundary conditions along $\hat{x}$. We set $L = 400$ and $w_{1L}=0,0.3,1$ from the left-hand side (in units of the bulk tunneling amplitude). }
\label{edgev}
\end{figure*}

In order to exploit Eq. \eqref{map}, the next step is to understand which bulk operator $\hat{o}(x,y)$ corresponds to $\eta^{\dagger}(x)$. To this purpose, we apply a continuous evolution of the Hamiltonian, with the aim of merging two disconnected gapless edges into a gapped bulk: this is the situation considered in Fig. \ref{edgev} in which the boundary conditions of a topological system are evolved from open to periodic. In particular, we refer to $\eta(x,y)$ to represent the low-energy field $\eta(x)$ embedded in the two-dimensional system with open boundaries. The annihilation operator $\eta(x,y)$ must evolve into a linear combination of annihilation operators $\hat{o}(x,y)$ of bulk particles, as the boundary conditions evolve continuously. In particular, for systems invariant under translation along $x$, the evolution can be clearly seen in the $k_x$ momentum basis $\eta(k_x,y)$.

To elaborate on this point, we focus again on the model in Eq. \eqref{BHZ_2D}. The spectral evolution between gapless edge modes for an open system and the bulk quasiparticle states  in the closed system is analyzed numerically in Fig. \ref{edgev}: the boundary conditions evolve continuously and the geometry changes from a cylinder to a torus.
{\color{black} When considering, for instance, $y=1$, this is done by allowing the hopping amplitude} $w_{1L}$, between the $L$-th and the first row of sites, to vary adiabatically between $0$ and $1$. When $w_{1L} = 0$, we find  two chiral counterpropagating edge states interpolating between the bulk bands. 
As  $w_{1L}$ increases towards 1,  these edge modes mix and separate in energy, with an increasing energy gap, around $k_x = 0$. At the same time, we checked that their eigenstate wavefunctions become progressively more distributed in the bulk. Therefore, we conclude that both the counterpropagating edge modes $\eta(k_x)$ for open boundaries are continuously connected with linear combinations of the bulk operators $\gamma^{(+)}_{{\bf k}}$ and $\gamma^{(-)}_{{\bf k}}$ around ${\bf k} = 0$ annihilating quasiparticles in the two energy bands. This correspondence holds also in Eq. \eqref{map}.

Hence, the two-point correlations of the edge theory, evolve, in general, into a linear combination of the strange correlators associated to the operators $\gamma^{(\pm)\dag}_{{\bf k}} \gamma^{(\pm)}_{{\bf k}}$ for the periodic system in momentum space {\color{black}, where the labels $(\pm)$ refer to the involved bulk bands. For translationally invariant systems, the} most significant contribution is provided by:
\beq
 \frac{\bra{\Omega} \gamma^{(+)\dag}_{{\bf k}} \gamma^{(-)}_{{\bf k}} \ket{\Psi}}{\bracket{\Omega}{\Psi}}\, .
 \label{strange2d}
\eeq 
Here the $\gamma^{(\pm)}_{{\bf k}}$ operators refer to the quasiparticle states in the system whose ground-state is $\ket{\Psi}$, such that this strange correlator {\color{black} describes the creation of a particle-hole pair excitation over $\ket{\Psi}$ associated to the momentum $\bf k$. The strange correlators in real space in Eq. \eqref{choice3} can be Fourier-decomposed in terms of the kind \eqref{strange2d} for translational invariant systems.} 

{\color{black} We observe that, besides the contribution in Eq. \eqref{strange2d}, the other quadratic combinations of the $\gamma^{\pm}_{{\bf k}}$ operators return trivial results:  $ {\bra{\Omega} \gamma^{(-)\dag}_{{\bf k}} \gamma^{(-)}_{{\bf k}} \ket{\Psi}}/{\bracket{\Omega}{\Psi}}=1$ since $\ket{\Psi}$ is the state in which all the negative energy quasiparticle states are filled. This also implies that $S[\gamma^{(-) \, \dagger} ,\gamma^{(-)} ]_L = \bar{S}[\gamma^{(-) \, \dagger} , \gamma^{(-)} ]_L =1$ 
[since $s[\gamma^{(-) \, \dagger} ,\gamma^{(-)} ]_{{\bf r},{\bf r^{\prime}}} =  \delta_{{\bf r} , {\bf r}^{\prime}}$, see Eq. \eqref{sumper} and App. \eqref{strasl}, Eq. \ref{stringa}]. 
Instead, $\gamma^{(+)}_{{\bf k}}$, annihilates $\ket{\Psi}$, making the strange correlator with operators $\gamma^{(\pm)\dag}_{{\bf k}} \gamma^{(+)}_{{\bf k}}$ vanish. }

Using the direct expressions for $a^{(-)}_{A,{\bf k}}$ and $a^{(-)}_{B,{\bf k}}$ for $\ket{\Omega}$  and $\ket{\Psi}$ in Eq. \eqref{trasf}, it is easy to check that Eq. \eqref{strange2d} displays a singularity at ${\bf k} = 0$, analogously to the ${\bf k} = 0$ component of the strange correlator in Eq. \eqref{sumper} (see also App. \ref{strasl}). 

{\color{black} Summarizing,} the strange correlators in Eq. \eqref{choicepart}, on one side, can be estimated through the correlation functions of the conformal field theory, thus returning the typical power law decay as $1/|x_1-x_2|^{2 \, \alpha}$, where $\alpha$ is the scaling dimension of the primary operator $\eta$; on the other, {\color{black} they are connected with the particle-hole excitations of the bulk system, as indicated in Eq. \eqref{strange2d} (see also App. \ref{strasl}, Eq. \ref{strexp}). In the discussion above,} we adopted the BHZ model as a paradigmatic example, but analogous considerations hold for general two-dimensional topological insulators and superconductors, and can be further extended to other dimensions. {\color{black} In the case of chiral superconductors, for instance, the main difference is the fact that the edge modes $\eta$ correspond to Bogoliubov (typically Majorana) modes. In this case, the operators $\hat{o}$ must be chosen as suitable Bogoliubov operators in order to have a good overlap with the creation operator for a particle-hole excitation over the state $\ket{\Psi}$ as indicated by Eq. \eqref{strange2d}. This corresponds, for example, to the choice in Eq. \eqref{corrMajo} for the Kitaev chain.}

We also comment that, instead of Eq. \eqref{strange2d}, more complex strange correlators of excitons (particle-hole pairs) could be considered (in real space): 
\beq
\frac{\bra{\Omega} \gamma^{(+)\dag} ({\bf{r}) \gamma^{(-)} ({\bf r}) \gamma^{(+)\dag} ({\bf{r}^{\prime}) \gamma^{(-)} ({\bf r}^{\prime}) \ket{\Psi}}}}{\bracket{\Omega}{\Psi}} \, .
\label{bilinear}
\eeq
However, Wick theorem and particle conservation allow to reduce Eq. \eqref{bilinear} to the strange correlator in Eq. \eqref{strange2d} (in momentum space).

{\color{black} Another generalization is provided by boundaries characterized by multiple chiral modes, which correspond to different primary operators in the conformal description; in this case, our analysis must be extended to strange correlators of the kind:} 
\beq
\frac{\langle \Omega | \eta^{(\alpha)\dagger} (x_1,y) \, \eta^{(\beta)} (x_2,y)|\Psi\rangle}{\langle \Omega |\Psi \rangle} \,.
\label{choice3}
\eeq
This expression generalizes indeed strange correlators to multiple bulk operators connected with different boundary fields.
Consequently, it is necessary to choose the reference state $\ket{\Omega}$ in order to fulfill all the symmetry requirements of the system and $\eta$ operators. The resulting scaling must then be derived by the suitable operator product expansion of the related primary operators \cite{mussardobook}.
Due to the bulk-boundary adiabatic continuation discussed above,  the $\eta^{(\alpha)} (x,y)$ operators are connected with independent linear combinations of the lowest-energy bulk excitations, when closed boundary conditions are assumed. Moreover, the specific constant coefficients in these combinations influence only the prefactors but do not affect the dominant scaling behaviour  of the corresponding strange correlators. This fact allows us to focus directly on the strange correlators of the low-energy bulk excitations, similar as in Eq. \eqref{strange2d}.  Finally, as in Section \ref{iqhe},  the scaling in $L$ of the 
largest eigenvalue of the matrix in Eq. \eqref{choice3} still yields a criterion for topology at zero temperature, 
independently from translational invariance, analogous to the Onsager-Penrose criterion and ODLRO \cite{PS}.
In the Section \ref{LRE}, a similar discussion will be performed for Laughlin fractional quantum Hall state where edge excitations with different charges can be considered. 

{\color{black} At the end of this general discussion, it is worth to stress that the described procedure to identify the optimal operators for the strange correlators does not rely at all on Lorentz invariance. Indeed, the bulk-boundary correspondence at the basis of the validity of the method, although
formalized in \cite{xu2014} by a conformal field theory (relying on Lorentz invariance) approach, does not rely at all on the same invariance.
}

What can we say about one- and three-dimensional systems? For several topological 1D and 3D models similar continuity arguments can be presented and we can check the validity of Eqs. \eqref{choice3} 
on specific examples, as those proposed in the previous Sections.

\begin{figure*} [t]
\includegraphics[scale=0.6]{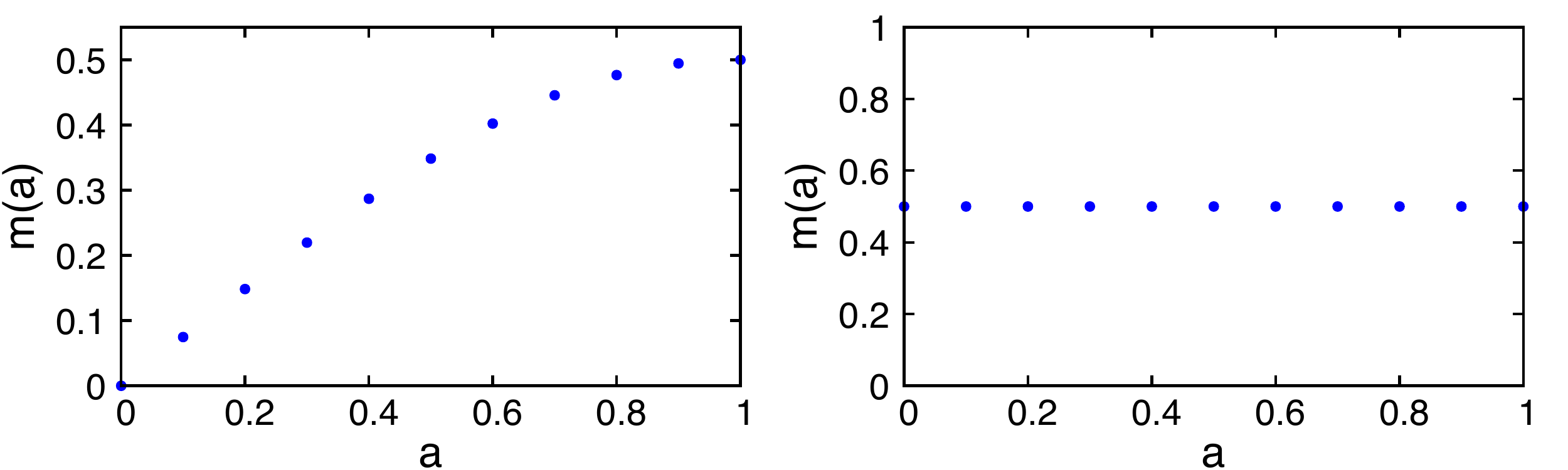}
\llap{\parbox[b]{17.0cm}{\color{black}(a)\\\rule{0ex}{3.5cm}}}
\llap{\parbox[b]{2.0cm}{\color{black}(b)\\\rule{0ex}{3.5cm}}}
\caption{Evolution of the energy $m$ of the lowest-energy excitation for the Kitaev chain in Eq. \eqref{kitaev} with $\delta = 10$, as the boundary conditions are changed adiabatically,
from antiperiodic ($a = 1$)  to open ($a =0$). We set $\mu = 0.5$ (in the topological regime) in the left panel and  $\mu = 1.5$ (in the non topological regime) in the right panel.}
\label{edgekit}
\end{figure*}

First of all, we focus on the Kitaev chain in Eq. \eqref{kitaev}. In Fig. \eqref{edgekit}, we report the evolution of the energy of the lowest excitation for the long-range Kitaev chain in Eq. \eqref{kitaev} with $\delta = 10$, as the boundary conditions are changed adiabatically,
from antiperiodic ($a = 1$)  to open ($a =0$). We set $\mu = 0.5$ (in the topological regime) in the left panel and  $\mu = 1.5$ (in the non topological regime) in the right panel.
In the topological phase, the excitation energy interpolates between $0$ and the bulk energy gap for $a$ going from open to antiperiodic boundary conditions, which is consistent with the existence of Majorana zero-modes and does not have a counterpart in the trivial state (right panel). Therefore, the adiabatic continuation between bulk and boundary modes is confirmed also in this one-dimensional case. 
Notably a similar behavior for the low-energy excitations is found also for the long-range Kitaev Hamiltonian with $\delta <1$, where, however, edge modes remain gapped in the thermodynamic limit. In this case, indeed, the validity of the mapping as in Eq. \eqref{map} is debated, so that in general we cannot assume an algebraic decay of the strange correlators associated to phases with gapped edge modes.

The analysis of strange correlators, however, is not restricted to fermionic models. In one dimension, this is easily shown by the analysis of the famous AKLT  spin-1 chain \cite{aklt},
\beq
\hat H_{\mathrm{AKLT}} = \sum_j {\bf s}_j \cdot {\bf s}_{j+1} + \frac{1}{3} ({\bf s}_j \cdot {\bf s}_{j+1})^2  \, .
\eeq
The ground-state of this Hamiltonian is known to be a valence bond solid, with a single bond connecting every neighboring pair of sites.
More in detail, each spin-1 can be decomposed in terms of two spins $\frac{1}{2}$ degrees of freedom, projected on a triplet state; in turn a  spin $\frac{1}{2}$ on a site is organized in a singlet, together with a spin $\frac{1}{2}$ on the neighbouring site.
For the case of periodic boundary conditions, the AKLT chain has a unique ground-state, $\ket{\mathrm{GS}}$. Instead, if open boundaries are assumed, the first and last spin-1 of the chain have only a single neighbor, leaving one of their constituent spin $\frac{1}{2}$ unpaired. As a result, the ends of the chain behave like free spins $\frac{1}{2}$.
For finite chains, these edge states mix in a singlet and a higher-energy triplet states, similarly as for the closed chain, while,
as the size increases, the edge states decouple exponentially, leading to a ground-state manifold that is four-fold degenerate. Therefore, the edge states are massless in the thermodynamic limit.
In this scheme, the ${\bf s}_j^{(\pm)}$ operators, applied at the end of a finite-size chain, create edge excitations with vanishing scaling dimension. Restoring adiabatically the closed boundary conditions, as described above for the model in Eq. \eqref{BHZ_2D}, it turns out that the same operators create the minimal energy excitations above the unique ground-states.
These facts explain the behaviour 
\beq
\langle \Omega | {\bf s}_i^{(+)} {\bf s}_j^{(-)} | \Psi \rangle \to \mathrm{constant} \, \, \, \, \,  \mathrm{as} \, \, \, \, \, |i-j| \to \infty 	\, , 
\eeq 
obtained for the closed chain in \cite{xu2014} and confirms our picture above, based on Eq. \eqref{choice3}. The same behaviour is actually similar to that for the Kitaev chain in Eq. \eqref{kitaev}, at $\delta >1$.

Finally, a similar discussion as for the model in Eq. \eqref{BHZ_2D} can be straightforwardly extended, especially numerically; to the three-dimensional case in Eq. \eqref{BHZ_3D}, although
the bulk-boundary analytical mapping \cite{qi2012}, provided by Eq. \eqref{map}, is not available. The same picture, encoded in Eq. \eqref{choice3}  can emerge.

Summing up, all the analyzed cases notably suggest a continuous evolution between lowest-energy excitations above the ground-state at closed boundary conditions and edge excitations, at least if massless. This behaviour is someway expected, since fortuitous energy crossings, in the absence of symmetries, are generally suppressed, especially in the absence of disorder. For the same reason, and as commented for the Kitaev chain at $\alpha <1$, the same spectrum evolution can hold also if edge states are gapped, although not leading to an algebraic decay for the strange correlators defined as in  Eq. \eqref{choice3}.
The same choices as in Eq. \eqref{choice3}  can hold for topological bosonic systems, also disordered. 

Finally, we comment that the mentioned construction in Ref. \cite{xu2014} {\color{black} can be applied to the more general case where $\ket{\Omega} $ and $\ket{\Psi}$ have different nontrivial topologies, as in Section \ref{iqhe}, but additional observables need to be considered to infer the nature of the phase transition between them, and the related change in their Chern number}. Moreover, in the presence of long-range couplings, as in Section \ref{kitaevLRs}, {\color{black} the standard strange correlator construction seems to fail as a reliable diagnostic tool for topological phase transitions, at least for the system sizes that we investigated ($L \le 160$)}. This is probably due to the fact that in this condition the separation between bulk- and boundary-states is generally not well stated and justified, as also suggested by the appearance of a mass gap for the edge states.

\section{Topological order in the Laughlin state}
\label{LRE}

The models analyzed in the previous Sections exemplify the behavior of strange correlators for symmetry-protected topological phases. Such topological states are characterized by short-range entanglement \cite{wen2010} and do not display genuine intrinsic topological order (thus degenerate ground-states on a torus geometry and anyonic excitations) and long-range entanglement. For two-dimensional systems, however, the strange correlators are not limited to short-range entangled systems with symmetry-protected topological phases, but can additionally be applied as a general tool to detect the onset of chiral gapless edge modes also in systems with intrinsic topological order and long-range entanglement \cite{wen2010}. In the following, we will discuss the behavior of the strange correlators in the paradigmatic example of the simplest fractional quantum Hall (FQH) state, namely the Laughlin state. In App. \ref{toricode}, instead, we show that the toric code does not display a power law decay of the strange correlators as expected for topologically order systems without gapless topological edge modes.

Similar to the symmetry-protected topological systems,
it is known \cite{kane2002} that also Abelian FQH phases can be described by arrays of one-dimensional quantum wires, which, essentially, constitute one-dimensional copies of the same conformal field theory that characterizes their edge modes. Analogous constructions have been obtained also for several non-Abelian FQH states \cite{teokane}. In the simplest scenario of the Laughlin state at filling $\nu=1/m$ with odd $m$, each one-dimensional subsystem can be effectively described by a Luttinger liquid with Hamiltonian:
\begin{equation} \label{luttingerham}
H_j = \frac{v}{2\pi} \int dx \, K \left(\partial_x \varphi_j \right)^2 + \frac{1}{K} \left(\partial_x \theta_j \right)^2\,,
\end{equation}
where the Luttinger parameter $K$ represents the effect of repulsive interactions, and the operators describing electrons in the bulk of the theory assume a bosonized form whose dominating contribution is provided by:
\begin{equation}
\Psi_j(x) \propto e^{-ik_Fx} e^{i\left(\varphi_j(x)+\theta_j(x)\right)} + e^{ik_Fx} e^{i\left(\varphi_j(x)-\theta_j(x)\right)}\,.
\end{equation}
Here $j$ labels the $y$ coordinate of the electrons, and $k_F=\pi \rho_{1D}$ is the effective Fermi momentum of the 1D subsystem, related to the 1D particle density $\rho_{1D}$. Each Hamiltonian of the kind \eqref{luttingerham} corresponds to a simple $c=1$ bosonic conformal field theory.
To obtain a Laughlin state, these 1D subsystems are gapped by interactions of the form \cite{kane2002,teokane}:
\begin{equation} \label{kaneint}
H_I \propto \sum_j e^{i\left[\frac{Bea}{\hbar} -2mk_F \right]x}e^{i\left[\varphi_{j+1} - \varphi_j - m \left(\theta_{j+1} +\theta_{j}\right)\right]} +{\rm H.c.}\,,
\end{equation}
where $a$ is the distance between neighboring 1D subsystem. Such interactions couple pairs of fields with opposite chirality in neighboring edges; they can become relevant in the renormalization group sense, and, consequently, open a bulk gap, only if $Bea/\hbar = 2\pi \rho_{1D} m$, thus setting the value of the filling factor $\nu= 1/m$. In particular, the Laughlin state can be mapped into a non-interacting state of composite fermions \cite{jainbook} and its onset corresponds  to interactions such that $K=1/m$.

The previous interaction, however, does not affect two counter-propagating modes lying at the edges of the 2D system: under the previous assumptions, indeed, the combination of fields $\varphi_1 - m \theta_1$ and $\varphi_L + m \theta_L$ are not affected for systems with open boundaries. This provides a description of the Laughlin edge modes based on chiral Luttinger liquids \cite{fradkinbook} of the form:
\begin{equation} \label{chiralLutt}
H_{b} = \frac{vm}{2\pi} \int dx \left(\partial_x \varphi_{e^*} \pm \partial_x \theta_{e^*}\right)^2\,.
\end{equation}
The pair of dual bosonic field $\varphi_{e^*}=\varphi/m$ and $\theta_{e^*}=\theta$ are associated to the chiral (left or right) vertex operators describing fractional quasiparticles with charge $e^*=e/m$:
\begin{equation}\label{qplaughlin}
\Psi_{e^*}^{(L/R)}\propto e^{i\left(\varphi_{e^*}\pm \theta_{e^*}\right)},
\end{equation}
and their equal-time commutation relation is given by:
\begin{equation}
\left[\varphi_{e^*}(x_1),\theta_{e^*}(y)\right]=-i \frac{\pi}{m} \Theta\left(y-x\right),
\end{equation}
such that, for $m=1$, they describe chiral fermionic modes, consistently with integer quantum Hall states. The electron annihilation operators in the Laughlin edge states are instead given by $\Psi_{e}^{(L/R)}\propto e^{i\left(m\varphi_{e^*}\pm \theta_{e^*}\right)}$.

Following the Section \ref{choice}, the suitable strange correlators to expose topology can be identified exploiting the mapping between the two-point functions of the conformal field theory defined by Eqs. \eqref{luttingerham} and \eqref{chiralLutt} and the two-point strange correlators in the bulk \cite{xu2014}. When considering the $\nu=1/3$ Laughlin state, the creation operators of the elementary $e/3$ fractional excitations correspond to the conformal vertex operators \eqref{qplaughlin}, and determine the following scaling through Eq. \eqref{map}:
\begin{equation}
\frac{\left\langle \Omega (N) \right| \Psi_{e^*}^{L\dag}({\bf r}_1)  \Psi_{e^*}^{L}({\bf r}_2)\left| \Psi_{\rm L} (N)\right\rangle}{\bra{\Omega (N)}\Psi_{\rm L}(N)\rangle} \propto \left|{\bf r}_1-{\bf r}_2\right|^{-1/3}\,,
\end{equation}
where $\ket{\Omega(N)}$ represents a non-topological gapped state of $N$ electrons, whereas $\ket{\Psi_{\rm L}(N)}$ corresponds to the Laughlin state (or a state in the same universality class). The exponent $1/3$ results from the scaling dimension $\eta_{e^*}=1/6$ of the quasiparticle operators in Eq. \eqref{qplaughlin}, which can be derived by the Hamiltonians \eqref{luttingerham} or \eqref{chiralLutt}. The related sum $\bar{S}$ scales as $L^{5/3}$.

In an analogous way, we can derive the decay of the two-point strange correlator associated with the electronic operators $\Psi_e$. It results
\begin{equation}
\frac{\left\langle \Omega (N) \right| \Psi_{e}^{L\dag}({\bf r}_1)  \Psi_{e}^{L}({\bf r}_2)\left| \Psi_{\rm L} (N)\right\rangle}{\bra{\Omega (N)}\Psi_{\rm L}(N)\rangle} \propto \left|{\bf r}_1-{\bf r}_2\right|^{-5/3}\,,
\end{equation}
consistent with a scaling dimension $\eta_{e}=(1/4K)+(K/4)=5/6$ for interacting electrons with $K=1/3$. The related sum $\bar{S}$ scales as $L^{1/3}$.

The wire construction, recalled in Section \ref{choice}, of topological phases of matter shows that the onset of gapped two-dimensional bulk states hosting protected gapless edge modes, described by conformal field theories, is signaled by the power-law decay of the related strange correlators \cite{xu2014}.

We emphasize, however, that this is not true in general for topologically ordered 
states without protected gapless edge modes.   The main counterexample is indeed provided by Kitaev's toric code \cite{kitaev2003} 
and its generalization to a pure $\mathbb{Z}_2$ lattice gauge theory \cite{fradkinbook}, in which two-point strange correlators of local operators cannot discriminate between the topological and the non-topological phase. In App. \ref{toricode}, we show indeed that two-point strange correlators display a trivial behavior for the ground-states of the toric code model. This is verified when considering single-qubit operators. It is known, however, that the creation of pairs of excitations in the toric code and lattice gauge theories is associated in general with string-operators. We also verified that a straightforward extension of the two-point strange correlators to string-like strange correlators does not provide either a diagnostic tool for the onset of topological order in the toric code.

Finally, we observe that the construction of the strange correlators can be extended to two-dimensional critical lattice models \cite{vanhove2021} and to non-chiral states with true topological order (as in the case of string-net models \cite{wen2005}) based on projected entangled pair states, with suitable matrix product operator (MPO) symmetries on the virtual level of these tensor network states \cite{vanhove2018}. The construction of these strange correlators, however, is based on the introduction of virtual anyonic defect lines, which, in turn, result in the introduction of anyons in the bulk of these topological systems \cite{haegeman2015}. This can be achieved by acting with suitable open MPO strings on the virtual level of these tensor network states. On the physical level, the so-generalized strange correlators would no longer be simple two-point functions, but they would rather correspond to the application of suitable string operators in the bulk of the system. The results of \cite{vanhove2018} therefore suggest that it could be possible to generalize the strange correlators in Eq. \eqref{dec} to a string operator construction, without breaking the mapping into suitable conformal field theories. This would provide a useful tool to investigate the physics of anyonic models or lattice gauge theories.

\section{Conclusions}  

In this work, we analyzed the scaling behavior of the strange correlators, for several examples of topological phases of matter, characterized by gapless protected boundary modes. 
The paradigmatic examples that we examined include both models that display short-range entanglement (symmetry-protected topology), as topological insulators and superconductors in various dimensions, and models with long-range entanglement (intrinsic topological order). We also addressed long-range coupling effects.

 In order to improve the efficiency in the diagnosis of topological states, as well as to reduce the impact of finite-size effects and disorder, we considered
the sum of the strange correlators; and we showed, in particular, that the sum of their moduli enables to detect the universal scaling dimensions of the operators defining the related gapless boundary theories.  It is interesting that our results do not depend qualitatively on the assumed boundary conditions.

Furthermore, we discussed a general strategy 
for the optimal choice of the operators entering the definition of the strange correlators, which must be based on the bulk-boundary correspondence of the models under scrutiny. 
 Our results clarify and extend the validity of the strange correlators approach to reveal the onset of topological phases with protected gapless edge modes described by a conformal field theory. Furthermore,  they provide a tool to discern these phases, already on small-enough systems, as those relevant in typical numerical simulations.
 {\color{black} Overall, our work contributes relevantly to elaborate a general strategy to probe (symmetry-protected and genuine) topology in finite size systems, possibly with different boundary conditions and disorder, and by local operators.}

Our discussion pointed out the analogy of the strange correlators approach with the analysis of off-diagonal long-range order, used to define Bose-Einstein condensation. 
While this work focused on systems at zero temperature, we think that an interesting point to be clarified is whether one can actually define a Penrose-Onsager criterion (or the equivalent of it) criterion for topological phases at finite temperature.

 The dependence of the strange correlators on the number and kind of edge modes (in turn related with the values of the topological invariants), also requires further investigation, as well as
the formal generalization of the construction in \cite{xu2014} to the more general case where $\ket{\Omega} $ and $\ket{\Psi}$ have different nontrivial topologies. Instead, the validity of the same construction in the presence of long-range couplings results debated from our analysis. Finally, we leave open the general issue whether a direct strategy can be drawn to measure the Chern numbers via the strange correlators. \\

{\bf Acknowledgements --} 
We thank warmly Maria Luisa Chiofalo, Miguel Angel Martin-Delgado, Andrea Nava,  Luca Pezz\'{e}, and Augusto Smerzi for useful discussions, correspondence, and support. {\color{black} We also thank Stavros Athanasiou and Matteo Michele Wauters for sharing their data on several topological models with disorder.}   

L. L. acknowledges financial support by the Qombs Project [FET Flagship on Quantum Technologies grant n. 820419]. M.B. is supported by the Villum foundation (research grant n.~25310). 
L. L. also acknowledges financial support by a project funded under the National Recovery and Resilience Plan (NRRP), Mission 4 Component 2 Investment 1.3 - Call for tender No. 341 of 15/03/2022 of Italian Ministry of University and Research funded by the European Union – NextGenerationEU, award number PE0000023, Concession Decree No. 1564 of 11/10/2022 adopted by the Italian Ministry of University and Research, CUP D93C22000940001, Project title "National Quantum Science and Technology Institute" (NQSTI).
S. P. acknowledges financial support from the University of L'Aquila by the internal project "Classical and Quantum Entropy".

\newpage


\begin{appendix}
\onecolumngrid

\section{Details on the scaling of the sums of strange correlators}
\label{appscal}

In this Appendix, we comment more on the scaling of the sums of strange correlators, mentioned in Sec. \ref{scalings}.
Focusing at first on translational invariant lattice systems, we can fix conventionally a "zero-point" (which all the distances are measured from) site and write:
\beq
\bar{S}[ \hat{o} , \hat{o}^{\prime}]_L = \sum_{{\bf r}}  \, |s[ \hat{o}, \hat{o}^{\prime}]_{{\bf r},0}| \, .
\eeq
We assume first a power law scaling $|s[ \hat{o},  \hat{o}^{\prime}]_{{\bf r},{\bf r^{\prime}}}| \sim |{\bf r}-{\bf r^{\prime}}|^{-2\alpha}$,  and we compare the sum of the moduli of the strange correlators at different sizes,
$\bar{S}[ \hat{o}, \hat{o}^{\prime}]_{L+1}$  and $\bar{S}[ \hat{o} , \hat{o}^{\prime}]_L$, related to two lattices with $L^d$ and $(L+1)^d$ sites respectively,
in the $L \to \infty$ limit. The two quantities differ for the \emph{dominant} contributions of $\sim L^{d-1}$ sites around the boundaries of the bigger lattice, having distance $\sim L$ from the zero-point site. Therefore we obtain, at the leading order:
\beq
\bar{S}[ \hat{o} , \hat{o}^{\prime}]_{L+1} - \bar{S}[ \hat{o} , \hat{o}^{\prime}]_L \sim L^{d-1} \, L^{-2 \alpha} = L^{d-1-2 \alpha} \, ,
\label{incr}
\eeq
such that $\bar{S}[ \hat{o} , \hat{o}^{\prime}]_L \sim L^{d-2 \alpha}$ if $\alpha < \frac{d}{2}$.  Instead, 
if $\alpha = \frac{d}{2}$, then $\bar{S}[ \hat{o} , \hat{o}^{\prime}]_L \sim \mathrm{ln} L$.
Finally, if $\alpha >\frac{d}{2}$,
$\bar{S}[ \hat{o} , \hat{o}^{\prime}]_L \sim c + L^{d-2 \alpha}$, $c$ being a dominating constant.

Vice versa, if we assume the scaling of the sums $\bar{S}[ \hat{o} , \hat{o}^{\prime}]_L\sim L^{d-2 \alpha}$, we directly obtain
Eq. \eqref{incr}. In turn, because of the same counting of the lattice sites as above, if $\alpha < \frac{d}{2}$, it results immediately that, for $L \to \infty$, 
\beq
|s[ \hat{o}, \hat{o}^{\prime}]_{L {\bf n} ,0}| \sim L^{-2 \alpha}  
\label{scal}
\eeq
($0< \alpha \leq 1$ and ${\bf n}$ being a unit vector), plus possible subdominant terms. Similar arguments immediately hold for the case $\alpha \geq \frac{d}{2}$.
We stress that no assumption is made about scale invariance of the lattice, then 
on the possible presence of a mass gap in its spectrum.
 In the absence of translational invariance,  the demonstration proceeds along similar lines as above, at least if the asymptotical
 behaviour for $|{\bf r}-{\bf r^{\prime}}| \to \infty$ can be defined. 
 If $\ket{\Psi} = \ket{\Omega}$, the strange correlator reduces to a standard correlator. In this case, the scaling of $s[ \hat{o} , \hat{o}^{\prime}]_L$ has been exploited recently, for scale-invariant, i. e. critical, lattice systems in \cite{hauke2016} and gapped one-dimensional ones in \cite{pezze2017,sengupta2014bis}.

If  $|s[ \hat{o} , \hat{o}^{\prime}]_{{\bf r}, {\bf r^{\prime}}}|$ decays exponentially, as for short-range gapped systems, then $\bar{S}[ \hat{o} , \hat{o}^{\prime}]_L$ tends  to a constant  \cite{hauke2016,pezze2017},  similarly to the case $\alpha > \frac{d}{2}$ above. Indeed, from Eq. \eqref{incr} and in the presence of an exponential decay, characterized by the decay length $\xi \ll L$, we obtain that
\beq
\bar{S}[ \hat{o} , \hat{o}^{\prime}]_{L+1} - \bar{S}[ \hat{o} , \hat{o}^{\prime}]_L \sim L^{d-1} \, \frac{e^{-\frac{L}{\xi}}}{L^{2 \alpha}} \to 0 \;\text{as}\;  L \to \infty,
\label{incr2}
\eeq
such that $\bar{S}[ \hat{o} , \hat{o}^{\prime}]_L \sim c^{\prime} + \frac{e^{-\frac{L}{\xi}}}{L^{1+2 \alpha-d}}$  (the constant  $c^{\prime}$ being possibly vanishing) with $L \to \infty$. Instead, Eq. \eqref{incr2} implies a subalgebraic scaling for $s[ \hat{o} , \hat{o}^{\prime}]_{{\bf r}, {\bf r^{\prime}}}$. Notice that, if $\xi \to \infty$ (at the gapless points and with infinite $L$), the scaling in Eq. \eqref{incr} is recovered. Finally, in the finite-size gapless regime we have $\xi = L$, then we still recover the scaling in Eq. \eqref{incr}, plus possible subdominant finite-size terms.

\section{Derivation of the strange correlators for a periodic Kitaev chain} 
\label{app:Kitaev}
From the definition of the eigenmodes of the Kitaev chain in Eq. 
\eqref{eigenmodes} 
on a generic Bogoliubov ground-state $\ket{GS}$, we obtain:
\begin{equation} \label{cdagcdag}
\left\langle GS \left|c^\dag_k c_k\right| SS\right\rangle=|v_k|^2\,,\quad \left\langle GS \left|c^\dag_k c^\dag_{-k}\right| GS \right\rangle=v^*_ku_k\,.
\end{equation} 
Therefore, the overlap between two normalized ground-states of the Kitaev chain in Eq. 
\eqref{Hkitmom} 
can be easily derived by considering the form of two ground-states  $\ket{\Omega}$ and $\ket{\Psi}$:
\begin{widetext}
\begin{multline} \label{kitaevoverlap2}
\bracket{\Omega}{\Psi}= \bra{0} \left[\prod_{0\le k \le \pi}\left(u^{\prime *}_{k} + v^{\prime *}_{k}c_{-k}c_k\right)\right]\left[\prod_{0\le k \le \pi}\left(u_{k} + v_{k}c_{k}^\dag c^\dag_{-k}\right)\right] \ket{0}=\\
\prod_{0\le k < \pi}\left(u^{\prime *}_k u_k + v^{\prime *}_k v_k\right) =
\prod_{0\le k < \pi} \left[\cos\frac{\theta(k)}{2}\cos\frac{\theta'(k)}{2} + \sin\frac{\theta(k)}{2}\sin\frac{\theta'(k)}{2} \right]  = 
\prod_{0\le k < \pi} \left[\cos\frac{\theta(k) - \theta'(k)}{2} \right] ,
\end{multline}

where all the primed coefficients are related to the ground-state $\ket{\Omega}$.

Also concerning the strange correlators, the terms in their numerators can be straightforwardly derived considering the form of the ground-states in Eq. \eqref{Gauss}.
We the following convention for the Fourier transform of the operators $c^\dag_x = \sum_q e^{iqx} / \sqrt{L}$, and we obtain:
\begin{equation}
\bra{\Omega}c^\dag_x c^\dag_y \ket{\Psi} = \mathcal{N}'  \mathcal{N} \sum_{qq'} \frac{e^{i(qx+q'y)}}{L}\bra{0} \left(\prod_{0\le k < \pi} e^{g^{\prime *}_kc_{-k} c_k}\right)c^\dag_q c^\dag_{q'} \left(\prod_{0\le k < \pi} e^{g_kc^\dag_k c^\dag_{-k}}\right)\ket{0} .
\end{equation}
The only non-vanishing terms come for $q=-q'$, because in the other cases two different Cooper pairs are broken. Then we split the sum over $q$ into negative and positive values and we obtain:
\begin{multline}
\bra{\Omega}c^\dag_x c^\dag_y \ket{\Psi} = \\
\mathcal{N}' \mathcal{N} \sum_{q>0} \frac{1}{L} \bra{0}\left(1+g^{\prime *}_q c_{-q}c_q\right) \left(e^{iq(x-y)} c^\dag_{q}c^\dag_{-q} + e^{-iq(x-y)}c^\dag_{-q}c^\dag_{q} \right) \ket{0} \bra{0} \prod_{ k\neq q,-q} e^{g^{\prime *}_kc_{-k} c_k} e^{g_kc^\dag_k c^\dag_{-k}}\ket{0} = \\
\sum_{q>0} 2i g^{\prime *}_q \frac{\sin q(x-y)}{L} u'_q u_q \prod_{k \neq q} \left(u_k'u_k + v^{\prime *}_k v_k\right)=
\left[\prod_{0\le k < \pi} \left(u_k'u_k + v^{\prime *}_k v_k\right) \right]\sum_{q>0} \frac{2iu_qv^{\prime *}_q \sin q (x-y)}{L\left(u'_qu_q + v^{\prime *}_qv_q\right)} \, ,
\label{cccorr}
\end{multline}
\end{widetext}
where we used that $u_q$ is real. For $\ket{\Omega}=\ket{\Psi}$, this is consistent with Eq. \eqref{cdagcdag}.
The term in the square brackets matches exactly Eq. \eqref{kitaevoverlap2}. Therefore, the resulting strange correlator $s\left[\ket{\Omega},\ket{\Omega},c^\dag,c^\dag\right]_{x,y}$ is given by:
\begin{multline} 
\frac{\bra{\Omega}c^\dag_x c^\dag_y \ket{\Psi}}{\bracket{\Omega}{\Psi}} = \sum_{q>0} \frac{2iu_qv^{\prime *}_q \sin q (x-y)}{L\left(u'_qu_q + v^{\prime *}_qv_q\right)} = \\
-\sum_{q>0} \frac{2\sin q(x-y) \cos \frac{\theta(q)}{2}\sin\frac{\theta'(q)}{2}}{L \cos \frac{\theta(q)-\theta'(q)}{2}}\,.
\label{Kcdcd_bis}
\end{multline}
The calculation for $s\left[c,c\right]_{x,y}$ is analogous.
Also the number conserving strange correlators $s\left[\ket{\Omega},\ket{\Omega},c^\dag,c\right]_{x,y}$ are derived in a similar way:
\begin{widetext}
\begin{multline}
\bra{\Omega}c^\dag_x c_y \ket{\Psi} = \\
\mathcal{NN}' \sum_{q} \frac{1}{L} \bra{0}\left(1+g^{\prime *}_q c_{-q}c_q\right) e^{iq(x-y)} c^\dag_{q}c_{q} \left(1+g_q c^\dag_{q}c^\dag_{-q}\right) \ket{0} \bra{0} \prod_{ 0\le k \neq q,-q} e^{g^{\prime *}_kc_{-k} c_k} e^{g_kc^\dag_k c^\dag_{-k}}\ket{0} = \\
\sum_{q>0} 2 g^{\prime *}_q g_q \frac{\cos q(x-y)}{L} u'_q u_q \prod_{0\le k \neq q} \left(u_k'u_k + v^{\prime *}_k v_k\right)=
\left[\prod_{0\le k < \pi} \left(u_k'u_k + v^{\prime *}_k v_k\right) \right]\sum_{q>0} \frac{2v^{\prime *}_q v_q \cos q (x-y)}{L\left(u'_qu_q + v^{\prime *}_qv_q\right)}\,,
\end{multline}
\end{widetext}
such that:
\begin{multline} 
\frac{\bra{\Omega}c^\dag_x c_y \ket{\Psi}}{\bracket{\Omega}{\Psi}} = \sum_{q>0} \frac{2v^{\prime *}_q v_q \cos q (x-y)}{L\left(u'_qu_q + v^{\prime *}_qv_q\right)} =
\\
 \sum_{q>0} \frac{2\cos q(x-y) \sin \frac{\theta(q)}{2}\sin\frac{\theta'(q)}{2}}{L\cos \frac{\theta(q)-\theta'(q)}{2}}\,.
\label{Kcdc2}
\end{multline}

\section{Strange correlators in real space for generic BCS states} \label{app:BCS}

To deal with systems without translational invariance in real space, we consider the following definition of the eigenmodes with positive energy, similarly to Ref. \cite{mbeng2020}:
\beq \label{eigenmodes2}
\eta_{l} = \sum_i \left(U_{li}  \, c_i - V_{li}  \, c_i^{\dagger}\right)\,,
\eeq
which generalizes Eq. \eqref{eigenmodes}. Here. the indices may include also the spin degree of freedom.

In Ref. \cite{mbeng2020} it is shown that:
\begin{equation} \label{product}
\left|\bracket{\Psi'}{\Psi}\right|^2 = \left|\det\left(UU^{\prime \dag} + VV^{\prime \dag}\right) \right|,
\end{equation}
which generalizes Eq. \eqref{kitaevoverlap2}.
In the same paper, the authors found a useful relation for the non-trivial strange correlator of the eigenmodes $\eta$ of one of the two BCS states. In particular, we must define first:
\begin{equation}
\mathbb{U} = UU^{\prime\dag} + VV^{\prime \dag}\,, \qquad \mathbb{V}= V^*U^{\prime\dag}+U^*V^{\prime \dag}\,.
\end{equation}
Then we define:
\begin{equation}
Z=\left(\mathbb{U}^\dag\right)^{-1}\mathbb{V}^\dag\,.
\end{equation}
The $Z$ matrix is defined in such a way that:
\begin{equation}
\ket{\Psi'} = \tilde{\mathcal{N}} \exp\left[\frac{1}{2} \sum_{a,b} Z_{ab} \eta^\dag_a \eta^\dag_b\right] \ket{\Psi} \,,
\end{equation}
with the normalization $\tilde{|\mathcal{N}|}=\left|\bracket{\Psi'}{\Psi}\right|$ given by Eq. \eqref{product}.
The result in Ref. \cite{mbeng2020} can be recast into:
\begin{equation} \label{etacorr}
\frac{\bra{\Psi'} \eta^\dag_x \eta^\dag_y \ket{\Psi}}{\bracket{\Psi'}{\Psi}} = Z^{*}_{xy}\,.
\end{equation}
Indeed one has:
\begin{multline}
\bra{\Psi'} \eta^\dag_x \eta^\dag_y \ket{\Psi} = \\
\tilde{\mathcal{N}}^* \bra{\Psi} \exp\left[\frac{1}{2} \sum_{a,b} Z_{ab}^\dag \eta^\dag_a \eta^\dag_b\right]\eta^\dag_x \eta^\dag_y\ket{\Psi} =\\
\tilde{\mathcal{N}}^* \bra{\Psi} \left[\eta^\dag_x + \sum_a Z_{ax}^\dag \eta_a \right]\exp\left[\frac{1}{2} \sum_{a,b} Z_{ab}^\dag \eta_a \eta_b\right] \eta^\dag_y\ket{\Psi}
\\ = \tilde{\mathcal{N}}^* \bra{\Psi}  Z_{yx}^\dag\exp\left[\frac{1}{2} \sum_{a,b} Z_{ab}^\dag \eta_a \eta_b\right] \ket{\Psi} =Z_{yx}^\dag \bracket{\Psi'}{\Psi}
\end{multline}
where we used that $Z$ is antisymmetric, that $\bra{\Psi}\eta_p^\dag=0$, and the property:
\begin{multline}
\exp\left[\frac{1}{2} \sum_{a,b} Z_{ab}^\dag \eta^\dag_a \eta^\dag_b\right]\eta^\dag_x = 
\\ 
\left(\eta_x^\dag + \left[\frac{1}{2} \sum_{a,b} Z_{ab}^\dag \eta_a \eta_b,\eta_x^\dag\right]\right)\exp\left[\frac{1}{2} \sum_{a,b} Z_{ab}^\dag \eta_a \eta_b\right]\,.
\end{multline}
Besides the main equation \eqref{etacorr}, we need also to consider:
\begin{equation} \label{etacorr2}
\frac{\bra{\Psi'} \eta_x \eta_y \ket{\Psi}}{\bracket{\Psi'}{\Psi}}= \frac{\bra{\Psi'} \eta^\dag_x \eta_y \ket{\Psi}}{\bracket{\Psi'}{\Psi}}=0\,, \frac{\bra{\Psi'} \eta_x \eta_y^\dag \ket{\Psi}}{\bracket{\Psi'}{\Psi}}= \delta_{xy}.
\end{equation}
From Eqs. \eqref{etacorr} and \eqref{etacorr2}, we can derive all the strange correlators, by inverting Eq. \eqref{eigenmodes2}:
\begin{equation}
\begin{pmatrix}c \\ c^\dag \end{pmatrix} = \begin{pmatrix} U^\dag & -V^{T} \\ -V^\dag & U^{T}\end{pmatrix} \begin{pmatrix}\eta \\ \eta^\dag \end{pmatrix} \,.
\end{equation}
From all the previous equations, we get (repeated indices are summed):
\begin{multline}
\frac{\bra{\Psi'} c^\dag_x c^\dag_y \ket{\Psi}}{\bracket{\Psi'}{\Psi}} = \\
\frac{\bra{\Psi'} -V_{xa} U^{T}_{yd} \eta_a \eta^\dag_d \ket{\Psi} + \bra{\Psi'}  U^{T}_{xb} U^{T}_{yd} \eta^\dag_b \eta^\dag_d \ket{\Psi}}{\bracket{\Psi'}{\Psi}} \\
= - \left(V^\dag U\right)_{xy} + \left(U^{T}Z^{*}U\right)_{xy}\,,
\end{multline}
which generalizes Eq. \eqref{Kcdcd}
(in the diagonal case in momentum space $V^\dag U=0$ because of the canonical conditions), and:
\begin{multline}
\frac{\bra{\Psi'} c^\dag_x c_y \ket{\Psi}}{\bracket{\Psi'}{\Psi}} =\\
 \frac{\bra{\Psi'} V^\dag_{xa} V^{T}_{yd} \eta_a \eta^\dag_d \ket{\Psi} - \bra{\Psi'}  U^{T}_{xb} V^{T}_{yd} \eta^\dag_b \eta^\dag_d \ket{\Psi}}{\bracket{\Psi'}{\Psi}} \\
= \left(V^\dag V\right)_{xy} - \left(U^{T}Z^{*}V\right)_{xy}\,,
\end{multline}
which generalizes Eq. 
\eqref{Kcdc}.

\section{Strange correlators for the translational invariant lattices in Section \ref{sec:transl}
}
\label{strasl}

\subsection{Two-dimensional Chern insulator in Subsection 
\ref{iqhe}
}
In this subsection, the sums $S[ c^{\dagger}_f , c_{f^{\prime}}]_L$ of the strange correlators $s[ c^{\dagger}_f , c_{f^{\prime}}]_{{\bf r} , {\bf r^{\prime}}}$, for the 2D BHZ model in Eq. \eqref{BHZ_2D}
, are considered. We define the Fourier transform of the strange correlators as:
\beq
s[ c^{\dagger}_f , c_{f^{\prime}}]_{{\bf r} , {\bf r^{\prime}}} = \frac{1}{L^d}  \sum_{{\bf k}} \tilde{s}[ c^{\dagger}_f , c_{f^{\prime}}]_{{\bf k}} \, \, e^{i{\bf k} \cdot ({\bf r} - {\bf r^{\prime}})}   \, , 
\label{trasfb}
\eeq
where we considered that $s[ c^{\dagger}_f , c_{f^{\prime}}]_{{\bf r} , {\bf r^{\prime}}}$ depends only on ${\bf r} - {\bf r^{\prime}}$ in translationally invariant systems.
The following analytical expressions are easy to be derived:
\begin{widetext}
\begin{multline}
S[ c^{\dagger}_f , c_{f^{\prime}}]_L   = 
\frac{1}{L^d} \sum_{{\bf r} -  {\bf r^{\prime}}} \sum_{{\bf k}} \tilde{s}[ c^{\dagger}_f , c_{f^{\prime}}]_{{\bf k}} \, \, e^{i{\bf k} \cdot ({\bf r} - {\bf r^{\prime}})} =  
 \frac{1}{L^d}  \sum_{{\bf k}}  \tilde{s}[ c^{\dagger}_f , c_{f^{\prime}}]_{{\bf k}} \, \times \\
\times  \Bigg[\frac{1-e^{ik_xL}}{1-e^{ik_x}}\frac{1-e^{ik_yL}}{1-e^{ik_y}}  \,   (1-  \delta_{k_x,0})(1-  \delta_{k_y,0}) \,    
+ L  \, \frac{1-e^{ik_yL}}{1-e^{ik_y}}  \,  \delta_{k_x,0} \, (1-  \delta_{k_y,0})  + L \, \frac{1-e^{ik_x L}}{1-e^{ik_x}}    \,  \delta_{k_y,0} \,  (1-  \delta_{k_x,0})  +  L^d \,  \delta_{k_x,0} \, \delta_{k_y,0} \Bigg] \, ,
\label{stringa}
\end{multline}
\end{widetext}
where, from Eq. \eqref{trasf}, the Fourier transform $\tilde{s}[c^{\dagger}_f , c_{f^{\prime}}]_{{\bf k}}$ results:
\beq
 \tilde{s}[c^{\dagger}_f , c_{f^{\prime}}]_{{\bf k}} = \frac{a^{(\Omega , -) \, *}_{f,{\bf k}} \, a^{(\Psi , -)}_{f^{\prime},{\bf k}}}{a^{(\Omega , -) \, *}_{A,{\bf k}}\,a^{(\Psi , -)}_{A,{\bf k}}+ a^{(\Omega , -) \, *}_{B,{\bf k}}\,a^{(\Psi , - )}_{B,{\bf k}}} \, .
 \label{strexp}
\eeq
For periodic boundaries, we have $e^{ik_jL}=1$, and we obtain:
\begin{equation}
S[ c^{\dagger}_f , c_{f^{\prime}}]_L  =   \tilde{s}[ c^{\dagger}_f , c_{f^{\prime}}]_{{\bf 0}} \, ,
\label{sumperapp}
\end{equation}
also derivable directly from the definition of Fourier transform in \eqref{stringa}. For antiperiodic boundaries, we have $e^{ik_jL}=-1$, and  we obtain:
\begin{equation}
S[ c^{\dagger}_f , c_{f^{\prime}}]_L  = \frac{1}{L^d} \, \sum_{{\bf k}} \tilde{s}[ c^{\dagger}_f , c_{f^{\prime}}]_{{\bf k}} \, \, \frac{4 }{\left(1-e^{ik_x}\right)\left(1-e^{ik_y}\right)} \, .
\end{equation}

We stress that, when periodic boundary conditions are assumed, so that Eq. 
\eqref{sumper} 
is fulfilled, possible divergences in ${\bf k} = 0$ may arise due to the orthogonality of single-particle states \cite{xu2014}. In order to regularize them, we evaluate the sums at ${\bf k} = \frac{2 \pi}{L} (1,1)$, instead that strictly at ${\bf k} = 0$, as in 
Eq. \eqref{sumper}.

\subsection{Three-dimensional BHZ model in Subsection 
\ref{BHZ}
}
In this subsection, we derive the expressions for some strange correlators for translational invariant  $4$ by $4$ model in Eq. 
\eqref{BHZ_3D}.
As there, $A$ and $B$ are general indices, not changed by time-reversal symmetry. The  related Hamiltonian is
\beq
H=\sum_{{\bf k}}\left(\begin{array}{cccc}
c_{A\uparrow{\bf k}}^{\dagger} & c_{B\uparrow{\bf k}}^{\dagger} & c_{A\downarrow{\bf k}}^{\dagger} & c_{B\downarrow{\bf k}}^{\dagger}\end{array}\right)H\left({\bf k}\right)\left(\begin{array}{c}
c_{A\uparrow{\bf k}}\\
c_{B\uparrow{\bf k}}\\
c_{A\downarrow{\bf k}}\\
c_{B\downarrow{\bf k}}
\end{array}\right),
\eeq
where $H\left({\bf k}\right)$ is a $4\times4$ matrix depending on
the parameter ${\bf k}$. $H$ is diagonalized,
\[
U^{\dagger}\left({\bf k}\right)H\left({\bf k}\right)U\left({\bf k}\right)=\left(\begin{array}{cccc}
\epsilon_{1}\left({\bf k}\right) & 0 & 0 & 0\\
0 & \epsilon_{2}\left({\bf k}\right) & 0 & 0\\
0 & 0 & \epsilon_{3}\left({\bf k}\right) & 0\\
0 & 0 & 0 & \epsilon_{4}\left({\bf k}\right)
\end{array}\right),
\]
by the unitary transformation
\[
\left[U\left({\bf k}\right)\right]_{ij}=u_{ij}\left({\bf k}\right),\:\left[U^{\dagger}\left({\bf k}\right)\right]_{ij}=u_{ji}^{*}\left({\bf k}\right),
\]
where
\[
H\left({\bf k}\right)\left(\begin{array}{c}
u_{1j}\\
u_{2j}\\
u_{3j}\\
u_{4j}
\end{array}\right)=\epsilon_{j}\left({\bf k}\right)\left(\begin{array}{c}
u_{1j}\\
u_{2j}\\
u_{3j}\\
u_{4j}
\end{array}\right) ,
\quad
U\left(\begin{array}{c}
\gamma_{1}\\
\gamma_{2}\\
\gamma_{3}\\
\gamma_{4}
\end{array}\right)=\left(\begin{array}{c}
c_{A\uparrow}\\
c_{B\uparrow}\\
c_{A\downarrow}\\
c_{B\downarrow}
\end{array}\right) .
\] Therefore
\[
c_{i{\bf k}}=\sum_{j=1}^{4}U_{ij}\gamma_{j{\bf k}}=\sum_{j=1}^{4}u_{ij}\left({\bf k}\right)\gamma_{j{\bf k}}\, , \]
with $1=A\uparrow,\ 2=B\uparrow,\;3=A\downarrow,\ 4=B\downarrow$.
The corresponding operators in the direct lattice are
\begin{widetext}
\begin{align*}
&c_{j{\bf r}}  =  \frac{1}{L^{\frac{d}{2}}}\sum_{{\bf k}}c_{j{\bf k}}e^{-i{\bf r}\cdot{\bf k}}=\frac{1}{L^{\frac{d}{2}}}\sum_{{\bf k},l}u_{jl}\gamma_{l{\bf k}}e^{-i{\bf r}\cdot{\bf k}} \, ,\\
&\gamma_{i{\bf k}}=\sum_{j=1}^{4}U_{ij}^{\dagger}c_{j{\bf k}}=u_{1i}^{*}\left({\bf k}\right)c_{A\uparrow{\bf k}}+u_{2i}^{*}\left({\bf k}\right)c_{B\uparrow{\bf k}}+u_{3i}^{*}\left({\bf k}\right)c_{A\downarrow{\bf k}}+u_{4i}^{*}\left({\bf k}\right)c_{B\downarrow{\bf k}}=\sum_{j=1}^{4}u_{ji}^{*}\left({\bf k}\right)c_{i{\bf k}} \ ,
\end{align*}
\end{widetext}
and
\[
H=\sum_{j=1}^{4}\sum_{{\bf k}}\epsilon\left({\bf k}\right)\gamma_{j}^{\dagger}\gamma_{j} \, .
\]
As described in the main text, the model displays two degenerate bands, symmetric with the respect to 
zero energy.
Suppose now to fill only the lower bands $\epsilon_{1}\left({\bf k}\right)$
and $\epsilon_{2}\left({\bf k}\right)$, so that  the ground-state reads:
\[
\ket{\Psi}=\prod_{{\bf k}}\gamma_{2{\bf k}}^{\dagger}\gamma_{1{\bf k}}^{\dagger}\ket{0} \, .
\]
Letting now the Hamiltonian to depend on a parameter $M$, we denote as
\[
\ket{\Psi}=\prod_{{\bf k}}\gamma_{2{\bf k}}^{\dagger}\gamma_{1{\bf k}}^{\dagger}\ket{0},
\]
the ground-state of $H\left(M\right)$, and as
\[
\ket{\Omega}=\prod_{{\bf k}}\tilde{\gamma}_{2{\bf k}}^{\dagger}\tilde{\gamma}_{1{\bf k}}^{\dagger}\ket{0},
\]
the ground-state of $H\left(\tilde{M}\right)$. The strange correlator
\[
S[ c^{\dagger} , c^{\dagger}]_{{\bf r},{\bf r^{\prime}}}  =\frac{\braket{\Omega|c_{j{\bf r}}^{\dagger}c_{l{\bf r^{\prime}}}|\Psi}}{\braket{\Omega|\Psi}}
\]
is calculated as follows:
\begin{widetext}
\begin{eqnarray*}
\braket{\Omega|\Psi} & = & \braket{0|\prod_{{\bf k},{\bf q}}\tilde{\gamma}_{1{\bf k}}\tilde{\gamma}_{2{\bf k}}\gamma_{2{\bf q}}^{\dagger}\gamma_{1{\bf q}}^{\dagger}|0}=\braket{0|\prod_{{\bf k}}\tilde{\gamma}_{1{\bf k}}\tilde{\gamma}_{2{\bf k}}\gamma_{2{\bf k}}^{\dagger}\gamma_{1{\bf k}}^{\dagger}|0}=\\
 & = & \braket{0|\prod_{{\bf k}}\left(\sum_{j=1}^{4}\tilde{u}_{j1}^{*}\left({\bf k}\right)c_{j{\bf k}}\right)\left(\sum_{j=1}^{4}\tilde{u}_{j2}^{*}\left({\bf k}\right)c_{j{\bf k}}\right)\left(\sum_{j=1}^{4}u_{j2}\left({\bf k}\right)c_{j{\bf k}}^{\dagger}\right)\left(\sum_{j=1}^{4}u_{j1}\left({\bf k}\right)c_{j{\bf k}}^{\dagger}\right)|0}=\\
 & = & \braket{0|\prod_{{\bf k}}\sum_{l,j=1}^{4}\left(\tilde{u}_{l1}^{*}\tilde{u}_{j2}^{*}u_{l2}u_{j1}c_{l{\bf k}}c_{j{\bf k}}c_{l{\bf k}}^{\dagger}c_{j{\bf k}}^{\dagger}+\tilde{u}_{j1}^{*}\tilde{u}_{l2}^{*}u_{l2}u_{j1}c_{j{\bf k}}c_{l{\bf k}}c_{l{\bf k}}^{\dagger}c_{j{\bf k}}^{\dagger}\right)|0}=\\
 & = & \braket{0|\prod_{{\bf k}}\sum_{l\neq j=1}^{4}\left(1-n_{j{\bf k}}\right)\left(1-n_{l{\bf k}}\right)\left(\tilde{u}_{l2}^{*}u_{l2}\tilde{u}_{j1}^{*}u_{j1}-\tilde{u}_{j2}^{*}u_{j1}\tilde{u}_{l1}^{*}u_{l2}\right)|0}\\
 & = & \prod_{{\bf k}}\left[\sum_{l\neq j=1}^{4}\left(\tilde{u}_{l2}^{*}\left({\bf k}\right)u_{l2}\left({\bf k}\right)\tilde{u}_{j1}^{*}\left({\bf k}\right)u_{j1}\left({\bf k}\right)-\tilde{u}_{j2}^{*}\left({\bf k}\right)u_{l2}\left({\bf k}\right)\tilde{u}_{l1}^{*}\left({\bf k}\right)u_{j1}\left({\bf k}\right)\right)\right].
\end{eqnarray*}
Moreover, starting from
\begin{eqnarray*}
c_{l{\bf r^{\prime}}}\ket{\Psi} & = & \frac{1}{L^{\frac{d}{2}}}\sum_{{\bf k},j}u_{lj}\left({\bf k}\right)\gamma_{j{\bf k}}e^{-i{\bf r^{\prime}}\cdot{\bf k}}\ket{\Psi}=\\
 & = & \frac{1}{L^{\frac{d}{2}}}\sum_{{\bf k}}e^{-i{\bf r^{\prime}}\cdot{\bf k}}\left(-1\right)^{h({\bf k})}\left[u_{l1}\left({\bf k}\right)\gamma_{2{\bf k}}^{\dagger}-u_{l2}\left({\bf k}\right)\gamma_{1{\bf k}}^{\dagger}\right]\prod_{{\bf q}\neq{\bf k}}\gamma_{2{\bf q}}^{\dagger}\gamma_{1{\bf q}}^{\dagger}\ket{0},
\end{eqnarray*}
\begin{eqnarray*}
c_{j{\bf r}}\ket{\Omega} & = & \frac{1}{L^{\frac{d}{2}}}\sum_{{\bf k}}e^{-i{\bf r}\cdot{\bf k}}\left(-1\right)^{h({\bf k})}\left[\tilde{u}_{j1}\left({\bf k}\right)\tilde{\gamma}_{2{\bf k}}^{\dagger}-\tilde{u}_{j2}\left({\bf k}\right)\tilde{\gamma}_{1{\bf k}}^{\dagger}\right]\prod_{{\bf q}\neq{\bf k}}\tilde{\gamma}_{2{\bf q}}^{\dagger}\tilde{\gamma}_{1{\bf q}}^{\dagger}\ket{0},
\end{eqnarray*}
we obtain
\begin{eqnarray*}
\braket{\Omega|c_{j{\bf r}}^{\dagger}c_{l{\bf r^{\prime}}}|\Psi} & = & \frac{1}{L^{d}}\sum_{{\bf k}}e^{-i\left({\bf r}-{\bf r^{\prime}}\right)\cdot{\bf k}}\left[\tilde{u}_{j1}^{*}\left({\bf k}\right)u_{l1}\left({\bf k}\right)\bra{0}\tilde{\gamma}_{2{\bf k}}\gamma_{2{\bf k}}^{\dagger}\left(\prod_{{\bf q}\neq{\bf k}}\tilde{\gamma}_{1{\bf q}}\tilde{\gamma}_{2{\bf q}}\gamma_{2{\bf q}}^{\dagger}\gamma_{1{\bf q}}^{\dagger}\right)\ket{0}+\right.\\
 &  & +\tilde{u}_{j2}^{*}\left({\bf k}\right)u_{l2}\left({\bf k}\right)\bra{0}\tilde{\gamma}_{1{\bf k}}\gamma_{1{\bf k}}^{\dagger}\left(\prod_{{\bf q}\neq{\bf k}}\tilde{\gamma}_{1{\bf q}}\tilde{\gamma}_{2{\bf q}}\gamma_{2{\bf q}}^{\dagger}\gamma_{1{\bf q}}^{\dagger}\right)\ket{0}+\\
 &  & -\tilde{u}_{j1}^{*}\left({\bf k}\right)u_{l2}\left({\bf k}\right)\bra{0}\tilde{\gamma}_{2{\bf k}}\gamma_{1{\bf k}}^{\dagger}\left(\prod_{{\bf q}\neq{\bf k}}\tilde{\gamma}_{1{\bf q}}\tilde{\gamma}_{2{\bf q}}\gamma_{2{\bf q}}^{\dagger}\gamma_{1{\bf q}}^{\dagger}\right)\ket{0}+\\
 &  & \left.-\tilde{u}_{j2}^{*}\left({\bf k}\right)u_{l1}\left({\bf k}\right)\bra{0}\tilde{\gamma}_{1{\bf k}}\gamma_{2{\bf k}}^{\dagger}\left(\prod_{{\bf q}\neq{\bf k}}\tilde{\gamma}_{1{\bf q}}\tilde{\gamma}_{2{\bf q}}\gamma_{2{\bf q}}^{\dagger}\gamma_{1{\bf q}}^{\dagger}\right)\ket{0}\right]=
\end{eqnarray*}
\begin{eqnarray*}
\bra{0}\tilde{\gamma}_{2{\bf k}}\gamma_{2{\bf k}}^{\dagger}\left(\prod_{{\bf q}\neq{\bf k}}\tilde{\gamma}_{1{\bf q}}\tilde{\gamma}_{2{\bf q}}\gamma_{2{\bf q}}^{\dagger}\gamma_{1{\bf q}}^{\dagger}\right)\ket{0} & = & \bra{0}\left(\sum_{j=1}^{4}\tilde{u}_{j2}^{*}\left({\bf k}\right)c_{j{\bf k}}\right)\left(\sum_{j=1}^{4}u_{j2}\left({\bf k}\right)c_{j{\bf k}}^{\dagger}\right)\left(\prod_{{\bf q}\neq{\bf k}}\tilde{\gamma}_{1{\bf q}}\tilde{\gamma}_{2{\bf q}}\gamma_{2{\bf q}}^{\dagger}\gamma_{1{\bf q}}^{\dagger}\right)\ket{0}=\\
 & = & \sum_{j=1}^{4}\tilde{u}_{j2}^{*}u_{j2}\bra{0}\left(1-n_{j{\bf k}}\right)\left(\prod_{{\bf q}\neq{\bf k}}\tilde{\gamma}_{1{\bf q}}\tilde{\gamma}_{2{\bf q}}\gamma_{2{\bf q}}^{\dagger}\gamma_{1{\bf q}}^{\dagger}\right)\ket{0}=\\
 & = & \sum_{j=1}^{4}\tilde{u}_{j2}^{*}u_{j2}\bra{0}\left(\prod_{{\bf q}\neq{\bf k}}\tilde{\gamma}_{1{\bf q}}\tilde{\gamma}_{2{\bf q}}\gamma_{2{\bf q}}^{\dagger}\gamma_{1{\bf q}}^{\dagger}\right)\ket{0},
\end{eqnarray*}
\begin{eqnarray*}
\bra{0}\tilde{\gamma}_{1{\bf k}}\gamma_{1{\bf k}}^{\dagger}\left(\prod_{{\bf q}\neq{\bf k}}\tilde{\gamma}_{1{\bf q}}\tilde{\gamma}_{2{\bf q}}\gamma_{2{\bf q}}^{\dagger}\gamma_{1{\bf q}}^{\dagger}\right)\ket{0} & = & \bra{0}\left(\sum_{j=1}^{4}\tilde{u}_{j1}^{*}\left({\bf k}\right)c_{j{\bf k}}\right)\left(\sum_{j=1}^{4}u_{j1}\left({\bf k}\right)c_{j{\bf k}}^{\dagger}\right)\left(\prod_{{\bf q}\neq{\bf k}}\tilde{\gamma}_{1{\bf q}}\tilde{\gamma}_{2{\bf q}}\gamma_{2{\bf q}}^{\dagger}\gamma_{1{\bf q}}^{\dagger}\right)\ket{0}=\\
 & = & \sum_{j=1}^{4}\tilde{u}_{j1}^{*}u_{j1}\bra{0}\left(\prod_{{\bf q}\neq{\bf k}}\tilde{\gamma}_{1{\bf q}}\tilde{\gamma}_{2{\bf q}}\gamma_{2{\bf q}}^{\dagger}\gamma_{1{\bf q}}^{\dagger}\right)\ket{0},
\end{eqnarray*}
\begin{eqnarray*}
\bra{0}\tilde{\gamma}_{1{\bf k}}\gamma_{2{\bf k}}^{\dagger}\left(\prod_{{\bf q}\neq{\bf k}}\tilde{\gamma}_{1{\bf q}}\tilde{\gamma}_{2{\bf q}}\gamma_{2{\bf q}}^{\dagger}\gamma_{1{\bf q}}^{\dagger}\right)\ket{0} & = & \bra{0}\left(\sum_{j=1}^{4}\tilde{u}_{j1}^{*}\left({\bf k}\right)c_{j{\bf k}}\right)\left(\sum_{j=1}^{4}u_{j2}\left({\bf k}\right)c_{j{\bf k}}^{\dagger}\right)\left(\prod_{{\bf q}\neq{\bf k}}\tilde{\gamma}_{1{\bf q}}\tilde{\gamma}_{2{\bf q}}\gamma_{2{\bf q}}^{\dagger}\gamma_{1{\bf q}}^{\dagger}\right)\ket{0}=\\
 & = & \sum_{j=1}^{4}\tilde{u}_{j1}^{*}u_{j2}\bra{0}\left(1-n_{j{\bf k}}\right)\left(\prod_{{\bf q}\neq{\bf k}}\tilde{\gamma}_{1{\bf q}}\tilde{\gamma}_{2{\bf q}}\gamma_{2{\bf q}}^{\dagger}\gamma_{1{\bf q}}^{\dagger}\right)\ket{0}=\\
 & = & \sum_{j=1}^{4}\tilde{u}_{j1}^{*}u_{j2}\bra{0}\left(\prod_{{\bf q}\neq{\bf k}}\tilde{\gamma}_{1{\bf q}}\tilde{\gamma}_{2{\bf q}}\gamma_{2{\bf q}}^{\dagger}\gamma_{1{\bf q}}^{\dagger}\right)\ket{0},\\
\bra{0}\tilde{\gamma}_{2{\bf k}}\gamma_{1{\bf k}}^{\dagger}\left(\prod_{{\bf q}\neq{\bf k}}\tilde{\gamma}_{1{\bf q}}\tilde{\gamma}_{2{\bf q}}\gamma_{2{\bf q}}^{\dagger}\gamma_{1{\bf q}}^{\dagger}\right)\ket{0} & = & \sum_{s=1}^{4}\tilde{u}_{j2}^{*}u_{j1}\bra{0}\left(\prod_{{\bf q}\neq{\bf k}}\tilde{\gamma}_{1{\bf q}}\tilde{\gamma}_{2{\bf q}}\gamma_{2{\bf q}}^{\dagger}\gamma_{1{\bf q}}^{\dagger}\right)\ket{0},
\end{eqnarray*}
 
\begin{eqnarray*}
\braket{\Omega|c_{j{\bf r}}^{\dagger}c_{l{\bf r^{\prime}}}|\Psi} & = & \frac{1}{L^{d}}\sum_{{\bf k}}\sum_{s=1}^{4}e^{-i\left({\bf r}-{\bf r^{\prime}}\right)\cdot{\bf k}}\left[\tilde{u}_{j1}^{*}u_{l1}\tilde{u}_{s2}^{*}u_{s2}+\tilde{u}_{j2}^{*}u_{l2}\tilde{u}_{s1}^{*}u_{s1}-\tilde{u}_{j1}^{*}u_{l2}\tilde{u}_{s2}^{*}u_{s1}-\tilde{u}_{j2}^{*}u_{l1}\tilde{u}_{s1}^{*}u_{s2}\right]\\
 &  & \bra{0}\left(\prod_{{\bf q}\neq{\bf k}}\tilde{\gamma}_{1{\bf q}}\tilde{\gamma}_{2{\bf q}}\gamma_{2{\bf q}}^{\dagger}\gamma_{1{\bf q}}^{\dagger}\right)\ket{0}=\\
 & = & \frac{1}{N^{d}}\sum_{{\bf k}}\sum_{s=1}^{4}e^{-i\left({\bf r}-{\bf r^{\prime}}\right)\cdot{\bf k}}\left[\left(\tilde{u}_{j2}^{*}\tilde{u}_{s1}^{*}-\tilde{u}_{j1}^{*}\tilde{u}_{s2}^{*}\right)\left(u_{l2}u_{s1}-u_{l1}u_{s2}\right)\right]\bra{0}\left(\prod_{{\bf q}\neq{\bf k}}\tilde{\gamma}_{1{\bf q}}\tilde{\gamma}_{2{\bf q}}\gamma_{2{\bf q}}^{\dagger}\gamma_{1{\bf q}}^{\dagger}\right)\ket{0} .
\end{eqnarray*}
Therefore, finally:
\[
S[ c^{\dagger} , c^{\dagger} ]_{{\bf r},{\bf r^{\prime}}} =\frac{1}{L^{d}}\sum_{{\bf k}}\frac{\sum_{s=1}^{4}\left[\left(\tilde{u}_{j2}^{*}\tilde{u}_{s1}^{*}-\tilde{u}_{j1}^{*}\tilde{u}_{s2}^{*}\right)\left(u_{l2}u_{s1}-u_{l1}u_{s2}\right)\right]e^{-i\left({\bf r}-{\bf r^{\prime}}\right)\cdot{\bf k}}}{\sum_{l\neq j=1}^{4}\left(\tilde{u}_{l2}^{*}u_{l2}\tilde{u}_{j1}^{*}u_{j1}-\tilde{u}_{j2}^{*}u_{l2}\tilde{u}_{l1}^{*}u_{j1}\right)} \, .
\]
\end{widetext}

\section{Strange correlators for general hopping models in real space} \label{app:gen}
In this Appendix, we derive an expression for some strange correlators of fermionic operators involving lattice ground-states and real lattice variables, as in Eq. 
\eqref{matper}.
We consider a lattice and with $N$ sites with a purely hopping Hamiltonian depending
on a parameter $x$:
\begin{multline}
H\left(x\right)=\sum_{ij}t_{ij}\left(x\right)c_{i}^{\dagger}c_{j}+H.c.=
\\
\left(\begin{array}{cccc}
c_{1}^{\dagger} & c_{2}^{\dagger} & c_{3}^{\dagger} & \ldots\end{array}\right)\mathbf{T}\left(x\right)\left(\begin{array}{c}
c_{1}\\
c_{2}\\
c_{3}\\
\vdots
\end{array}\right)=\mathbf{c}^{\dagger}\mathbf{T}\left(x\right)\mathbf{c} \, .
\end{multline}
Let \textbf{$\mathbf{T}_{D}\left(x\right)$ }be the diagonal matrix
whose elements are the eigenvalues of $\mathbf{T}\left(x\right)$
(i.e. the single-particle energies), sorted by an increasing order 
\[
\left(T_{D}\right)_{11}=\epsilon_{1}\left(x\right)\leq\left(T_{D}\right)_{22}=\epsilon_{2}\left(x\right)\leq\left(T_{D}\right)_{33}=\epsilon_{3}\left(x\right)\ldots \, ,
\]
and $\mathbf{U}\left(x\right)$ the unitary matrix diagonalizing
$\mathbf{T}\left(x\right)$ i.e. $\mathbf{U}\left(x\right)\mathbf{T}\left(x\right)\mathbf{U}^{\dagger}\left(x\right)=\mathbf{T}_{D}\left(x\right)$.
This matrix acts on the $\mathbf{c}$ vector as
\[
\mathbf{U}\left(x\right)\mathbf{c}=\mathbf{w}\left(x\right)=\left(\begin{array}{c}
w_{1}\left(x\right)\\
w_{2}\left(x\right)\\
w_{3}\left(x\right)\\
\vdots
\end{array}\right),
\]
so the Hamiltonian can be written in the diagonal form
\begin{multline}
H\left(x\right)=\mathbf{c}^{\dagger}\mathbf{U}^{\dagger}\left(x\right)\mathbf{U}\left(x\right)\mathbf{T}\left(x\right)\mathbf{U}^{\dagger}\left(x\right)\mathbf{U}\left(x\right)\mathbf{c}=\\
\mathbf{w}^{\dagger}\left(x\right)\mathbf{T}_{D}\left(x\right)\mathbf{w}\left(x\right)=\sum_{j}\epsilon_{j} \, w_{j}^{\dagger}w_{j} \, .
\end{multline}
For fixed number of particles $M$, the ground-state is
\[
\ket{g\left(x\right)}=\prod_{k=1,\ldots,M}w_{k}^{\dagger}\left(x\right)\ket{0}.
\]
Notice that
\[
\mathbf{c}=\mathbf{U}^{\dagger}\left(x\right)\mathbf{w}\left(x\right)\Rightarrow c_{j}=\sum_{l}U_{lj}^{*}\left(x\right)w_{l}\left(x\right),
\]
\begin{multline*}
\mathbf{w}\left(y\right)=\mathbf{U}\left(y\right)\mathbf{c}=\mathbf{U}\left(y\right)\mathbf{U}^{\dagger}\left(x\right)\mathbf{w}\left(x\right)=\mathbf{D}\left(x,y\right)\mathbf{w}\left(x\right)\\
\Rightarrow w_{k}\left(y\right)=\sum_{l}D_{kl}\left(x,y\right)w_{l}\left(x\right),
\end{multline*}
\[
\mathbf{w}^{\dagger}\left(y\right)=\mathbf{w}^{\dagger}\left(x\right)\mathbf{D}^{\dagger}\left(x,y\right)\Rightarrow w_{k}^{\dagger}\left(y\right)=\sum_{l}D_{kl}^{*}\left(x,y\right)w_{l}^{\dagger}\left(x\right)  .
\]
The matrix $\mathbf{D}\left(x,y\right)$ is defined as
\[
\mathbf{D}\left(x,y\right)=\mathbf{U}\left(y\right)\mathbf{U}^{\dagger}\left(x\right)\Rightarrow D_{kl}\left(x,y\right)=\sum_{r}U_{kr}\left(y\right)U_{lr}^{*}\left(x\right).
\]
It also holds:
\begin{multline*}
c_{j}\ket{g\left(x\right)}  =  \sum_{l=1}^{N}U_{lj}^{*}\left(x\right)w_{l}\left(x\right)\prod_{k=1,\ldots,M}w_{k}^{\dagger}\left(x\right)\ket{0}=\\
  = \sum_{l=1}^{M}\left(-1\right)^{l+1}U_{lj}^{*}\left(x\right)\prod_{k=1,\ldots,l-1,l+1,\ldots,M}w_{k}^{\dagger}\left(x\right)\ket{0},
\end{multline*}
and
\[
\braket{g\left(x\right)|c_{j}^{\dagger}c_{i}|g\left(x\right)}=\sum_{l=1}^{M}U_{lj}\left(x\right)U_{li}^{*}\left(x\right).
\]
Choosing another parameter $y$, the ground-state can be expressed in
terms of $\mathbf{w}^{\dagger}\left(x\right)$, as 
\begin{eqnarray*}
\ket{g\left(y\right)} & = & \prod_{k=1,\ldots,M}w_{k}^{\dagger}\left(y\right)\ket{0}=\\
 & = & \prod_{k=1,\ldots,M}\sum_{s=1}^{N}D_{ks}^{*}\left(x,y\right)w_{s}^{\dagger}\left(x\right)\ket{0}=\\
 & = & \sum_{\left\{ k_{r}\right\} _{M}}\Delta_{\left\{ k_{r}\right\} _{M}}\prod_{r=1}^{M}w_{k_{r}}^{\dagger}\left(x\right)\ket{0} \, ,
\end{eqnarray*}
where we adopted the following notation:
\begin{itemize}
\item $\left\{ k_{r}\right\} _{M}=\left\{ k_{1},k_{2}\ldots k_{M}\right\} $
is a string of $M$ elements; 
\item $\left\{ k_{r}\right\} _{M}^{1,\ldots,N}$means all the possible strings
of $M$ elements selected from $\left\{ 1,\ldots,N\right\} $;
\item $\Delta_{\left\{ k_{r}\right\} }=\det\left[\mathbf{D}^{*}\right]_{\left\{ k_{r}\right\} },$
where $\left[.\right]_{\left\{ k_{r}\right\} }$is the submatrix obtained
selecting the first $M$ lines and the columns corresponding to the
string $\left\{ k_{r}\right\} .$
\end{itemize}
Fixed the normalization
\[
\braket{g\left(y\right)|g\left(x\right)}=\Delta_{\left\{ 1,2,\ldots,M\right\} }^{*}=\det\left[\mathbf{D}\left(x,y\right)\right]_{\left\{ 1,2,\ldots,M\right\} } \, ,
\] 
we have
\begin{widetext}
\begin{eqnarray*}
c_{j}\ket{g\left(y\right)} & = & \sum_{l=1}^{N}U_{lj}^{*}\left(x\right)w_{l}\left(x\right)\prod_{k=1,\ldots,M}w_{k}^{\dagger}\left(y\right)\ket{0}=\\
 & = & \sum_{l=1}^{N}U_{lj}^{*}\left(x\right)w_{l}\left(x\right)\prod_{k=1,\ldots,M}\sum_{s=1}^{N}D_{ks}^{*}\left(x,y\right)w_{s}^{\dagger}\left(x\right)\ket{0}=\\
 & = & \sum_{l=1}^{N}U_{lj}^{*}\left(x\right)w_{l}\left(x\right)\sum_{\left\{ k_{r}\right\} _{M}^{1,\ldots,N}}\Delta_{\left\{ k_{r}\right\} }\prod_{r=1}^{M}w_{k_{r}}^{\dagger}\left(x\right)\ket{0}=\\
 & = & \sum_{l=1}^{M}U_{lj}^{*}\left(x\right)w_{l}\left(x\right)\Delta_{\left\{ k_{r}\right\} _{M}^{1,\ldots,M}}\prod_{r=1}^{M}w_{r}^{\dagger}\left(x\right)\ket{0}+\\
 &  & +\sum_{l=M+1}^{N}U_{lj}^{*}\left(x\right)w_{l}\left(x\right)\sum_{\left\{ k_{r}\right\} _{M-1}^{1,\ldots,M}\cup\left\{ l\right\} }\Delta_{\left\{ k_{r}\right\} _{M-1}^{1,\ldots,M}\cup\left\{ l\right\} }\left(\prod_{r=1}^{M-1}w_{k_{r}}^{\dagger}\left(x\right)\right)w_{l}^{\dagger}\left(x\right)\ket{0}+\ket{R} \, ,
\end{eqnarray*}
being $\ket{R}$ a remaining term of the combination that gives zero
overlap with $c_{i}\ket{g\left(x\right)}$. The latter expression can be managed as follows:
\begin{eqnarray*}
c_{j}\ket{g\left(y\right)} & = & \sum_{l=1}^{M}\left(-1\right)^{l+1}U_{lj}^{*}\left(x\right)\Delta_{\left\{ k_{r}\right\} _{M}^{1,\ldots,M}}\prod_{r=1,\ldots,l-1,l+1\ldots M}w_{r}^{\dagger}\left(x\right)\ket{0}+\\
 &  & +\left(-1\right)^{M+1}\sum_{l=M+1}^{N}U_{lj}^{*}\left(x\right)\sum_{\left\{ k_{r}\right\} _{M-1}^{1,\ldots,M}\cup\left\{ l\right\} }\Delta_{\left\{ k_{r}\right\} _{M-1}^{1,\ldots,M}\cup\left\{ l\right\} }\left(\prod_{r=1}^{M-1}w_{k_{r}}^{\dagger}\left(x\right)\right)\ket{0}+\ket{R}=\\
 & = & \sum_{l=1}^{M}\left(-1\right)^{l+1}U_{lj}^{*}\left(x\right)\Delta_{\left\{ 1,\ldots,M\right\} }\prod_{r=1,\ldots,l-1,l+1\ldots M}w_{r}^{\dagger}\left(x\right)\ket{0}+\\
 &  & +\left(-1\right)^{M+1}\sum_{l=1}^{M}\left[\sum_{s=M+1}^{N}U_{sj}^{*}\left(x\right)\Delta_{\left\{ 1,\ldots,l-1,l+1\ldots M\right\} \cup\left\{ s\right\} }\right]\prod_{r=1,\ldots,l-1,l+1\ldots M}w_{r}^{\dagger}\left(x\right)\ket{0}+\ket{R}=\\
 & = & \sum_{l=1}^{M}\eta_{l}^{(j)}\prod_{r=1,\ldots,l-1,l+1\ldots M}w_{r}^{\dagger}\left(x\right)\ket{0}+\ket{R},
\end{eqnarray*}
with
\[
\eta_{l}^{(j)}=-\left(-1\right)^{l}U_{lj}^{*}\left(x\right)\Delta_{\left\{ 1,\ldots,M\right\} }-\left(-1\right)^{M}\sum_{s=M+1}^{N}U_{sj}^{*}\left(x\right)\Delta_{\left\{ 1,\ldots,l-1,l+1\ldots M\right\} \cup\left\{ s\right\} } \, ,
\]
so that
\[
\braket{gs\left(y\right)|c_{j}^{\dagger}c_{i}|gs\left(x\right)}=-\sum_{l=1}^{M}\left(\eta_{l}^{(j)}\right)^{*}\left(-1\right)^{l}U_{li}^{*}\left(x\right) \, ,
\]
\[
\braket{gs\left(y\right)|c_{j}^{\dagger}c_{i}|gs\left(x\right)}=\Delta_{\left\{ 1,\ldots,M\right\} }^{*}\sum_{l=1}^{M}U_{lj}\left(x\right)U_{li}^{*}\left(x\right)+\sum_{l=1}^{M}\left(-1\right)^{M+l}U_{li}^{*}\left(x\right)\sum_{s=M+1}^{N}\Delta_{\left\{ 1,\ldots,l-1,l+1\ldots M\right\} \cup\left\{ s\right\} }^{*}U_{sj}\left(x\right) \, ,
\]
 and finally
\[
\frac{\braket{gs\left(y\right)|c_{j}^{\dagger}c_{i}|gs\left(x\right)}}{\braket{g\left(y\right)|g\left(x\right)}}=\sum_{l=1}^{M}U_{lj}\left(x\right)U_{li}^{*}\left(x\right)+\frac{1}{\Delta_{\left\{ 1,2,\ldots,M\right\} }^{*}}\sum_{l=1}^{M}\left(-1\right)^{M+l}U_{li}^{*}\left(x\right)\sum_{s=M+1}^{N}\Delta_{\left\{ 1,\ldots,l-1,l+1\ldots M\right\} \cup\left\{ s\right\} }^{*}U_{sj}\left(x\right) \ .
\]
\end{widetext}
If $x=y$, then $\mathbf{D}=\mathbf{I}$, so all $\Delta_{\left\{ 1,\ldots,M\right\} }=1$
and the others are $\Delta_{\left\{ k_{r}\right\} }=0$. Then we obtain
again:
\[
\braket{gs\left(x\right)|c_{j}^{\dagger}c_{i}|gs\left(x\right)}=\sum_{l=1}^{M}U_{lj}\left(x\right)U_{li}^{*}\left(x\right).
\]

\section{Strange correlators for the ground-state of the toric code} 
\label{toricode}
Kitaev's toric code \cite{kitaev2003} is the paradigmatic example of a model with topological order, thus degenerate ground-states on the torus and anyonic excitations.
For the purpose of estimating its strange correlators, we consider its surface code extension to a system with smooth open boundaries \cite{fowler2012}. The system is defined by spin $1/2$ degrees of freedom located on the boundaries of an open square lattice, and its Hamiltonian is dictated by:
\begin{equation}
H_{\textrm{TC}}=-J_A\sum_s\hat{A}_s-J_B\sum_p\hat{B}_p\, ,
\label{ham}
\end{equation}
where the  operators $\hat{A}_s\equiv\prod_{i \in s}\hat{\sigma}_i^x$ and
$\hat{B}_p\equiv\prod_{i\in p}\hat{\sigma}_i^z$ both contain four spins belonging to a star $s$ and a
plaquette $p$, respectively. 
The model can be solved exactly, since
the star and plaquette operators commute with each other, i.e., $[\hat{A}_s,\hat{B}_p]=0$ for $\forall \,  s,p$; its ground-state is given by:
\begin{equation}
\ket{\Psi_{TC}} = \prod_p  \left( \frac{1 + \hat{B}_p}{\sqrt{2}}\right) \ket{\Omega}\,,
\end{equation}
where $\ket{\Omega}=\ket{+++\ldots}$ is a trivial product state of the positive eigenstates $\ket{+}$ of $\hat{\sigma}_x$. In the language of lattice gauge theories, we can consider the operators $\hat{A}_s$ as local $\mathbb{Z}_2$ gauge symmetries associated to the lattice vertices. In order to exemplify the strange correlators in the universality class of the toric code and $\mathbb{Z}_2$ lattice gauge theories, we consider the gauge-invariant states $\ket{\Psi_{TC}}$, which represents the topological phase, and $\ket{\Omega}$, which is a trivial product state representing the so-called confined phase in the corresponding gauge theory. Both $\ket{\Psi_{TC}}$ and $\ket{\Omega}$ are gauge invariant states, being eigenstates of eigenvalue $1$ for all the star operators $\hat{A}_s$. Therefore, any strange correlator involving operators which are not gauge invariant vanishes and cannot be adopted to detect topological order. For instance, we have $\bra{\Omega} \hat{\sigma}^z_i \hat{\sigma}^z_j \ket{\Psi_{TC}} \propto \delta_{ij}$.

Concerning gauge-invariant operators, instead, the related strange correlators may be unsuitable to distinguish topological and non-topological states. Let us, in particular, consider the following example:
\begin{align}
&\frac{\bra{\Omega} \hat{\sigma}^x_i \hat{\sigma}^x_j \ket{\Psi_{TC}}}{\bra{\Omega} \Psi_{TC} \rangle} = 1\,, \label{strtc1}\\
&\bra{\Omega} \hat{\sigma}^x_i \hat{\sigma}^x_j \ket{\Omega} = 1\,.\label{strtc2}
\end{align}
 $\hat{\sigma}^x_i$ is a gauge-invariant operator which creates two plaquette excitations when applied to $\ket{\Psi_{TC}}$, so that, in general, $\hat{\sigma}^x_i\hat{\sigma}^x_j$ creates four plaquette excitations.
However, the strange correlator in Eq. \eqref{strtc1} is equal to 1 because $\ket{\Omega}$ is an eigenstate for the $\hat{\sigma}^x$ operators. Furthermore, the same is trivially true for the two-point correlation function in Eq. \eqref{strtc2}. Hence, in both cases the result doe not depend on the positions $i$ and $j$, and this suggests that these strange correlators do not display a different behavior for target states $\ket{\Psi}$ in the topological and non-topological phase.

These conclusions on the strange correlators can be generalized to the pure $\mathbb{Z}_2$ lattice gauge theory Hamiltonian defined by:
\begin{equation}
H_{LGT}(h) = H_{\textrm{TC}} -h \sum_l \hat{\sigma}^x_l\,,
\end{equation}
where we take $J_A \to \infty$. For increasing $h$, this Hamiltonian interpolates between the topological and non-topological phases: for $h=0$, its ground-state is $\ket{\Psi_{TC}}$, and for $h\to \infty$ the ground-state is given by $\Omega$. One can consider, in general, the ground-state of the model $\ket{\Psi(h)}$. Also in this case, however, the two-point strange correlators that do not fulfill gauge invariance vanish, whereas for the $\hat{\sigma}^x$ correlators we obtain again:
\begin{equation}
\frac{\bra{\Omega} \hat{\sigma}^x_i \hat{\sigma}^x_j \ket{\Psi(h)}}{\bra{\Omega} \Psi(h) \rangle} = 1\,,
\end{equation}
independently on $h$. 

Finally, we observe that similar observations concern the possibility of introducing string-like strange correlators. Each Pauli operator $\hat{\sigma}^a$, with $a=x,z$, creates indeed two excitations in the toric code (either electric or magnetic). Therefore the previous correlators involve four excitations in total. In order to evaluate strange correlators introducing only pairs of excitations, we may extend the two point correlation function to Wilson strings $\prod_{j \in \Gamma} \hat{\sigma}^z_j $ over a line $\Gamma$ of the square lattice, or 't Hooft strings $\prod_{j \in \tilde{\Gamma}} \hat{\sigma}^x_j $ over a line $\tilde{\Gamma}$ on the dual lattice. In both cases, however, also these string-like strange operators cannot be adopted to determine the onset of topological order: the Wilson strings, as in the case of the $\hat{\sigma}^z$ operators, violate the local gauge invariance in its end points; the 't Hooft strings, instead, acquire a trivial value on the state $\ket{\Omega}$,  as the the $\hat{\sigma}^x$ operators.

The negative results of the strange correlators concerning the toric code, may indeed be related to the fact that such state does not possess gapless edge modes. Therefore the conformal field theory construction adopted to derive the relation in Eq. 
\eqref{map}
cannot be applied in this non-chiral state.
\end{appendix}

\end{document}